\documentclass[aps,pra,onecolumn,superscriptaddress,amsmath,amssymb,eqsecnum,nofootinbib]{revtex4-1}
\usepackage[breaklinks=true]{hyperref}
\hypersetup{
  colorlinks   = true, 
  urlcolor     = blue, 
  linkcolor    = blue, 
  citecolor   =  red 
}

\usepackage{todonotes}

\makeatletter
\def\@bibdataout@aps{%
 \immediate\write\@bibdataout{%
  @CONTROL{%
   apsrev41Control,author="08",editor="1",pages="0",title="0",year="1",eprint="1"%
  }%
 }%
 \if@filesw
  \immediate\write\@auxout{\string\citation{apsrev41Control}}%
 \fi
}%
\makeatother 

\usepackage{graphicx,color}
\usepackage{epstopdf,bbm}
\usepackage{cooltooltips}
\usepackage{epsfig,tikz}
\usepackage{bm}
\usetikzlibrary{3d,calc,shapes,shapes.geometric}
\usepackage{MnSymbol}

\usepackage{lipsum}

\usepackage[caption=false,subrefformat=simple,labelformat=simple,listofformat=subsimple]{subfig}

\usepackage{dsfont} 
\usepackage[capitalise]{cleveref}

\DeclareMathOperator{\Tr}{Tr}
\newcommand{\prproj}[1]{\Pi^{(\text{pr})}_{#1}}

\newcommand{\pr}{\Pr}

\newcommand{\defined}{\mathrel{\mathop:}=}
\newcommand{\defines}{=\mathrel{\mathop:}}
\newcommand{\tpm}{\mathbin{\widetilde{\pm}}}
\newcommand{\tmp}{\mathbin{\widetilde{\mp}}}
\newcommand{\tplus}{\mathbin{\widetilde{+}}}
\newcommand{\tminus}{\mathbin{\widetilde{-}}}
\newcommand{\sqrtdt}{\sqrt{\!\dt}\,}

\newcommand{\comm}[2]{\left[#1,#2\right]}

\newcommand{\plant}{\ensuremath{_\text{sys}}}
\newcommand{\probe}{\ensuremath{_\text{pr}}}
\newcommand{\interaction}{\ensuremath{_\text{interaction}}}

\newcommand{\D}[1]{\mathcal{D}\sq{#1}}
\newcommand{\Hc}[1]{\mathcal{H}\sq{#1}}
\newcommand{\G}[1]{\mathcal{G}\sq{#1}}

\newcommand{\dg}{^\dagger}
\newcommand{\smallfrac}[2]{\mbox{$\frac{#1}{#2}$}}
\newcommand{\bra}[1]{\left\langle{#1}\right\vert}
\newcommand{\ket}[1]{\left\vert{#1}\right\rangle}
\newcommand{\oprod}[2]{\left\vert{#1}\middle\rangle\middle\langle{#2}\right\vert}
\newcommand{\expt}[1]{\left\langle{#1}\right\rangle}
\newcommand{\Expt}[1]{\mathbb{E}\left[#1\right]}
\newcommand{\sand}[3]{\left\langle{#1}\middle\vert{#2}\middle\vert{#3}\right\rangle}
\newcommand{\nsand}[3]{\langle{#1}\vert{#2}\vert{#3}\rangle}
\newcommand{\iprod}[2]{\left\langle{#1}\middle\vert{#2}\right\rangle}
\newcommand{\sch}{Schr\"odinger}

\newcommand{\half}{\smallfrac{1}{2}}

\newcommand{\sq}[1]{\left[{#1}\right]}
\newcommand{\tr}[1]{\Tr\sq{{#1}}}
\newcommand{\ptr}[2]{\Tr_{#1}\sq{{#2}}}
\newcommand{\df}{d} 
\newcommand{\Id}{\mathds{1}}
\newcommand{\abs}[1]{\left\vert#1\right\vert}
\newcommand{\dt}{\Delta\tau} 
\newcommand{\ie}{\emph{i.e.}}
\newcommand{\eg}{\emph{e.g.}}
\newcommand{\innovation}{\mathcal{I}}
\newcommand{\BigO}{\mathcal{O}}

\begin{document}

\title{Qubit models of weak continuous measurements}
\author{Jonathan A. Gross}\email{jagross@unm.edu}
\affiliation{Center for Quantum Information and Control, University of New Mexico,
Albuquerque NM 87131-0001, USA}
\author{Carlton M. Caves}\email{ccaves@unm.edu}
\affiliation{Center for Quantum Information and Control, University of New Mexico,
Albuquerque NM 87131-0001, USA}
\author{Gerard J. Milburn}\email{milburn@physics.uq.edu.au}
\affiliation{Centre for Engineered Quantum Systems, School of Mathematics and
Physics, The University of Queensland, St.~Lucia QLD 4072, Australia}
\author{Joshua Combes}\email{joshua.combes@gmail.com}
\affiliation{Center for Quantum Information and Control, University of New Mexico,
Albuquerque NM 87131-0001, USA}
\affiliation{Centre for Engineered Quantum Systems, School of Mathematics and
Physics, The University of Queensland, St.~Lucia QLD 4072, Australia}
\affiliation{Institute for Quantum Computing, Department of Applied Mathematics, University of Waterloo, Waterloo, ON, Canada}
\affiliation{Perimeter Institute for Theoretical Physics, 31 Caroline St.~N, Waterloo, Ontario, Canada N2L 2Y5}
\date{\today}

\begin{abstract}
In this paper we approach the theory of continuous measurements and the
associated unconditional and conditional (stochastic) master equations from the
perspective of quantum information and quantum computing.  We do so by showing
how the continuous-time evolution of these master equations arises from
discretizing in time the interaction between a system and a probe field and by
formulating quantum-circuit diagrams for the discretized evolution.  We then
reformulate this interaction by replacing the probe field with a bath of qubits,
one for each discretized time segment, reproducing all of the standard
quantum-optical master equations. This provides an economical formulation of the
theory, highlighting its fundamental underlying assumptions.
\end{abstract}
\maketitle

\section{Introduction and Motivation}\label{sec:intro_motivation}

The strength of a projective measurement is made known in weakness. Although we
are taught from youth that quantum measurements project the target system onto
an eigenstate of the measured observable as an irreducible action, closer
inspection reveals a more nuanced reality. Measurements involve coupling quantum
systems to macroscopic devices via finite-energy interactions, these devices
have finite temporal resolution, and a host of imperfections lead to encounters
with the classical world that violate unitarity without conforming to the
projective-measurement mold.

Of course, in many scenarios these discrepancies are fleeting and the projective
description is all that is needed---and sometimes all that can be observed!
Modern experiments, however, show projective measurements for what they are, and
if we are to glory in this revelation, we need tools like the theory of quantum
trajectories, which generalizes measurement projection to weak, continuous
monitoring of a quantum system.

Consider using a transition-edge sensor to detect photons. As a photon is
absorbed by the detector, the output current begins to drop. At first it is
difficult to tell the difference between a photon and thermal fluctuations, but
as the current continues to drop we become more and more confident of the
detection prognosis until we have integrated enough current deficiency to
announce a detection.  The accumulating current deficiency is the result of a
continuous sequence of \emph{weak measurements\/} (sometimes called
\emph{gentle\/} or \emph{fuzzy\/}), where the name signifies that each
measurement outcome (in this case output current integrated over a short time
interval) contains little information about the system being measured and
consequently only gently disturbs that system. Many repetitions of such weak
measurements, however, do have an appreciable effect upon the system, sometimes
as dramatic as a projective measurement.  Because these measurements are nearly
continuous, differential equations are used to track the cumulative effect on
the system, and because quantum theory tells us the measurement results are
random, these differential equations are stochastic.

The system's time-dependent state (or some expectation value thereof)
conditioned on a continuous measurement record is called a \emph{quantum
trajectory\/} in the continuous-measurement literature. Physically, this
continuous measurement record is written on successive probes that interact
weakly with the system. The stochastic differential equations that generate
quantum trajectories take a variety of forms, going by names such as
\emph{stochastic \sch\ equations}, \emph{quantum-filtering equations}, or in
this paper \emph{stochastic master equations\/} (SMEs).  A great deal of
attention has been devoted both to deriving stochastic equations for and to
observing trajectories in a variety of physical systems, \eg, cavity QED
\cite{HoodChapLynn98}, circuit QED (superconducting
systems)~\cite{GambBlaiBois08,BoisGambBlai08,MurcWebeMack13}, fermionic
systems~\cite{GoanMilb01,OxtoWarsWise05,OxtoGambWise08}, and mechanical
systems~\cite{MilbJacoWall94,HopkJacoHabi03,RuskSchwKoro05,JacoLougBlen07}.

The ability to resolve these subprojective effects opens up many possibilities,
including feedback protocols and continuous-time parameter estimation. An
example of feedback control is continuous-time quantum error correction.
\citet{AhnDoheLand02} investigated using continuous-time quantum measurements
for this purpose, thus pioneering a fruitful line of research
\cite{SaroAhnJaco04,AhnWiseJaco04,GregWern04,HandMabu05,ChasLandGere08,Mabu09a,LidaBrun13,NguyHillHoll15,HsuBrun16}.
Feedback control additionally allows one to view weak measurements as building
blocks for constructing other generalized measurements, as explored by Brun and
collaborators~\cite{OresBrun05,VarbBrun07,FlorBrun15,FlorBrun14}. Continuous
weak measurements have also been pressed into service for parameter and state
estimation
\cite{Mabu96,GambWise01,ChasGere09,GammMolm13,KiilMolm14,KiilMolm16,GongCui17}.
One notable example is the single shot tomography of an ensemble of identically
prepared qubits \cite{CookRiofDeut14}.

Error correction, parameter estimation, and state tomography are important
subjects in quantum computation and information. Unfortunately, much of the
literature on continuous weak measurements, which would otherwise be of interest
to this community, suffers from needlessly arcane terminology and
interpretations. We take the refiner's fire to trajectory theory, revealing a
foundation of finite-dimensional probe systems, unitary gates between the system
and successive probes, and quantum operations to describe the system state after
the probe is measured---all three familiar to the quantum information scientist
of today. This process also distills the essence of trajectory theory from its
origins in field-theoretic probes, yielding insights that can be appreciated
even by veterans of the subject.  A particularly useful tool that arises
naturally within our approach is the quantum circuit diagram, and we take pains
throughout our presentation to illustrate relevant principles with this tool.

Of all the prior work on this subject, our paper is most related to---and indeed
inspired by---Brun's elegant work on qubit models of quantum
trajectories~\cite{Brun02}. In \cref{sec:meas,sec:qubit-vac} we describe the
connection between his work and ours. Looking further back to the origin of this
line of research, one might identify an important precedent in the
work~\cite{SculLamb67} of the great theorists, Scully and Lamb (Lamb also did
experiments), in which they considered systems interacting with a spin bath. The
mathematics literature has a related body of work that studies approximating
Fock spaces with chains of qubits known as ``toy Fock
spaces''~\cite{Meye86,Atta02,GougSobo04,Goug04,AttaPaut05,AttaPaut06,BELT08,BoutHandJame09,AttaNech11}.

The physics and mathematical-physics communities have a rich history of deriving
the stochastic equations of motion for a system subject to a continuous
measurement. So rich, in fact, that these equations have been discovered and
rediscovered many times. Historically, the theory was developed in the 80's and
early 90's by a number of authors: \citet{Mens79,Mens93}, \citet{Bela80},
\citet{SrinDavi81}, \citet{BragBragKhal95}, \citet{BarcLanzPros82},
\citet{Gisi84}, \citet{Dios86,Dios88}, Caves~\cite{Cave86,Cave87}, Caves and
Milburn~\cite{CaveMilb87}, Milburn~\cite{WiseMilb93}, \citet{Carm93a},
\citet{DaliCastMolm92}, and Wiseman~\cite{WiseMilb93}.
Most recently \citet{Koro99,Koro01} developed a treatment of continuous measurements;
this treatment, which Korotkov dubbed ‘quantum Bayesian theory,’  is notable in that it is based on baths and detection processes that are characteristic of condensed systems.

Many other good references on the topic are available for the interested reader.
We recommend the following articles: \citet{Brun02}, \citet{JacoStec06},
Wiseman's PhD thesis \cite{Wise94}, and for the mathematically inclined
reader,~\citet{BoutHandJame09,BoutHandJame07}.  Helpful books include \cite{Carm93a},
\cite{Carm08}, \cite{WiseMilb10}, and \cite{Jaco14}.

This paper is structured as follows: \Cref{sec:notation} lays out our notational
conventions. \Cref{sec:meas} gives a unified description of strong and weak
measurements via ancilla-coupled measurements, followed by quantum-circuit
depictions of the iterated interactions that limit to continuous quantum
measurements and their relation to Markovicity. \Cref{sec:qbittraj}
develops the continuous-measurement theory in terms of a system undergoing
successive weak interactions with a probe field.

\Cref{sec:qubit-vac,sec:gaussian-states,sec:interaction-rand} are the heart of
the paper: they show how to replace a probe field with probe qubits in
constructing quantum trajectories, and they explore the consequences of changing
the parameters of the formalism, i.e., using different probe initial states,
different interaction unitaries, and different measurements on the probes.
\Cref{sec:qubit-vac} contains the first derivation of a SME in our model,
focusing on the vacuum SMEs.  These arise when probe initial states are vacuum
(ground state for probe qubits); the probe undergoes a weak interaction with the
system and then experiences one of several kinds of measurements, which lead to
different quantum trajectories.  The theme of \cref{sec:qubit-vac} is thus
exploring the effect of different kinds of measurements on the probes.
\Cref{sec:gaussian-states} considers the Gaussian SMEs, in which a probe field
starts in a Gaussian state, undergoes a weak interaction with the system, and
then is subjected to homodyne measurements.  The theme of this section is thus
exploring the effect of different probe initial states, but the main
contribution of this section is a technique to accommodate all the Gaussian
field states in probe qubits and to show that, since qubits have too small a
Hilbert space to achieve this by only changing the initial state, one must also
modify aspects of the weak system/probe interaction unitary.
\Cref{sec:interaction-rand} explores a radical departure that allows
interactions between the probe qubits and the system that are strong, but occur
randomly.

As an aid to intuition, Sec.~\ref{sec:examples-and-code} presents visualizations
of numerical solutions to some of the SMEs derived in the previous sections.
Finally, Sec.~\ref{sec:discuss-conclusion} summarizes lessons learned from our
approach and suggests promising related approaches.

\section{Notational Conventions}\label{sec:notation}

Confusion can arise when denoting the states of quantum-field modes and
two-level systems (qubits) in the same context.  In particular, that
$a\ket{n}=\sqrt{n}\ket{n-1}$ and thus $a\ket{1}=\ket{0}$, yet
$\sigma_-\ket{0}=\ket{1}$, can lead to momentary confusion and even persistent
perplexity.  The standard qubit states are the eigenstates of
$\sigma_z=\oprod{0}{0}-\oprod{1}{1}=\sum_{a=0,1}(-1)^a\oprod{a}{a}$; since the qubit
Hamiltonian is often proportional to $\sigma_z$---this is why one chooses
$\ket0$ and $\ket1$ to be the standard states---it is natural to regard $\ket1$
(eigenvalue $-1$ of $\sigma_z$) as the ground state and $\ket0$ (eigenvalue $+1$)
as the excited state.  In doing so, one is allowing the multiplicative label $(-1)^a$
to trump the bitwise label $a$, which gives an opposite hint for what should
be labeled ground and excited.

\begin{figure}[ht!]
  \begin{center}
    \includegraphics[scale=0.9]{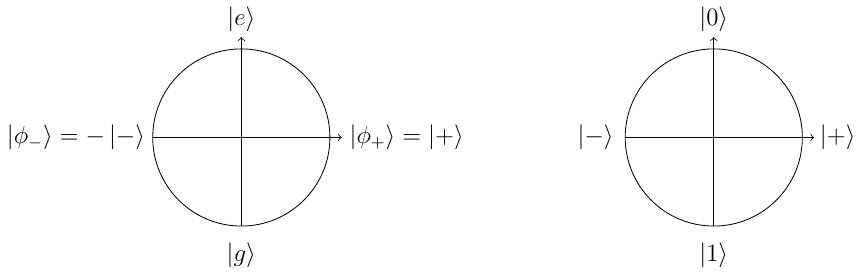}
  \end{center}
  \caption{Bloch-sphere illustration ($z$ the vertical axis, $x$ the horizontal axis, $y$ direction suppressed) of our convention for qubit states (left) and the conventional quantum-information notation (right).  In conventional notation,
  the eigenstates of $\sigma_x$ with eigenvalue $\pm1$ are denoted by $\ket\pm=(\ket0\pm\ket1)/\sqrt2$, but in our qubit notation, we use the eigenstates $\ket{\phi_\pm}=(\ket g\pm\ket e)/\sqrt2=(\ket1\pm\ket0)/\sqrt2=\pm\ket\pm$; \ie, we change the sign of the eigenstate with eigenvalue $-1$.}
  \label{fig:eg-vs-01-convention}
\end{figure}

To allay this confusion, one good practice would be to label the standard qubit
states by the eigenvalue, $(-1)^a$, of $\sigma_z$, but instead we choose the
more physical labeling of $\ket g=\ket 1$ as the ``ground state'' and
$\ket e=\ket0$ as the ``excited state.''   In this notation, $\sigma_-\ket{e}=\ket{g}$,
as expected; this notation plays well with the correspondence we develop between
field modes and two-level systems.  Our notation is illustrated in
\cref{fig:eg-vs-01-convention}.  As a further check on confusion, we often label
the vacuum state of a field mode as $\ket{\rm vac}$ instead of $\ket0$.

Some useful relations between qubit operators are given below:
\begin{align}
\begin{split}
  \sigma_+&=\oprod{e}{g}=\oprod{0}{1}=\half(\sigma_x+i\sigma_y)\,,\\
  \sigma_-&=\oprod{g}{e}=\oprod{1}{0}=\half(\sigma_x-i\sigma_y)\,,\\
  \sigma_x&=\sigma_++\sigma_-=\oprod{e}{g}+\oprod{g}{e}\,,\\
  \sigma_y&=-i\sigma_++i\sigma_-=-i\oprod{e}{g}+i\oprod{g}{e}\,,\\
  \sigma_z&=\oprod{e}{e}-\oprod{g}{g}\,,\\
  [\sigma_-,\sigma_+]&=\half i[\sigma_x,\sigma_y]=-\sigma_z\,.
  \label{eq:operator-relations}
\end{split}
\end{align}

When writing qubit operators and states in their matrix representations, we
order the rows and columns starting from the top and left with $\ket{e}=\ket0$
followed by $\ket{g}=\ket1$.  Thus $\sigma_-=\oprod{g}{e}$ has the
representation
\begin{align}
\bordermatrix{~       & \bra{e} & \bra{g} \cr
              \ket{e} & 0       & 0       \cr
              \ket{g} & 1       & 0       \cr}\,.
\end{align}

The first place our notation has the potential to confuse is in how we denote
the eigenstates of $\sigma_x$.  These eigenstates are conventionally written as
$\ket\pm=(\ket0\pm\ket1)/\sqrt2$, but we choose to denote them by
\begin{align}\label{eq:Xbasis-def}
\ket{\phi_\pm}\defined\frac{1}{\sqrt2}\big(\ket{g}\pm\ket{e}\big)
=\frac{1}{\sqrt2}\big(\ket1\pm\ket0)=\pm\ket\pm\,;
\end{align}
\ie, we change the sign of the eigenstate with eigenvalue $-1$.  This notation
is illustrated in \cref{fig:eg-vs-01-convention}.

In our circuit diagrams, each wire corresponds to an individual system; a
collection of those wires corresponds to a tensor product of the systems.  To
keep track of the various systems when moving between circuit and algebraic
representations, the tensor-product order equates systems left-to-right in
equations with the systems bottom-to-top in the circuits.  We also reserve the
leftmost/bottom position for the system in our discussions, putting the probe
systems to the right/above.  True to conventional quantum-circuit practice,
single wires carry quantum information (\ie, systems in quantum states), whereas
double wires carry classical information (typically measurement outcomes).

We use the notation $\Expt{\Delta A}$ to denote the expectation value of a
classical random variable $\Delta A$, which need not correspond to a Hermitian
observable. Typically $\Delta A$ can be thought of as a map from measurement
outcomes to numbers, in which case sampling from $\Delta A$ involves performing
said measurement and mapping the outcome to the appropriate value. For a
measurement defined by a POVM $\{E_j\}$ (see Sec.~\ref{sec:kraus}) and
corresponding random-variable values denoted by $\Delta A_j$, the expectation
value evaluates to
\begin{align}
  \Expt{\Delta A}&=\sum_j\Delta A_j\tr{\rho E_j}\,.
  \label{eq:exptDeltaA}
\end{align}
The implicit dependence on quantum state $\rho$ and measurement POVM $\{E_j\}$
should be clear from context.

\section{Measurements and the quantum-circuit depiction}\label{sec:meas}
\subsection{Indirect and weak measurements}\label{sec:indie_meas}

The instantaneous direct measurement of quantum systems, still the staple of
many textbook discussions of quantum measurement, is only a convenient fiction.
As discussed in the Introduction, one typically makes a measurement by coupling
the system of interest to an ancillary quantum system prepared in a known state
and then measuring the ancilla.  This is called an \emph{indirect\/} or
\emph{ancilla-coupled\/} measurement.  For brevity we refer to the system of
interest as the \emph{system}.  Although the ancillary system goes by a variety
of names in the literature, we refer to such systems here as \emph{probes} to
evoke the way they approach the system to interrogate it and depart to report
their findings.  When additional clarity is helpful, we use subscripts to
identify states with various systems, so $\ket{\psi}\plant$ and $\rho\plant$
designate system states and $\ket{\phi}\probe$ and $\sigma\probe$ designate
probe states.

\begin{figure}[ht!]
  \vspace{2em}
  \begin{center}
    \includegraphics[scale=.75]{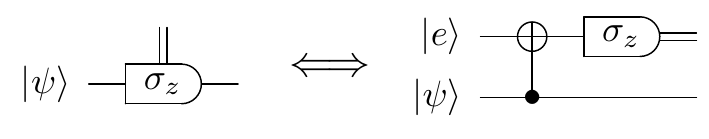}
  \end{center}
  \caption{Equivalence between a direct (left) and ancilla-coupled (right)
  measurement of $\sigma_z$.  Note that for the CNOT gate in the
  ancilla-coupled measurement, the application of the NOT gate to the probe is
  controlled on $\ket{g}=\ket1$, as shown algebraically in Eq.~(\ref{eq:CNOT}).
  The single wires carry systems in quantum states, while the double wires
  carry classical information. In both the direct and the ancilla-coupled
  measurement, the double wire emerging from the measurement apparatus carries
  the result of the measurement, either $e$~(0) or $g$~(1).  After the
  measurement, the system is left in the corresponding state, $\ket e$~($\ket0$)
  or $\ket g$~($\ket1$); this state is carried by the system wire emerging from
  the right of the measurement apparatus in the direct measurement and by the
  system wire proceeding to the right in the ancilla-coupled version.}
  \label{fig:dir-ind-meas-equiv}
\end{figure}

Ancilla-coupled measurements can be used to effect any generalized measurement,
including the direct measurements of textbook lore. Suppose one wants to measure
$\sigma_z$ on a qubit system. This can be accomplished by preparing a probe
qubit in the state $\ket{e}$, performing a controlled-NOT (CNOT) gate from the
system to the probe, and finally measuring $\sigma_z$ directly on the probe. The
CNOT gate is defined algebraically as
\begin{align}\label{eq:CNOT}
\begin{split}
  \mathrm{CNOT}&\defined\oprod{e}{e}\otimes\Id+\oprod{g}{g}\otimes\sigma_x \\
  &=\oprod{0}{0}\otimes\Id+\oprod{1}{1}\otimes\sigma_x\,.
\end{split}
\end{align}
Doing nothing when the probe is in the excited state might feel strange, but
this convention is chosen to harmonize with the quantum-information notation
that is shown in the second form of Eq.~(\ref{eq:CNOT}), in which the NOT gate
($\sigma_x$) is applied to the probe when the system is in the state $\ket1=\ket
g$; this is called control on $\ket1$ or, in this context, control on $\ket g$.
\Cref{fig:dir-ind-meas-equiv} depicts in quantum circuits the equivalence
between a direct measurement of $\sigma_z$  and the ancilla-coupled
measurement.

For an arbitrary initial system state
\begin{align}
\ket{\psi}\plant\defined\alpha\ket{g}\plant+\beta\ket{e}\plant\,,
\end{align}
the joint state of the system and probe after the interaction is
\begin{equation}
\begin{split}
  \ket{\Psi}&\defined\mathrm{CNOT}\ket{\psi}\plant\otimes\ket{e}\probe\\
  &=\alpha\ket{gg}+\beta\ket{ee}\,.
\end{split}\label{eq:joint-state}
\end{equation}
Local $\sigma_z$ measurements on the probe are described by the projectors
$\prproj{g}\defined\Id\otimes\oprod{g}{g}$ and
$\prproj{e}\defined\Id\otimes\oprod{e}{e}$ (the superscript indicates
projection only on the probe).  These measurements give the following
probabilities and post-measurement system states:
\begin{align}
  \pr(g)&=\nsand{\Psi}{\prproj{g}}{\Psi}=\abs{\alpha}^2\;,
  &\frac{\ptr{\textrm{pr}}{\prproj{g}\ket{\Psi}\!\bra{\Psi}\prproj{g}}}{\pr(g)}
  &=\ket{g}\plant\!\bra{g}\,,\\
  \pr(e)&=\nsand{\Psi}{\prproj{e}}{\Psi}=\abs{\beta}^2\;,
  &\frac{\ptr{\textrm{pr}}{\prproj{e}\ket{\Psi}\!\bra{\Psi}\prproj{e}}}{\pr(e)}
  &=\ket{e}\plant\!\bra{e}\,.
  \label{eq:proj}
\end{align}
These are the same probabilities and post-measurement system states as for a
direct measurement of $\sigma_z$ on the system. This equivalence comes about
because the CNOT gate produces perfect correlation in the standard
qubit basis.

More general interactions between the system and probe do not produce
perfect correlation. A specific example of an imperfectly correlating interaction,
\begin{align}\label{eq:theta_CNOT}
\begin{split}
  U_\text{CNOT}(\theta)&\defined\exp\left(-i\theta\,\mathrm{CNOT}\right) \\
  &=\cos\theta\,\Id\otimes\Id-i\sin\theta\,\mathrm{CNOT}\,,
\end{split}
\end{align}
was presented by Brun~\cite{Brun02}; $\theta=0$ gives the identity, \ie, no correlation between
system and probe, and $\theta=\pi/2$ gives (up to the global phase $-i$) CNOT, \ie, perfect correlation
between system and probe. For $0<\theta<\pi/2$ the probe becomes partially
correlated with the system. This kind of partial CNOT can be constructed
because the CNOT gate is Hermitian as well as unitary, and therefore generates
unitary transformations. The joint state of the system/probe after the
interaction is
\begin{align}
\begin{split}
  \ket{\Psi_\theta}
  &\defined U_\text{CNOT}(\theta)\ket{\psi}\plant\otimes\ket{e}\probe \\
  &=\cos\theta\ket{\psi}\plant\otimes\ket{e}\probe-i\sin\theta\ket\Psi \\
  &=\beta e^{-i\theta}\ket{ee}+\alpha\cos\theta\ket{ge}
  -i\alpha\sin\theta\ket{gg}\,.
\end{split}
\label{eq:weak-int-joint-state}
\end{align}

A projective measurement on the probe after the interaction gives only partial
information about the system and thus only partially projects the system state.
As explained in the Introduction, such measurements have been called
\emph{weak}, \emph{fuzzy}, or \emph{gentle}. These measurements should not be
equated with \emph{weak values}~\cite{AharAlbeVaid88,DuckStevSuda89}, a
derivative concept utilizing weak measurements but with no additional relation
to the continuous-measurement schemes we consider. The outcome probabilities
and post-measurement system states are
\begin{align}
  \pr(g)&=\nsand{\Psi_\theta}{\prproj{g}}{\Psi_\theta}
  =\abs{\alpha}^2\sin^2\!\theta\;,
  &\frac{\ptr{\textrm{pr}}{\prproj{g}\ket{\Psi_\theta}\!\bra{\Psi_\theta}
  \prproj{g}}}{\pr(g)}&=\ket{g}\plant\!\bra{g}\,,\\
  \pr(e)&=\nsand{\Psi_\theta}{\prproj{e}}{\Psi_\theta}
  =\abs{\beta}^2+\abs{\alpha}^2\cos^2\!\theta\;,
  &\frac{\ptr{\textrm{pr}}{\prproj{e}\ket{\Psi_\theta}\!\bra{\Psi_\theta}
  \prproj{e}}}{\pr(e)}&=\ket{\chi}\plant\!\bra{\chi}\,,
  \label{eq:weak-post-meas}
\end{align}
where
\begin{align}
  \ket{\chi}\plant&=\frac{\alpha\cos\theta\ket{g}\plant
  +\beta e^{-i\theta}\ket{e}\plant}{\sqrt{\abs{\alpha}^2\cos^2\!\theta
  +\abs{\beta}^2}}\,.
\end{align}

For $\theta\ll1$, we can expand these results to second order in $\theta$ to
see more clearly what is going on in the case of a weak measurement. The
outcome $e$ is very likely, occurring with probability
$\pr(e)\simeq1-\abs{\alpha}^2\theta^2$, and when this outcome is observed, the
post-measurement state of the system is almost unchanged from the initial state
\begin{align}
  \ket{\chi}\plant\simeq\alpha\big(1-\half\abs{\beta}^2\theta^2\big)
  \ket{g}\plant+\beta\big(1-i\theta-\half\abs{\beta}^2\theta^2\big)
  \ket{e}\plant\,.
\end{align}
In contrast, the outcome $g$ is very unlikely, occurring with probability
$\pr(g)\simeq\abs{\alpha}^2\theta^2$, and when this outcome is observed, the
system is projected into the state $\ket{g}\plant$, which can be very different
from the initial state. This kind of weak measurement can be thought of as
usually providing very little information about the system, but occasionally
determining that the system is in the ground state~\cite{ZollMartWall87}.

\begin{figure}[ht!]
  \begin{center}
    \subfloat[]{
        \includegraphics{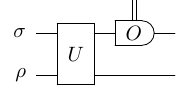}
        \label{fig:gen-anc-meas-O}
    }
    \qquad\qquad
    \subfloat[]{
        \includegraphics{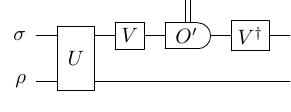}
        \label{fig:gen-anc-meas-Oprime}
    }\\
    \subfloat[]{
        \includegraphics{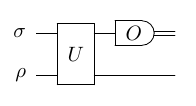}
        \label{fig:gen-anc-meas-O-simplified}
    }
    \qquad\qquad
    \subfloat[]{
        \includegraphics{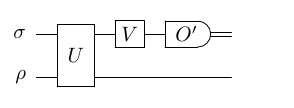}
        \label{fig:gen-anc-meas-Oprime-simplified}
    }
  \end{center}
  \caption{General ancilla-coupled measurement.  System in initial state $\rho$
  and probe in initial state $\sigma$ are subjected to an interaction unitary
  $U$. (a)~Probe is measured in the eigenbasis of an observable $O$;
  (b)~equivalently, by including a basis-changing unitary $V$ in the circuit,
  the measurement of $O$ is replaced by a measurement in the eigenbasis of a
  standard observable $O'$ related to the original observable by $O=V\dg O'V$;
  (c)~same as (a), except that the post-measurement state of the probe is
  discarded, there being no further use for the probe; (d)~same as (b), except
  that the post-measurement state of the probe is discarded.}
  \label{fig:gen-anc-meas}
\end{figure}

\subsection{Quantum-circuit description of measurements}\label{sec:qc}

In the most general ancilla-coupled-measurement scheme, the system is initially
in a (possibly mixed) state $\rho$ and the probe begins in the (possibly mixed)
state $\sigma$. System and probe interact via an interaction unitary $U$ and
then the probe is measured in the eigenbasis of an observable $O$.  We
illustrate and elaborate on this scheme in \cref{fig:gen-anc-meas}.

Because a weak measurement extracts partial information and thus only partially
projects the system onto an observed eigenstate, we can learn more about the
system by performing repeated weak measurements (contrast this with a projective
measurement, where one gains no new information by immediately repeating the
measurement).  One method of extracting all the available information about the
system is to repeat a weak measurement many times. Such iterated weak
measurements are explored in more detail in Sec.~\ref{sec:qbittraj}.

\begin{figure}[ht!]
  \begin{center}
    \subfloat[]{
      \includegraphics{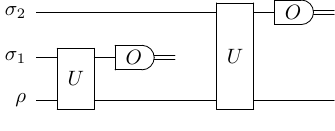}
      \label{fig:iter-meas-clumsy}
    }
    \qquad
    \subfloat[]{
      \includegraphics{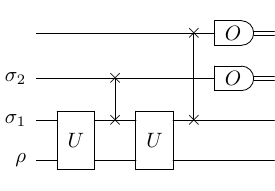}
      \label{fig:iter-meas-neat}
    }
        \qquad
    \subfloat[]{
      \includegraphics{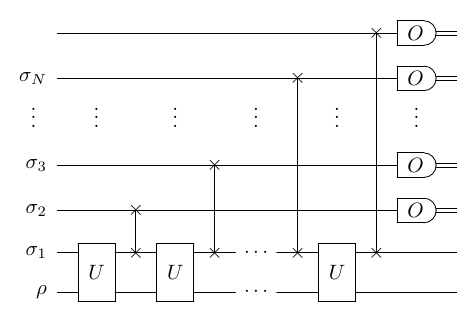}
      \label{fig:iter-meas-itt}
    }
  \end{center}
  \caption{Circuit representations of repeated measurements. (a)~This
  straightforward representation quickly becomes unwieldy as more probes are
  added to the diagram.  (b)~The straightforward depiction is cleaned up by
  using a SWAP gate to move the probe destined to interact next with the system
  onto the wire closest to the system for the interaction and then, after the
  probe's interaction, another SWAP gate to move it onto the wire just above
  its initial wire, ready to be measured.  (c)~Use of the SWAP-gate trick
  allows one easily to depict the repetitive interaction of $N$ probes with the
  system.  Readers familiar with circuit diagrams might find this usage
  confusing at first, but with a little practice, will come to appreciate both
  its convenience and its manifestly iterative depiction of the initial probe
  states, of the probes' interactions with the system, and of the measurements
  on the probes.  Indeed, (c)~depicts clearly the essential elements of
  Markovian system evolution: the separate probe states on the left, the
  separate probe interactions on the bottom two wires, and the separate probe
  measurements on the right.}
  \label{fig:iter-meas}
\end{figure}

We introduce a circuit convention in \cref{fig:iter-meas} that makes it easy to
depict iterated measurements. The na\"ive depiction,
\cref{fig:iter-meas-clumsy}, is clumsy and distracts from the repetitive
character of the probe interactions. For the remainder of the paper, we employ a
cleaner convention by reserving one probe wire (usually the one nearest to the
system) for all interactions with the system. We then use SWAP gates to bring
probes into and out of contact with the system as necessary.  Thus the circuit
in \cref{fig:iter-meas-clumsy} transforms to \cref{fig:iter-meas-neat}.
Generally, the SWAP trick leads to circuit diagrams like
\cref{fig:iter-meas-itt}. The SWAPs in all cases are purely formal and used
only for convenience.

The SWAP trick works because our system is distinct from the probes in an
important way.  We are assuming that the system is persistent and not directly
accessible---\ie, we cannot directly measure or swap the state of the
system---while the probes are transient, interacting with the system once and
then flying away to be measured.  In \cref{fig:iter-meas} we have included
subscripts to individuate the probes, although we often omit these
designations since the circuit wire already contains this information---\eg, in
a circuit diagram, we can drop the probe designation $n$ from $\sigma_n$ since
the diagram tells us which probe this density operator describes.

Under the repetitive measurements depicted in \cref{fig:iter-meas-itt}, the
system undergoes a conditional dynamics, where the conditioning is on the
results of the measurements on the probes.  Discarding the results of the
measurements on the probe is equivalent to not doing any measurements on the
probe, and then the system dynamics are the unconditional open-system dynamics
that come from tracing out the probes after they interact with the system.

The circuit diagram in \cref{fig:iter-meas-itt} can be thought of as depicting
probes that successively and separately scatter off the system and then are
measured to extract the information picked up from the system in the scattering
event.  Indeed, the diagrams highlight the essential assumptions behind the
Markovian system evolution that comes with this sort of scattering.  Each
probe, in its own state, uncorrelated with the other probes, scatters off the
system and then flies away, never to encounter the system again; this happens,
for example, when a vacuum or thermal field scatters off the system and
propagates away to infinity.  The result is Markovian unconditional evolution;
to get Markovian conditional evolution, one requires in addition that the probes
be measured independently.  Markovian evolution is usually thought of in the
context of continuous time evolution, in which the interaction unitaries $U$
correspond to repetitive Hamiltonian evolution for infinitesimal time intervals
and thus are necessarily weak interactions that give rise to weak, continuous
measurements on the system.  Despite the importance of continuous time evolution and
continuous measurements, which are the focus of this paper, the circuit diagram
in \cref{fig:iter-meas-itt} allows one to see clearly what is involved in
Markovian evolution even for finite-time interaction events: the separate probe
states on the left, the separate probe interactions on the bottom two wires,
and the separate probe measurements on the right.  The circuit diagrams for
infinitesimal-time interactions are the foundation for the Markovian
input-output theory of quantum optics, which we consider in
Sec.~\ref{sec:gaussian-setup-fields}.

\begin{figure}[ht!]
  \begin{center}
    \subfloat[]{
      \includegraphics[scale=1]{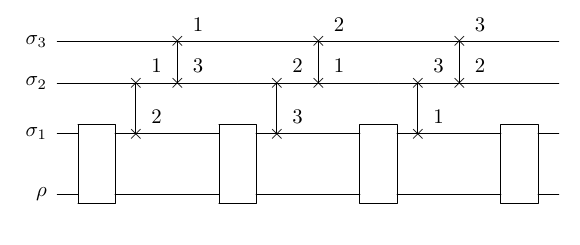}
      \label{fig:finite-environment-ckt}
    }
        \qquad
     \subfloat[]{
      \includegraphics[scale=1]{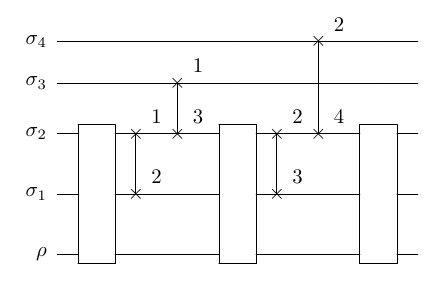}
      \label{fig:overlapping-probes-ckt}
    }
  \end{center}
  \caption{Two different scenarios that produce non-Markovian system dynamics by
  changing the probe-system interactions away from the Markovian pattern in
  \cref{fig:iter-meas-itt}: (a)~A finite environment (here consisting of three
  probes) forces the probes to return and interact repeatedly with the system.
  (b)~Successive probes simultaneously interact with the system, making it
  impossible to separate the environment into disjoint modes that individually
  interact with the system.}
  \label{fig:nonMarkov}
\end{figure}

Various modifications to the circuit diagram of \cref{fig:iter-meas} give
non-Markovian evolution.  One modification is to initialize the probes in a
correlated state, either via classical correlations or via the quantum
correlations of entanglement.  A second kind of modification, depicted in
\cref{fig:nonMarkov}, is to allow the system to interact with each probe
multiple times, by having a probe return and interact yet again after other
probes have interacted with the system, as in \cref{fig:finite-environment-ckt},
or to have a time window in which multiple probes interact with the system, as
in \cref{fig:overlapping-probes-ckt}.  The first of these is the general
situation when a finite environment interacts with the system; environment
``modes'' acting as probes never exit cleanly, so a mode can interact with the
system more than once.  We note that the methods developed
in~\cite{Cave86,Cave87} allow probes to overlap in the same time window and thus
might provide an avenue to describing non-Markovian dynamics.  Finally,
conditional evolution can be non-Markovian when one makes joint measurements,
instead of independent measurements, on the probes after they depart from the
system. This occurs when modeling finite detector bandwidth as discussed
in~\cite[Sec.~4.8.4]{WiseMilb10}.

\subsection{Conditional evolution and Kraus operators}
\label{sec:kraus}

Suppose that, as is depicted in \cref{fig:iter-meas-itt}, we cause the system,
initially in pure state $\rho=\oprod{\psi}{\psi}$, to interact sequentially with
$N$ probes, initially in the product state
$\sigma_1\otimes\cdots\otimes\sigma_N$, where we assume, for the moment, that
the initial probe states are pure, \ie, $\sigma_n=\ket{\phi}_n\!\bra{\phi}$.
The interaction of the $n$th probe with the system is described by the unitary
operator $U^{(n)}$, and after the interaction, we measure the observable $O$ on
each probe, obtaining outcomes $o_{j_1},\ldots,o_{j_N}$.  We want to calculate
probabilities for obtaining different sequences of measurement outcomes, as well
as the conditional quantum state of the system after observing a particular
sequence of outcomes.  These probabilities can be derived in a variety of ways,
some of which were explored in~\cite{Cave86,Cave87}, producing the following
expressions~\cite{CaveMilb87}: the probability for the outcome sequence is
\begin{align}\label{eq:post-iter-prob}
\pr(o_{j_1},\ldots,o_{j_N}\vert\psi)
=\iprod{\tilde{\psi}_N}{\tilde{\psi}_N}\,,
\end{align}
where
\begin{align}
 \label{eq:post-iter-unnorm-state}
  \ket{\tilde{\psi}_N}=\bra{o_{j_1}}_1\otimes\cdots\otimes\bra{o_{j_N}}_N
  U^{(N)}\cdots U^{(1)}\ket{\psi}\otimes\ket{\phi}_1\otimes\cdots\otimes\ket{\phi}_N
\end{align}
is the unnormalized system state at the end of the entire process and
\begin{align}
 \label{eq:post-iter-state}
  \ket{\psi_N}=\ket{\tilde{\psi}_N}\Big\slash
  \sqrt{\iprod{\tilde{\psi}_N}{\tilde{\psi}_N}}
\end{align}
is the corresponding normalized state after the process.  As the number of
probes increases, these expressions become pointlessly unwieldy, since in the
Markovian situation of \cref{fig:iter-meas-itt} we should be able to deal
with the probes one at a time.  The most efficient way to write the results is
to use the system-only formalisms of \emph{positive-operator-valued measures\/}
(POVMs) and \emph{quantum operations}, which were historically introduced as
\emph{effects\/} and \emph{operations}.

The ingredient common to both POVMs and quantum operations that gives us this
system-only description is the \emph{Kraus operator}, which we define in the
standard way using partial inner products:
\begin{align}
  K_j\defined\sand{o_j}{U}{\phi}\,.
  \label{eq:anc-kraus-ops}
\end{align}
As usual, these Kraus operators give rise to POVM elements,
\begin{align}
  E_j&\defined K_j\dg K_j\,,
\label{eq:kraus_op}
\end{align}
and the POVM elements resolve the identity,
\begin{align}
  \sum_j E_j=\Id\,.
  \label{eq:kraus_id_resol}
\end{align}
The POVM elements specify the quantum statistics of a generalized measurement
on the system. The conditional (unnormalized) state of the quantum system after
observing a single outcome $o_j$ is
\begin{align}
  \label{eq:unnorm-cond-state}
  \tilde{\rho}=\sand{o_j}{U(\rho\otimes\oprod{\phi}{\phi})U\dg}{o_j}
  =\sand{o_j}{U}{\phi}\rho\sand{\phi}{U\dg}{o_j}
\end{align}
and is thus described by a quantum operation constructed from the single Kraus
operator~$K_j$,
\begin{align}
  \label{eq:quantum-operation}
  \tilde{\rho}=K_j\rho K_j\dg\,.
\end{align}

One can easily see that the unnormalized system
state~(\ref{eq:post-iter-unnorm-state}) after observing a particular outcome
sequence~is
\begin{align}
\ket{\tilde{\psi}_N}=K_{j_N}\cdots K_{j_1}\ket{\psi}\;.
\end{align}
Writing this in terms of the system's initial density operator---allowing us to
accommodate mixed initial system states---we get the unnormalized final system
state
\begin{align}
  \tilde\rho_N=K_{j_N}\cdots K_{j_1}\rho K_{j_1}\dg\cdots K_{j_N}\dg\,,
  \label{eq:post-iter-unnorm-state-simp}
\end{align}
the probability of the outcome sequence
\begin{align}
\pr(o_{j_1},\ldots,o_{j_N}\vert\rho)
=\tr{\tilde\rho_N}=\tr{K_{j_N}\cdots K_{j_1}\rho K_{j_1}\dg\cdots K_{j_N}\dg}\,,
\end{align}
and the normalized final state of the system,
\begin{align}
  \rho_N=\frac{K_{j_N}\cdots K_{j_1}\rho K_{j_1}\dg\cdots K_{j_N}\dg}
  {\tr{K_{j_N}\cdots K_{j_1}\rho K_{j_1}\dg\cdots K_{j_N}\dg}}\,.
  \label{eq:post-iter-state-simp}
\end{align}

The Markov nature of the model manifests itself algebraically as the
decomposition of the collective Kraus operator for all $N$ measurements into a
product of separate Kraus operators for each probe.  Indeed, the Kraus operators
for the $n$th probe, $K_{j_n}=\sand{o_{j_n}}{U^{(n)}}{\phi_n}$, neatly display
the elements of Markovian evolution: each probe has its own initial state, its
own interaction with the system, and its own measurement.  As a consequence, the
results for a sequence of measurements can be dealt with one probe at a time; in
particular, the system state after $n+1$ measurements is
\begin{align}
\label{eq:rho-nplus1-general}
\begin{split}
\rho_{n+1}
  &=\frac{K_{j_{n+1}}\cdots K_{j_1}\rho K_{j_1}\dg\cdots K_{j_{n+1}}\dg}
  {\pr(o_{j_1},\ldots,o_{j_{n+1}}\vert\rho)}\\
  &=K_{j_{n+1}}\rho_n K_{j_{n+1}}\dg
  \frac{\pr(o_{j_1},\ldots,o_{j_n}\vert\rho)}
  {\pr(o_{j_1},\ldots,o_{j_{n+1}}\vert\rho)}\\
  &=\frac{K_{j_{n+1}}\rho_n K_{j_{n+1}}\dg}{\pr(o_{j_{n+1}}\vert o_{j_1},\ldots,o_{j_n},\rho)}\,;
\end{split}
\end{align}
the final denominator here is the conditional probability for the $(n+1)$th outcome,
given the previous outcomes, which can be written as
\begin{align}
\begin{split}
\pr(o_{j_{n+1}}\vert o_{j_1},\ldots,o_{j_n},\rho)
&=\tr{\rho_n E_{j_{n+1}}}\,.
\end{split}
\end{align}
Notice that for consistency, we should denote the initial state as
$\rho=\rho_0$.

Quantum trajectories are usually formulated as difference equations,
\begin{align} \label{eq:delta-rho-sme}
  \Delta\rho_{n|j}&\defined\rho_{n+1|j}-\rho_n\,,
\end{align}
or, in the continuous-time limit, as the corresponding differential equation.
Here we have explicitly denoted the $(n+1)$th measurement outcome by $j$ and
left all prior measurement results implicit in the density operator $\rho_n$.
The object of this paper is to derive \cref{eq:delta-rho-sme} for different
choices of the elements that go into the Kraus
operator~(\ref{eq:anc-kraus-ops}), \ie, the measurement outcomes $\bra{o_j}$,
the interaction unitary $U$, and the initial state $\ket{\phi}$.

A final point that we need later on is how to find the Kraus operators when the
probes begin in a mixed state.  For a mixed probe initial state,
\begin{align}
\sigma=\sum_k\lambda_k\oprod{k}{k}\,,
\end{align}
the unnormalized post-measurement system state~(\ref{eq:unnorm-cond-state})
becomes
\begin{align}\label{eq:unnorm-cond-state-mixed}
\begin{split}
\tilde{\rho}&=\sand{o_j}{U\rho\otimes\sigma U\dg}{o_j}\\
  &=\sum_k\sqrt{\lambda_k}\sand{o_j}{U}{k}\rho\sand{k}{U\dg}{o_j}\sqrt{\lambda_k}\\
  &=\sum_k K_{jk}\rho K\dg_{jk}\,,
\end{split}
\end{align}
where the Kraus operators, defined by
\begin{align}
K_{jk}\defined\sqrt{\lambda_k}\sand{o_j}{U}{k}\,,
\label{eq:anc-kraus-ops-mixed}
\end{align}
act together to make up a quantum operation.

Armed with this language of Kraus operators, we can put forward alternative
descriptions of projective and weak measurements. A projective measurement is
one whose Kraus operators are one-dimensional projectors, and weakness
(or gentleness or fuzziness) is measured by the extent to which this is not
the case, by having Kraus operators that are either subunity multiples of
one-dimensional projectors or operators higher than rank one.  Typically,
what is meant by a weak measurement is a measurement whose Kraus operators are
mostly ``close'' to some multiple of the identity operator, corresponding to
outcomes that don't disturb the system much, although there might also be
some which are very ``small,'' corresponding to outcomes that might significantly
disturb the system, but that occur infrequently.

\subsection{Open-system dynamics}

We finally note that every conditional dynamics gives rise to an unconditional,
open-system dynamics that corresponds to throwing away information about
measurement outcomes. In the Markovian scenarios we are considering, throwing
away the probe information at timestep $n+1$ gives evolution described by a
quantum operation $\mathcal{A}$:
\begin{align}\label{eq:AKj}
  \mathcal{A}[\rho]&\defined\sum_jK_j\rho K\dg_j\,, \\
  \label{eq:open-system-update}
  \rho_{n+1}&=\mathcal{A}[\rho_n]\,.
\end{align}
Notice that for a mixed-state probe, the Kraus operators $K_{jk}$ of
Eq.~(\ref{eq:anc-kraus-ops-mixed}) go together in
Eq.~(\ref{eq:unnorm-cond-state-mixed}) to make an outcome-dependent quantum
operator $\mathcal{A}_j[\rho]\defined\sum_k K_{jk}\rho K_{jk}\dg$ that can be
thought of as coming from throwing away the information about the probe's
initial state.

The differential equation corresponding to the
evolution~(\ref{eq:open-system-update}) is known as the \emph{master equation}.
As is well-known~\cite{NielChua10}, the Kraus decomposition~(\ref{eq:AKj}) for
the quantum operation $\mathcal{A}$ is not unique.  Different Kraus
decompositions correspond to performing different measurements on the probes and
result in different system dynamics. In the trajectory literature, these
alternative stochastic dynamics are known by Carmichael's terminology of
\emph{unravelings}~\cite{Carm93a}. The relationship of the master equation to
\cref{eq:delta-rho-sme} is
\begin{align} \label{eq:delta-rho}
  \Delta\rho_{n}
  \defined \sum_j \Pr(j|\rho_n)\rho_{n+1|j}-\rho_n=\Expt{\Delta \rho_{n+1|j}}\,.
\end{align}

\section{Continuous measurements with probe fields}
\label{sec:qbittraj}

We have now presented circuit-model and algebraic representations of the
conditional evolution of a quantum system subjected to a sequence of weak
measurements. In this section we formally describe sequences of weak
interactions between a system and a probe field and discuss how the
approximations made in quantum input-output theory allow us to use the circuit
of \cref{fig:iter-meas-itt} to describe the quantum trajectories arising from
continuous measurement of the probe field. The probe field---and the probe
qubits we use \emph{in lieu\/} of a field---are often referred to as a reservoir
or a bath.

We begin by writing the combined Hamiltonian for the system coupled to the field
as
\begin{align}
  H&=H\plant+H_{\text{field}}+H\interaction\,.
  \label{eq:combined-plant-probe-hamiltonian}
\end{align}
For simplicity, we assume that the interaction Hamiltonian is linear in the
one-dimensional probe field~$a$,
\begin{align}\label{eq:Hint1}
H\interaction=i\sqrt\gamma\,\big(c\otimes a^\dagger-c^\dagger\otimes a\big)\,,
\end{align}
where $c$ is a system operator. An example discussed in the literature is $c=x$~\cite{DoheJaco99}.  Writing the interaction Hamiltonian in this
form uses the rotating-wave approximation (RWA) to keep only the
energy-conserving terms in the interaction.  Typical interaction terms involve
the product of a Hermitian system operator and a Hermitian field operator.
Writing these Hermitian operators as sums of positive- and negative-frequency
parts leads to four terms in the interaction Hamiltonian, only two of which
conserve energy when averaged over times much longer than the system's
characteristic dynamical time.  The RWA retains these two energy-conserving,
co-rotating terms and discards the two counter-rotating terms, leaving the
interaction Hamiltonian~(\ref{eq:Hint1}).  Making the RWA requires averaging
over times much longer than the system's dynamical time.  We say more about the
RWA below.

It is useful to work in the interaction picture, where the free time evolution
of the system and field (generated by $H_0\defined H\plant+H_{\text{field}}$)
is transformed into the operators, leaving a time-dependent interaction
Hamiltonian,
\begin{align}
\begin{split}
  H_I(t)\defined e^{iH_0t}H\interaction e^{-iH_0t}
  =i\sqrt\gamma\,\big[c(t)\otimes a^\dagger(t)-c^\dagger(t)\otimes a(t)\big]\,.
  \label{eq:Hint2}
\end{split}
\end{align}
In the interaction picture, the system operator $c$ acquires a free time
dependence; we assume now that the system has a single transition
(characteristic) frequency $\Omega$, so that $c(t)=c\,e^{-i\Omega t}$.   The
field operators also acquire a time dependence; each frequency mode of the
field oscillates at its angular frequency $\omega$, \ie, as $e^{-i\omega t}$.
Indeed, the positive-frequency part of the field appearing in Eqs.~(\ref{eq:Hint1})
and~(\ref{eq:Hint2}) is constructed from the frequency-mode annihilation
operators $a(\omega)$ and is given by
\begin{align}\label{eq:at}
a(t)=\int_0^\infty\frac{\df\omega}{2\pi}\,a(\omega)e^{-i\omega t}\,.
\end{align}
The field in \cref{eq:at} is written in photon-number units, by which we mean
it is the Fourier transform of the frequency-domain annihilation operators,
which obey the canonical commutation relations
\begin{align}
[a(\omega),a^\dagger(\omega')]=2\pi\delta(\omega-\omega')\,.
\end{align}
Writing the field in these units omits frequency-dependent factors in the
Fourier transform, and this omission is called the \emph{quasimonochromatic
approximation}, which assumes that the coupling of the field to the
system is weak enough, \ie, $\gamma\ll\Omega$, that only field frequencies near
the system transition frequency $\Omega$, \ie, those within a few linewidths
$\gamma$ of $\Omega$, are important.  This allows us to choose the averaging
time required by the RWA much longer than the system's characteristic time
$1/\Omega$, but much shorter than the inverse linewidth $1/\gamma$; \ie, the
averaging time is long enough to average away the counter-rotating,
energy-nonconserving parts of the interaction Hamiltonian, but short enough that
not much happens to the system during the averaging time.

It is convenient to introduce a new field operator,
\begin{align}
b(t)=e^{i\Omega t}a(t)=\int_{-\Omega}^\infty\frac{\df\epsilon}{2\pi}\,a(\Omega+\epsilon)e^{-i\epsilon t}\,,
\end{align}
which has its zero of frequencies shifted to the transition frequency $\Omega$.
Within the quasimonochromatic approximation, we can extend the integral over
$\epsilon$ to $-\infty$; introducing phantom modes at negative $\omega=\Omega+\epsilon$
doesn't make any difference because they don't participate in the narrow-bandwidth
coupling to the system.  This gives us
\begin{align}\label{eq:neg-freq-approx}
  b(t)=\int_{-\infty}^{\infty}\frac{\df\epsilon}{2\pi}\,a(\Omega+\epsilon)e^{-i\epsilon t}\,.
\end{align}
The advantage of extending the integral to $-\infty$ is that the field operators
$b(t)$ become instantaneous temporal annihilation operators, obeying the
canonical commutation relations,
\begin{align}\label{eq:white_noise_continuous}
[b(t),b^\dagger(t')]=\delta(t-t')\,.
\end{align}
These operators are often called ``white-noise operators'' because of their
delta commutator, which permits them to be delta-correlated in time like
classical white noise. The interaction Hamiltonian now assumes the following
continuous-time form:
\begin{align}
  H_I(t)
  =i\sqrt\gamma\,\big[c\otimes b^\dagger(t)-c^\dagger\otimes b(t)\big]\,.
  \label{eq:Hint3}
\end{align}
The essence of the quasimonochromatic approximation is the use of the
photon-units field operator~(\ref{eq:neg-freq-approx}).  The notion of creating
instantaneous photons at the characteristic frequency $\Omega$ clearly requires
a bit of cognitive dissonance: it is valid only if ``instantaneous'' is
understood to mean temporal windows that are broad compared to $1/\Omega$,
corresponding to a narrow bandwidth of frequencies near $\Omega$.

\begin{figure}[ht!]
  \begin{center}
    \subfloat[]{
      \includegraphics[scale=.6]{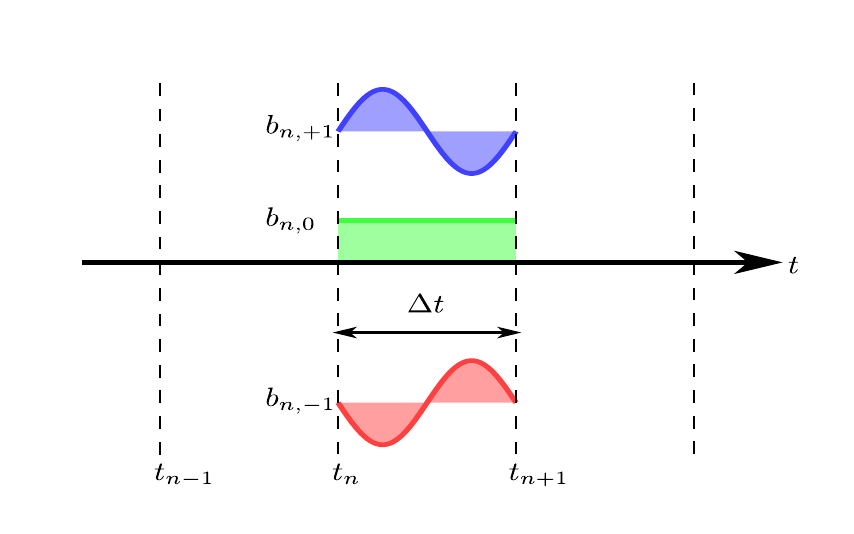}
      \label{fig:discrete-modes}
    }
    \qquad
    \subfloat[]{
      \includegraphics[scale=.6]{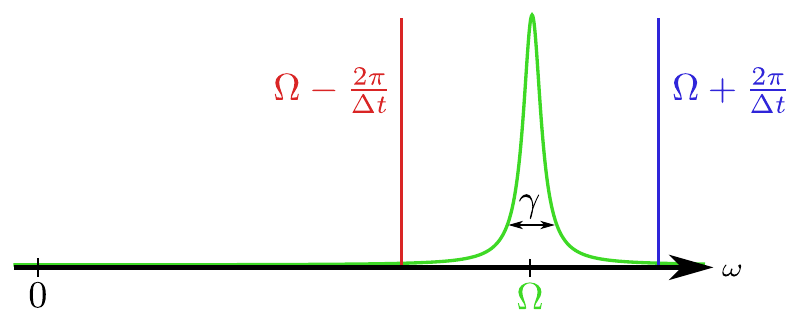}
      \label{fig:sidebands}
    }
  \end{center}
  \caption{(a)~On-resonance and first two sideband discrete temporal modes,
  represented in the interaction picture, where the on-resonance mode has
  frequency $\omega_0=\Omega$ ($\epsilon=0$) and the first two sideband modes
  have frequencies $\omega_{\pm}=\Omega\pm2\pi/\Delta t$
  ($\epsilon=\pm2\pi/\Delta t$).  (b)~Illustration of the case where
  the interaction is sufficiently weak that the first two sideband
  discrete modes---and, hence, all the other sidebands---are sufficiently
  off resonance to ignore; it is thus also true that sensitivity to low
  frequencies is small enough that we can introduce the phantom
  negative-frequency field modes of \cref{eq:neg-freq-approx},
  with frequencies $\omega=\Omega+\epsilon<0$, without altering
  the physics.  This diagram illustrates the essential assumptions
  for the RWA and the quasimonochromatic approximation:
  $\gamma\ll1/\Delta t\ll\Omega$.}
  \label{fig:quasimon}
\end{figure}

The discrete interactions in \cref{fig:iter-meas} arise from the continuous-time
interaction Hamiltonian~(\ref{eq:Hint3}) by dividing the field into probe
segments, starting at times $t_n=n\Delta t$, $n=-\infty,\ldots,\infty$, all of
duration $t_{n+1}-t_n\defined\Delta t$.  We assume, first, that $\Delta
t\gg\Omega^{-1}$ so that within each segment $\Delta t$, the interaction with
the probe field is averaged over many characteristic times of the system, as
required by the \hbox{RWA}, and, second, that $\Delta t\ll\gamma^{-1}$ so that
the probe/system interaction over the time $\Delta t$ is weak.  Instead of using
the frequency modes $a(\Omega+\epsilon)$ or the instantaneous temporal modes
$b(t)$, we now resolve the field into discrete temporal modes $b_{n,k}$ as
\begin{align}
b(t)=\sum_{n=-\infty}^\infty\sum_{k=-\infty}^\infty
\frac{1}{\sqrt{\Delta t}}\,b_{n,k}\Theta(t-t_n)e^{-i2\pi kt/\Delta t}\;,
\end{align}
where $\Theta(u)$ is the step function that is equal to 1 during the interval
$0<u<\Delta t$ and is 0 otherwise.  The discrete temporal modes are given by
\begin{align}
  \label{eq:discrete-field-op}
  \begin{split}
  b_{n,k}&\defined\frac{1}{\sqrt{\!\Delta t}}\int_{t_n}^{t_{n+1}}\df t\,
  e^{i2\pi kt/\Delta t}b(t)\\
  &=\sqrt{\!\Delta t}\int_{-\infty}^{\infty}\frac{\df\epsilon}{2\pi}\,
  a(\Omega+\epsilon)\exp\bigg[\!-i\bigg(\epsilon-\frac{2\pi k}{\Delta t}\bigg)
  \bigg(\frac{t_n+t_{n-1}}{2}\bigg)\bigg]\frac{\sin(\epsilon\Delta t-2\pi k)/2}
  {\epsilon\Delta t-2\pi k}\,.
  \end{split}
\end{align}
These modes obey discrete canonical commutation relations,
\begin{align}\label{eq:white_noise_discrete}
[b_{n,k},b_{m,l}^\dagger]=\delta_{nm}\delta_{kl}\,;
\end{align}
this is the discrete-time analogue of continuous-time white noise of
\cref{eq:white_noise_continuous}.  We now recall that the interaction is weak
enough, \ie, $\gamma\ll\Omega$, that only frequencies within a few $\gamma$ of
$\Omega$ need to be considered; given our assumption that $1/\Delta t\gg\gamma$,
this allows us to neglect all the discrete temporal modes with $k\ne0$, reducing
the probe field to
\begin{align}
b(t)=\sum_{n=-\infty}^\infty\frac{1}{\sqrt{\Delta t}}\,b_n\Theta(t-t_n)\;,
\end{align}
where
\begin{align}\label{eq:bn}
b_n\defined b_{n,0}=\frac{1}{\sqrt{\!\Delta t}}\int_{t_n}^{t_{n+1}}\df t\,b(t) \,.
\end{align}
The neglect of all the sideband modes is illustrated schematically in
\cref{fig:quasimon}. Plugging this expression for the probe field into the
Eq.~(\ref{eq:Hint3}) puts the interaction Hamiltonian in its final form,
\begin{align}\label{eq:Hint4}
H_I(t)=\sum_{n=-\infty}^\infty H_I^{(n)}\Theta(t-t_n)\,,
\end{align}
where
\begin{align}
  H_I^{(n)}&\defined i\sqrt{\frac{\gamma}{\Delta t}}
  \left( c\otimes b_n\dg-c\dg\otimes b_n \right)
  =i\left( \sqrt{\gamma}\,c\otimes\frac{b_n\dg}{\sqrt{\Delta t}}-
  \sqrt{\gamma}\,c\dg\otimes\frac{b_n}{\sqrt{\Delta t}}\right)
  \label{eq:interaction-picture-hamiltonian-discrete}
\end{align}
is the interaction Hamiltonian during the $n$th probe segment.  It is this
Hamiltonian that is used to generate the discrete unitaries in
\cref{fig:iter-meas}.

Before exploring the interaction unitary, however, it is good to pause to
review, expand, and formalize the assumptions necessary to get to the discrete
Hamiltonian~(\ref{eq:interaction-picture-hamiltonian-discrete}) that applies to
each time segment or, more generally, to get to the Markovian quantum circuit of
Fig.~\ref{fig:iter-meas}.  The restriction of the system-probe interaction to be
a sequence of joint unitaries between the system and a single probe segment is
often referred to as the \emph{first Markov approximation}.  This approximation
is valid when the spatial extent $\Delta x$ of the system is small with respect
to the spatial extent $c\Delta t$ of the discretized probes.  For many typical
scenarios (\eg, atomic systems), the time interval $\Delta t$ can be made quite
small, often even smaller than the characteristic evolution time $\Omega^{-1}$,
before the spatial extent of the probes becomes comparable to the spatial extent
of the system, which would force us to use a non-Markovian description like
\cref{fig:overlapping-probes-ckt}.  The reason we did not encounter this
assumption in the analysis above is that it is already incorporated in our
starting point, the interaction Hamiltonian~(\ref{eq:Hint1}).  A typical
interaction Hamiltonian involves a spatial integral over the extent of the
system.  In writing the interaction Hamiltonian~(\ref{eq:Hint1}), we have
already assumed that the system is small enough that the spatial integral can be
replaced by a point interaction.

The initial product state of the probes is often referred to as the \emph{second
Markov approximation}.  This approximation is valid when the correlation time
$\tau_c$ in the bath is much shorter than the duration $\Delta t$ of the
discrete probe segments.  This is often an excellent approximation, as baths
with even very low temperatures have very small correlation times.  For example,
the thermal correlation time $\tau_c=\hbar/2\pi kT\simeq10\,{\rm ps}/T$ given by
Eq.~(3.3.20) in~\cite{GardZoll04} is approximately $10\,\mathrm{ns}$ for a
temperature of $1\,\mathrm{mK}$.  On the other hand, the vacuum correlation time
$\tau_c\simeq1/2\pi\Omega$ at the characteristic frequency means that if vacuum
noise dominates, then the second Markov approximation requires that the probe
segments be much longer than the system's dynamical time, \ie, $\Delta
t\gg1/\Omega$. For a treatment of the nonzero correlation time of the vacuum in
an exactly solvable model, see~\cite{UnruhZurek89}.

The product measurements at the output of the circuit in
\cref{fig:iter-meas-itt} do not affect open-system dynamics, for which the
bath is not monitored, but they do enter into a Markovian description of
dynamics conditioned on measurement of the bath.  The product measurements are a
good approximation when the bandwidth of the detectors is sufficiently wide to
give temporal resolution much finer than the duration of the probe segments we
used to discretize the bath.

The remaining pair of closely related approximations, as we discussed
previously, are the RWA, which has to do with simplifying the form of the
interaction Hamiltonian, and the quasimonochromatic approximation, which has to
do with simplifying the description of the field so that each $\Delta t$ probe
segment has only one relevant probe mode.  The three important parameters in
these two approximations are the characteristic system frequency~$\Omega$, the
linewidth $\gamma$, and the duration of the time segments, $\Delta t$, and the
approximations require that $\gamma\ll1/\Delta t\ll\Omega$.

The approximations we make are summarized below:
\begin{align}
  &\Delta x\ll c\Delta t&\text{First Markov,} \\
  &\tau_c\ll\Delta t&\text{Second Markov,} \\
  \label{eq:quasi-mono-rwa}
  &\Omega^{-1}\ll\Delta t\ll\gamma^{-1}&\text{RWA and quasimonochromatic.}
\end{align}
We note that it is possible to model systems with several different,
well-separated transition frequencies by introducing separate probe fields
for each transition frequency, as long as it is possible to choose discrete
probe time segments in such a way that the above approximations are valid
for all fields introduced.  The several probe fields can actually be parts
of a single probe field, with each part consisting of the probe frequencies that are
close to resonance with a particular transition frequency.

The approximations now well in hand, we return to the
Hamiltonian~(\ref{eq:interaction-picture-hamiltonian-discrete}) for the $n$th
probe segment.  The associated interaction unitary between the system and the
$n$th probe segment is given by
\begin{equation}
  \label{eq:disc-int-unitary}
  U^{(n)}_I=e^{-iH^{(n)}_I\Delta t}=\Id\otimes\Id+
    \sqrt{\dt}\left(c\otimes b\dg_n-c\dg\otimes b_n\right)+
    \frac{1}{2}\dt\left(c\otimes b\dg_n-c\dg\otimes b_n\right)^2+
    \BigO\big( (\dt)^{3/2} \big)\,,
\end{equation}
where we define a dimensionless time interval,
\begin{align}
  \dt&\defined\gamma\Delta t\ll 1\,,
  \label{eq:dimensionless-time}
\end{align}
suitable for series expansions.  We only need to expand the unitary to
second order because we are only interested in terms up to order
$\dt$ for writing first-order differential equations. A comprehensive and related presentation of the issues discussed above can be found in the recent paper of \citet{FiscTrivRama17}.

Notice that we can account for an external Hamiltonian $H_\text{ext}$
applied to the system, provided it changes slowly on the characteristic
dynamical time scale $1/\Omega$ of the system and leads to slow
evolution of the system on the characteristic time scale (if such
a Hamiltonian is not slow, it should be included in the free system
Hamiltonian $H\plant$).  In the interaction picture, the external Hamiltonian
acquires a time dependence and becomes part of the interaction Hamiltonian; since
it is essentially constant in each time segment, its effect in each
time segment can be captured by expanding its effect to linear order
in $\Delta t$.  It is easy to see that the interaction
unitary~(\ref{eq:disc-int-unitary}) is then supplemented by an additional
term $-i\Delta t\,H_\text{ext}$; when we convert to the final differential
equation, this term introduces the standard commutator
$-i\,\df t\,[H_\text{ext},\rho]$ for an external Hamiltonian.

\section{Quantum trajectories for vacuum field and qubit probes}
\label{sec:qubit-vac}

We are now prepared to discuss the quantum trajectories arising from the
continuous measurement of a probe field coupled to a system as described in the
previous section. In this context we often drop the explicit reference to which
probe segment we are dealing with, since the Markovicity of
\cref{fig:iter-meas-itt} means we can consider each probe segment separately.
Unconditional open-system evolution follows from averaging over the quantum
trajectories or, equivalently, tracing out the probes.

Probe fields initially in the vacuum state are our concern in this section.
Because the interaction between individual probes and the system is weak, the
one-photon amplitude of the post-interaction probe segment is $\BigO(\sqrtdt)$,
the two-photon amplitude is $\BigO(\dt)$, and so on. Since these amplitudes are
squared in probability calculations, the probability of detecting a probe with
more than one photon is $\BigO(\dt^2)$ and can be ignored.  This suggests that
it is sufficient to model the probe segments with qubits, with $\ket{g}$
corresponding to the vacuum state of the field and $\ket{e}$ corresponding to
the single-photon state. We replace the discrete-field-mode annihilation
operator $b_n$ in Eq.~(\ref{eq:disc-int-unitary}) with the qubit lowering
operator $\sigma_-$ and $b_n^\dagger$ with $\sigma_+$:
\begin{subequations}
\label{eq:disc-int-qubit-unitary}
\begin{align}
  \label{eq:disc-int-qubit-unitary-unsimp}
  U_I&=\Id\otimes\Id
  +\sqrtdt\left( c\otimes\sigma_+ - c\dg\otimes\sigma_- \right)
  +\frac{1}{2}\dt\left(c\otimes\sigma_+ - c\dg\otimes\sigma_-\right)^2 \\
  \label{eq:disc-int-vac-qubit-unitary}
  &=\Id\otimes\Id+\sqrtdt\left( c\otimes\sigma_+-
  c\dg\otimes\sigma_- \right)-\frac{1}{2}\dt\left(
  c\,c\dg\otimes\oprod{e}{e}+c\dg c\otimes\oprod{g}{g} \right)\,.
\end{align}
\end{subequations}
With this replacement, the neglect of two-photon transitions in the probe-field
segments is made exact by the fact that $\sigma_+^2=\sigma_-^2=0$; these squared
terms thus do not appear in Eq.~(\ref{eq:disc-int-vac-qubit-unitary}).

In Secs.~\ref{sec:photon}--\ref{sec:meas-summary} we establish the
correspondence between this qubit model and \emph{vacuum SMEs}, where vacuum
refers to the state of the probe field. In particular, we present qubit
analogues of three typical measurements performed on probe fields: photon
counting, homodyne measurement, and heterodyne measurement.

We transcend the vacuum probe fields in Sec.~\ref{sec:gaussian-states} to
Gaussian probe fields and find that formulating a qubit model requires
additional tricks beyond just noting that weak interactions with the probe do
not lead to significant two-photon transitions. Nevertheless, we are able to
find qubit models that yield all the essential features of these Gaussian
stochastic evolutions.

While the qubit model we develop is meant to capture the behavior of a ``true''
field-theoretic model, it is important to note that there are scenarios where
qubits are the natural description. For example, in Haroche-style experiments
\cite{SayrDotsZhou11} a cavity interacts with a beam of atoms, accurately
described as a sequence of finite-dimensional quantum probes. Such scenarios
have been analyzed for their non-Markovian behavior \cite[Sec.~9.2]{SculZuba97},
and similar models are increasingly studied in the thermodynamics literature
\cite{Horo12,StraSchaBran17} and collisional
models~\cite{GiovPalm12,RybaFiliZima12,Cicc17}.

\subsection{Z basis measurement: Photon counting or direct detection}\label{sec:photon}

\begin{figure}[ht!]
  \begin{center}
    \includegraphics[scale=.75]{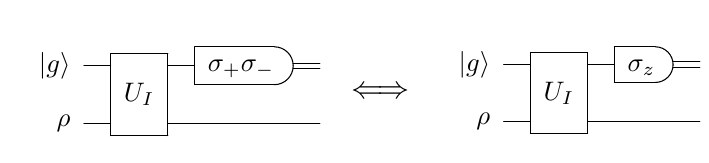}
  \end{center}
  \caption{Circuit depicting the system interacting with a vacuum probe (probe
  initially in the ground state) which is subsequently subjected to measurement
  of the qubit number operator $\sigma_+\sigma_-$, the qubit analogue of a
  photon-counting measurement.  The eigenvectors of
  $\sigma_+\sigma_-=\oprod{e}{e}=\half(\Id+\sigma_z)$ are identical to those of
  $\sigma_z$, thus allowing us to think instead of a measurement of the Pauli
  observable $\sigma_z$.}
  \label{fig:photon-counting-ckt}
\end{figure}

As a first example, we consider performing photon-counting measurements on the
probe field after its interaction with the system. We calculate the quantum
trajectory by first constructing the Kraus operators given by
\cref{eq:anc-kraus-ops}. For probes initially in the vacuum state we have
$\ket{\phi}=\ket{g}$, and our interaction unitary is given by
Eqs.~(\ref{eq:disc-int-qubit-unitary}).  What remains is to identify the
measurement outcomes $\bra{o_j}$. The qubit version of the number operator $b\dg
b$ is $\sigma_+\sigma_-=\oprod{e}{e}=\half(\Id+\sigma_z)$.  Measuring this
observable, as depicted in \cref{fig:photon-counting-ckt}, is equivalent to
measuring $\sigma_z$. The measurement outcomes are then $\bra{g}$ and $\bra{e}$
and give the Kraus operators
\begin{subequations}\label{eq:vac-pc-kraus-operators}
\begin{align}
  K_g &=\sand{g}{U_I}{g}=\Id-\half\dt\,c\dg c\,,\\
  K_e &=\sand{e}{U_I}{g}=\sqrtdt c\,.
\end{align}
\end{subequations}
The corresponding POVM elements (to linear order in $\dt$) are
\begin{subequations}\label{eq:vac-pc-POVM-elements}
\begin{align}
  E_g &=K_g\dg K_g=\Id-\dt\,c\dg c\,,\\
  E_e &=K_e\dg K_e=\dt\,c\dg c\,,
\end{align}
\end{subequations}
which trivially satisfy $E_g + E_e = \Id$.  We call the
Kraus operators~(\ref{eq:vac-pc-kraus-operators}) the
\emph{photon-counting Kraus operators}.  These operators are identical to those
derived for photon counting with continuous field modes~\cite[Eqs.~4.5 and
4.7]{WiseMilb10}, as we expected from the vanishing multi-photon probability
discussed earlier.

To calculate a quantum trajectory we need to describe the evolution of the
system conditioned on the outcomes of repeated measurements of this kind. The
state of the system after making a measurement and getting the result $g$ during
the $(n+1)$th time interval, \ie, between $t_n$ and $t_{n+1}$, is
\begin{align}
  \rho_{n+1|g}\defined\frac{K_g\rho_nK_g\dg}{\tr{\rho_n E_g}}
  =\frac{\rho_n-\half\dt\,\left(c\dg c\rho_n+\rho_nc\dg c\right)}
  {1-\dt\,\tr{\rho_nc\dg c}}\,.
\end{align}
The subscript $n+1$ on $\rho_{n+1|g}$ indicates, as in
Eq.~(\ref{eq:rho-nplus1-general}), that this is the state at the end of this
probe segment, after the measurement; the subscript $g$ indicates that this is
the state conditioned on the measurement outcome $g$.  The state $\rho_{n+1}$ is
conditioned on all previous measurement outcomes as well, but we omit all of
that conditioning, letting it be implicit in $\rho_n$.  Expanding the
denominator to first order in $\dt$ using the standard expansion
$(1+x\dt)^{-1}=1-x\dt+\BigO(\dt^2)$ allows us to calculate the difference
equation~(\ref{eq:delta-rho-sme}) when the measurement result is $g$:
\begin{align}
\begin{split}
  \Delta\rho_{n|g}&\defined \rho_{n+1|g}-\rho_n \\
  &=-\half\dt\,\Bigl(c\dg c\rho_n+\rho_nc\dg c-2\rho_n\tr{\rho_nc\dg c}\Bigr)
  +\BigO(\dt^2) \\
  &=-\half\dt\,\Hc{c\dg c}\rho_n\,,
\end{split}
\end{align}
where we employ the shorthand
\begin{equation}
  \Hc{X}\rho\defined X\rho+\rho X\dg-\rho\tr{\rho(X+X\dg)}\,.
  \label{eq:h-supop}
\end{equation}

Repeating the analysis for the case when the measurement result is $e$ gives
\begin{align}
  \rho_{n+1|e}\defined\frac{K_e\rho_nK_e\dg}{\tr{\rho_n E_e}}
  =\frac{c\rho_nc\dg}{\tr{\rho_nc\dg c}}\,.
\end{align}
The difference between the pre- and post-measurement system states when the
measurement result is $e$ is thus
\begin{align}
  \Delta\rho_{n|e}\defined\rho_{n+1|e}-\rho_n
  =\frac{c\rho_nc\dg}{\tr{\rho_nc\dg c}}-\rho_n
  =\G{c}\rho_n\,,
\end{align}
where we define
\begin{align}
  \label{eq:G-Super-op}
  \mathcal{G}[X]\rho\defined\frac{X\rho X\dg}{\tr{\rho X\dg X}}-\rho\,.
\end{align}

Having separate equations for the two measurement outcomes is not at all
convenient.  Fortunately, we can combine the equations by introducing a
random variable $\Delta N$ that represents the outcome of the measurement:
\begin{align}
  \Delta N&:g\mapsto0\;,e\mapsto1\,.
\end{align}
Since this random variable is a bit (\ie, a Bernoulli random variable) its
statistics are completely specified by its mean:
\begin{align}\label{eq:exptDeltaN}
\Expt{\Delta N}=0\cdot\tr{\rho E_g}+1\cdot\tr{\rho E_e}=\dt\,\tr{\rho c\dg c}.
\end{align}
We now combine the difference equations into a single stochastic equation using
the random variable $\Delta N$:
\begin{align}\label{eq:pc-sme-first-shot}
\begin{split}
  \Delta\rho_{n|\Delta N}&\defined\rho_{n+1|\Delta N}-\rho_n \\
  &=\Delta N\left(\frac{c\rho_nc\dg}{\tr{\rho_nc\dg c}}-\rho_n \right)
  -(1-\Delta N)\half\dt\,\Bigl(c\dg c\rho_n+\rho_nc\dg c-2\rho_n\tr{\rho_nc\dg c}\Bigr) \\
  &=\Delta N\G{c}\rho_n-(1-\Delta N)\half\dt\,\Hc{c\dg c}\rho_n\,.
\end{split}
\end{align}
It quickly becomes unnecessarily tedious to keep time-step indices around
explicitly, since everything in our equations now refers to the same time step,
so we drop those indices now.  Discarding $\Delta N\dt$, since it is
second-order in $\dt$ (see Eq.~(\ref{eq:exptDeltaN})
and~\cite[Chap.~4]{WiseMilb10}), we simplify Eq.~(\ref{eq:pc-sme-first-shot}) to
\begin{align}
\begin{split}
  \Delta\rho_{\Delta N}&=\Delta N\left(\frac{c\rho c\dg}{\tr{\rho c\dg c}}-
  \rho\right)-\half\dt\,\Bigl(c\dg c\rho+\rho c\dg c-2\rho\tr{\rho c\dg c}\Bigr) \\
  &=\dt\,\mathcal{D}[c]\rho+\Delta\innovation_D\G{c}\rho\,,
  \label{eq:pc-sme-with-innovation}
\end{split}
\end{align}
where we introduce the standard diffusion superoperator,
\begin{equation}
  \mathcal{D}[X]\rho\defined X\rho X\dg-\half(X\dg X\rho+\rho X\dg X)\,,
  \label{eq:diff-superop}
\end{equation}
and the \emph{photon-counting innovation},
\begin{align}\label{eq:innovationD}
\Delta\innovation_D\defined\Delta N-\Expt{\Delta N}\,,
\end{align}
which is the difference between the measurement result and the mean result
(\ie, it can be thought of as what is learned from the measurement).  The
subscript $D$ here plays off the fact that photon counting is often called
\emph{direct detection\/} and is used in place of $N$ because $N$ has too
many other uses in this paper.

By taking the limit $\dt\to\gamma\df t$ we obtain a stochastic differential
equation,
\begin{align}\label{eq:vacsmedirect}
\begin{split}
  \df\rho_D&=\df N\left(\frac{c\rho c\dg}{\tr{\rho c\dg c}}-\rho\right)
  -\half\gamma\df t\,\Bigl(c\dg c\rho+\rho c\dg c-2\rho\tr{\rho c\dg c}\Bigr) \\
  &=\df t\,\mathcal{D}[\sqrt{\gamma}\,c]\rho+
  \df\innovation_D\G{c}\rho\,,
\end{split}
\end{align}
where $\df N$ is a bit-valued random process, termed a point process, with mean
$\Expt{\df N}=\gamma\df t\,\tr{\rho c\dg c}$ and the innovation is given by
$\df\innovation_D=\df N-\Expt{\df N}$. This equation is called the \emph{vacuum
stochastic master equation (SME) for photon counting\/}; \ie, it is the
stochastic differential equation that describes the conditional evolution of a
system that interacts with vacuum probes that are subjected to photon-counting
measurements.

\Cref{eq:vacsmedirect} has no explicit system-Hamiltonian term. Although this
differs from other presentations our readers might be familiar with, it is merely
an aesthetic distinction. Recall from the discussion at the end of
\cref{sec:qbittraj} that well-behaved system Hamiltonians can be introduced by
including an additional commutator term in our differential equations. In this
case, the modification yields
\begin{align}\label{eq:vacsmedirect-ext-ham}
\begin{split}
  \df\rho_D&=-i\,\df t\,[H_{\text{ext}},\rho]
  +\df t\,\mathcal{D}[\sqrt{\gamma}\,c]\rho+\df\innovation_D\G{c}\rho\,.
\end{split}
\end{align}

It is important to stress that in practice, for numerical integration of these
equations, one uses the difference equation~(\ref{eq:pc-sme-with-innovation}),
not the differential equation~(\ref{eq:vacsmedirect}); \ie, what one uses in
practice is the difference equation that corresponds to the discrete-time
quantum circuit in Fig.~\ref{fig:iter-meas}(c).  One assigns to the system a
prior state $\rho_0$ that combines with the initial probe states to make an
initial product state on the full system/probe arrangement.  This prior state
describes the system at the moment coupling to the probes is turned on and
measurements begin. Each time a new measurement result is sampled,
\cref{eq:pc-sme-with-innovation} is used to update the description of the
system. If we describe our system by $\rho_n$ after collecting $n$ samples from
our measurement device, observing $\Delta N$ for sample $n+1$ leads to the
updated state $\rho_{n+1}=\rho_n+\Delta\rho_{\Delta N}$.

Another application of the difference equation is state/parameter
inference.  In the case of state inference, one has uncertainty regarding what
initial state $\rho_0$ to assign to the system.  General choices for $\rho_0$
will be incorrect, invalidating some of the properties described above.  In
particular, the innovation will deviate from a zero-mean random variable, and
these deviations observed for a variety of guesses for $\rho_0$ will yield
likelihood ratios that can be used to estimate the state, as was done in
\cite{CookRiofDeut14}.  One can also keep track of the trace of the unnormalized
state~(\ref{eq:quantum-operation}), which encodes the relative likelihood of the
trajectory given the evolution parameters, allowing one to judge different
parameter values against one another, as was implemented in \cite{ChasGere09}.

The differential equation that describes the unconditional evolution
corresponding to Eq.~(\ref{eq:vacsmedirect}) is called the master equation.  To
obtain the master equation, we simply average over measurement results in
\cref{eq:vacsmedirect}.  The only term that depends on the results is
$\df\innovation_D$, and its mean is zero, so the master equation is
\begin{align}\label{eq:vacme}
  \df\rho=\Expt{\df\rho_D}=\df t\,\D{\sqrt{\gamma}\,c}\rho\,.
\end{align}

Just as was the case for the SME~(\ref{eq:vacsmedirect}), \cref{eq:vacme} has no
explicit system-Hamiltonian term. The same reasoning that allowed us to add such
a term and arrive at \cref{eq:vacsmedirect-ext-ham} allows us to add the same
term to \cref{eq:vacme}:
\begin{align}\label{eq:vacme-ext-ham}
  \df\rho=\Expt{\df\rho_D}=-i\,\df t\,[H_\text{ext},\rho]
  +\df t\,\D{\sqrt{\gamma}\,c}\rho\,.
\end{align}
For the remainder of our presentation, such system-Hamiltonian terms are generally left
implicit.

\subsection{X basis measurement: Homodyne detection}\label{sec:homodyne}

\begin{figure}[ht!]
  \begin{center}
    \includegraphics[scale=.75]{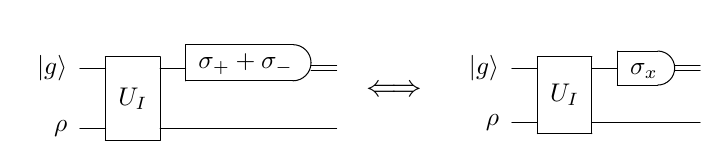}
  \end{center}
  \caption{Circuit depicting the system interacting with a vacuum probe (probe
  initially in the ground state) which is subsequently subjected to measurement
  of the qubit quadrature operator $\sigma_++\sigma_-=\sigma_x$, the qubit
  analogue of a homodyne measurement.  Just as for photon counting, the qubit
  measurement corresponds to a Pauli observable.}
  \label{fig:homodyne-ckt}
\end{figure}

We can produce, as in Brun's model \cite{Brun02}, a different system evolution
simply by measuring the probes in a different basis.  To be concrete, let us
consider measuring the $x$-quadrature of the field, $b\dg+b$.  In the
qubit-probe approach, this means measuring $\sigma_++\sigma_-=\sigma_x$ as shown
in \cref{fig:homodyne-ckt}.

This measurement projects onto the eigenstates
\begin{align}\label{eq:Xbasis}
\ket{\phi_\pm}\defined(\ket{g}\pm\ket{e})/\sqrt{2}
\end{align}
of $\sigma_x$.  The Kraus operators are linear combinations of the
photon-counting Kraus operators,
\begin{align}
  K_\pm&=\sand{\phi_\pm}{U_I}{g}=\frac{1}{\sqrt{2}}\left(K_g\pm K_e\right)
  =\frac{1}{\sqrt{2}}\left(\Id\pm\sqrtdt c-\half\dt\,c^\dagger c\right)\,,
  \label{eq:vac-homodyne-kraus-operators}
\end{align}
and the corresponding POVM elements (to linear order in $\Delta\tau$) are
\begin{align}
E_\pm=K\dg_\pm K_\pm=\frac12\Big(\Id\pm\sqrtdt(c+c\dg)\Big)\,,
\end{align}
which clearly satisfy $E_++E_-=\Id$.  We write the difference equation as
before, keeping terms up to order $\dt$,
\begin{align}
\begin{split}
  \label{eq:st-delrho}
  \Delta\rho_\pm
  &\defined\frac{K_\pm\rho K_\pm\dg}{\tr{\rho E_\pm}}-\rho \\
  &=\frac{\rho\pm\sqrtdt(c\rho+\rho c\dg)+\dt\,\D{c}\rho}
  {1\pm\sqrtdt\tr{\rho(c+c\dg)})}-\rho \\
  &=\left(\pm\sqrtdt-\dt\,\tr{(c+c\dg)\rho}\right)\Hc{c}\rho
  +\dt\,\D{c}\rho\,,
\end{split}
\end{align}
where we have again expanded the denominator using a standard series,
\begin{align}
  \frac{1}{1+x\sqrtdt}&=1-x\sqrtdt+x^2\dt+\BigO(\dt^{3/2})\,.
\end{align}

The dependence on the measurement result $\pm$ is reduced now to the
coefficient $\pm\sqrtdt$ in \cref{eq:st-delrho}. We rewrite this stochastic
coefficient as a random variable, $\Delta R$, again dependent on the
measurement outcome such that ${\Delta R:\pm\mapsto\pm\sqrtdt}$. The average
of this random variable to order $\dt$ is
\begin{align}\label{eq:homodyne_expect}
  \Expt{\Delta R}=\sqrtdt\tr{\rho E_+}-\sqrtdt\tr{\rho E_-}
  =\dt\,\tr{(c+c\dg)\rho}\,.
\end{align}
This is exactly the term subtracted from $\Delta R$ in the
coefficient of $\Hc{c}\rho$ in \cref{eq:st-delrho}; thus, defining the homodyne
version of the innovation as
\begin{align}\label{eq:innovationH}
\Delta\innovation_{H}\defined\Delta R-\Expt{\Delta R}\,,
\end{align}
we bring the homodyne difference equation into the form,
\begin{align}
  \label{eq:st-delrho-two}
  \Delta\rho_\pm
  =\dt\,\D{c}\rho
  +\Delta\innovation_{H}\Hc{c}\rho\,,
\end{align}
which is the difference equation one uses for numerical integration in the
presence of homodyne measurements.

Another simple calculation shows the second moment of $\Delta R$ to be
\begin{align}
  \Expt{(\Delta R)^2}=\dt\,\tr{E_+\rho}+\dt\,\tr{E_-\rho}=\dt\,.
\end{align}
By definition the innovation has zero mean, and its second moment is the
variance of $\Delta R$,
\begin{align}
\Expt{(\Delta\innovation_{H})^2}
=\Expt{(\Delta R)^2}-(\Expt{\Delta R})^2=\dt\,,
\end{align}
where again we work to linear order in $\dt$.  It is now trivial to write
the continuous-time stochastic differential equation that goes with the
difference equation~(\ref{eq:st-delrho-two}):
\begin{align}
  \df\rho_H&=\df t\,\D{\sqrt{\gamma}\,c}\rho+\df W\,\Hc{\sqrt{\gamma}\,c}\rho\,.
  \label{eq:vacsmehomo}
\end{align}
In the continuous limit, the innovation $\Delta\innovation_H$ becomes
$\sqrt{\gamma}\,\df W$, where $\df W$ is the Weiner process, satisfying
$\Expt{\df W}=0$ and $\Expt{\df W^2}=dt$.

Changing the measurement performed on the probes does not alter
the unconditional evolution of the system, so averaging over the homodyne
measurement results gives again the master equation~(\ref{eq:vacme}):
\begin{align}
  \df\rho=\Expt{\df\rho_H}=\df t\,\D{\sqrt{\gamma}\,c}\rho=\Expt{\df\rho_D}\,.
\end{align}

The results so far in this subsection are for homodyne detection of the probe
quadrature component $X=\sigma_++\sigma_-=\sigma_x$, \ie, measurement in
the basis~(\ref{eq:Xbasis}).  It is easy to generalize to measurement of an
arbitrary field quadrature $e^{i\varphi}b+e^{-i\varphi}b\dg$, which
for a qubit probe becomes a measurement of the spin component
\begin{align}\label{eq:Xvarphi}
X(\varphi)\defined e^{i\varphi}\sigma_-+e^{-i\varphi}\sigma_+=\sigma_x\cos\varphi+\sigma_y\sin\varphi\,.
\end{align}
This means measurement in the probe basis [eigenstates of $X(\varphi)$],
\begin{align}
\ket{\phi_{\pm}(\varphi)}\defined
  \frac{1}{\sqrt{2}}\big(\ket{g}\pm e^{-i\varphi}\ket{e}\big)\,,
\label{eq:phipmvarphi}
\end{align}
where we can also write
\begin{align}
\begin{split}
\ket{\phi_+(\varphi)}&=e^{-i\varphi/2}\big[\cos(\varphi/2)\ket{\phi_+}
+i\sin(\varphi/2)\ket{\phi_-}\big]\,,\\
\ket{\phi_-(\varphi)}&=e^{-i\varphi/2}\big[i\sin(\varphi/2)\ket{\phi_+}
+\cos(\varphi/2)\ket{\phi_-}\big]\,.
\end{split}
\end{align}
The resulting Kraus operators are
\begin{align}
\begin{split}
K_{\pm}(\varphi)&=\sand{\phi_\pm(\varphi)}{U_I}{g} \\
  &=\frac{1}{\sqrt2}\big(K_g\pm e^{i\varphi}K_e\big) \\
  &=\frac{1}{\sqrt{2}}\Big(\Id\pm\sqrtdt e^{i\varphi}c-\half\dt\,c\dg c\Big)\,,
\end{split}
\end{align}
with corresponding POVM elements
\begin{align}
E_\pm(\varphi)
=K_\pm\dg(\varphi)K_\pm(\varphi)
=\frac12\Big(\Id\pm\sqrtdt(e^{i\varphi}c+e^{-i\varphi}c\dg)\Big)\,.
\end{align}
We see that the results for measuring $X$ can be converted to those for measuring
$X(\varphi)$ by replacing $c$ with $c\,e^{i\varphi}$.  Thus the conditional difference
equation is
\begin{align}
  \label{eq:st-delrho-two-varphi}
  \Delta\rho_\pm
  =\dt\,\D{c}\rho
  +\Delta\innovation_H\Hc{c\,e^{i\varphi}}\rho\,,
\end{align}
and the vacuum SME becomes
\begin{align}
  \df\rho_H&=\df t\,\D{\sqrt{\gamma}\,c}\rho+\df W\,\Hc{\sqrt{\gamma}\,c\,e^{i\varphi}}\rho\,.
  \label{eq:vacsmehomo-arb-quad}
\end{align}

\subsection{Generalized measurement of X and Y: Heterodyne detection}\label{sec:heterodyne}

\begin{figure}[ht!]
  \begin{center}
    \includegraphics[scale=0.75]{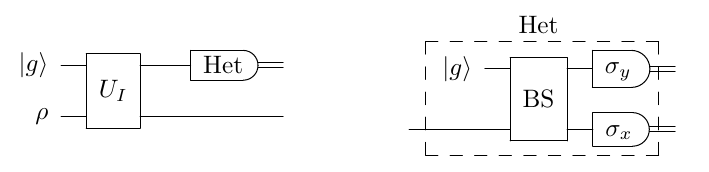}
  \end{center}
  \caption{Beamsplitter implementation of heterodyne measurement of
  $x$ and $y$ spin components of a qubit probe.}
  \label{fig:heterodyne-meas}
\end{figure}

Heterodyne measurement can be thought of as simultaneous measurement of two
orthogonal field quadrature components, \eg, $b+b\dg$ and $-i(b-b\dg)$.  In our
qubit model, this corresponds to simultaneously measuring along two orthogonal
axes in the $x$-$y$ plane of the Bloch sphere, \eg,
$\sigma_-+\sigma_+=\sigma_x=X$ and $i(\sigma_--\sigma_+)=\sigma_y=Y=X(\pi/2)$.
Obviously, it is not possible to measure these two qubit observables
simultaneously and perfectly, since they do not commute, but we can borrow a
strategy employed in optical experiments to measure two quadrature components
simultaneously.  The optical strategy makes two ``copies'' of the field mode to
be measured, by combining the field mode with vacuum at a 50-50 beamsplitter;
this is followed by orthogonal homodyne measurements on the two copies.  This
strategy works equally well for our qubit probes, once we define an appropriate
beamsplitter unitary for two~qubits,
\begin{align}\label{eq:BSeta}
\begin{split}
  \mathrm{BS}(\eta)
  &\defined\exp\Big[i\big(\eta\sigma_-\otimes\sigma_+ +\eta^*\sigma_+\otimes\sigma_-\big)\Big] \\
  &=\oprod{gg}{gg}+\oprod{ee}{ee}+\cos|\eta|\,\big(\oprod{ge}{ge}+\oprod{eg}{eg}\big)
  +i\sin|\eta|\,\big(e^{i\delta}\oprod{ge}{eg}+e^{-i\delta}\oprod{eg}{ge}\big)\,,
\end{split}
\end{align}
where $\eta=|\eta|e^{i\delta}$.  Specializing to $\eta=-i\pi/4$ yields
\begin{align}\label{eq:BSstates}
  \mathrm{BS}\defined\mathrm{BS}(-i\pi/4)
  =\oprod{gg}{gg}+\oprod{ee}{ee}
  +\frac{1}{\sqrt2}\big(\oprod{ge}{ge}+\oprod{eg}{eg}
  +\oprod{ge}{eg}-\oprod{eg}{ge}\big)\,.
\end{align}
This ``beamsplitter'' behaves rather strangely when excitations are fed to both
input ports, but this isn't an issue since the second (top) port of the
beamsplitter is fed the ground state, as illustrated in
\cref{fig:heterodyne-meas}.

It is useful to note here, for use a bit further on, that the beamsplitter
unitary, when written in terms of Pauli operators, factors into two commuting
unitaries,
\begin{align}\label{eq:BSfactored}
\mathrm{BS}=\exp\bigg(i\frac{\pi}{8}\sigma_x\otimes\sigma_y\bigg)
\exp\bigg(\!-i\frac{\pi}{8}\sigma_y\otimes\sigma_x\bigg)\,.
\end{align}
This factored form is easy to work with and leads to
\begin{align}\label{eq:BSsigmas}
\begin{split}
\mathrm{BS}&=
\Big[\Id\otimes\Id\cos(\pi/8)+i\sigma_x\otimes\sigma_y\sin(\pi/8)\Big]
\Big[\Id\otimes\Id\cos(\pi/8)-i\sigma_y\otimes\sigma_x\sin(\pi/8)\Big]\\
&=\frac{1}{2}\big(\Id\otimes\Id+\sigma_z\otimes\sigma_z\big)
+\frac{1}{2\sqrt2}\big(\Id\otimes\Id-\sigma_z\otimes\sigma_z
+i\sigma_x\otimes\sigma_y-i\sigma_y\otimes\sigma_x\big)\,,
\end{split}
\end{align}
which immediately confirms \cref{eq:BSstates}.

To calculate the Kraus operators for heterodyne measurement, we project
the first probe qubit onto the eigenstates of the spin component $X=\sigma_x$
and second probe qubit onto the eigenstates of the spin component $Y=\sigma_y$.
Before proceeding to that, we deal with a notational
point for the eigenstates of $\sigma_y=Y=X(\pi/2)$, analogous
to the notational convention for $\sigma_x$ that is summarized in
\cref{fig:eg-vs-01-convention}.  The conventional quantum-information
notation for the $\pm1$ eigenstates of $\sigma_y$ is
$\ket{\pm i}=(\ket0\pm i\ket1)/\sqrt2$, whereas as we introduced in
 Eq.~(\ref{eq:phipmvarphi}), we are using eigenstates that differ by a
 phase factor of $\mp i$:
\begin{align}
\ket{\phi_{\pm i}}\defined\ket{\phi_\pm(\pi/2)}
=\frac{1}{\sqrt{2}}\big(\ket{g}\mp i\ket{e}\big)
=\mp i\frac{1}{\sqrt2}\big(\ket0\pm i\ket1\big)
=\mp i\ket{\pm i}\,.
\end{align}
When all this is accounted for, the Kraus operators come out to be
\begin{align}
\begin{split}
  K_{\pm,\tpm}
  &=(\bra{\phi_\pm}\otimes\bra{\phi_{\,\tpm\,i}})(\Id\otimes\mathrm{BS})(U_I\otimes\Id)\ket{gg}\\
  &=\frac{1}{\sqrt2}\big\langle\phi_\pm(\pm\tpm\pi/4)\big|U_I\big|g\big\rangle\\
  &=\frac{1}{2}\big(K_g\pm e^{\pm\tpm i\pi/4}K_e\big)\\
  &=\frac{1}{2}\left(\Id\pm e^{\pm\tpm i\pi/4}\sqrt{\dt}c-\frac{1}{2}\dt\,c\dg c\right)\,,
\end{split}
\label{eq:kraus-hetro}
\end{align}
where we have introduced two binary variables, $\pm$ and $\tilde\pm$, to account
for the four measurement outcomes.  The juxtaposition of these two variables,
$\pm\tpm$, denotes their product, \ie, the parity of the two bits.  We see this
notation at work in
\begin{align}
\pm e^{\pm\tpm i\pi/4}=\pm\frac{1}{\sqrt2}(1\,{\pm\tpm}\,i)=\frac{1}{\sqrt2}(\pm1\tpm i)\,.
\end{align}
The POVM elements that correspond to the Kraus operators~(\ref{eq:kraus-hetro})
are
\begin{align}
  E_{\pm,\tpm}
  &=\frac{1}{4}\left(\Id\pm\sqrtdt\big(e^{\mp\tpm i\pi/4}c\dg+e^{\pm\tpm i\pi/4}c\big)\right)\\
  &=\frac{1}{4}\left[\Id+\sqrtdt\left(\pm\frac{c+c\dg}{\sqrt2}\tpm\frac{i(c-c\dg)}{\sqrt2}\right)\right]\,.
  \label{eq:het-povm}
\end{align}

\begin{figure}[ht!]
  \begin{center}
    \includegraphics[scale=1.2]{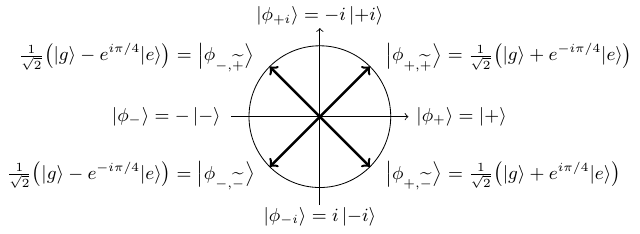}
  \end{center}
  \caption{The four states, $\ket{\phi_{\pm,\tpm}}=\ket{\phi_\pm(\pm\tpm\pi/4)}$,
  whose scaled projectors on the primary probe qubit make up the heterodyne POVM,
  as viewed in the $x$-$y$ plane of the Bloch sphere, shown relative to the
  positive and negative eigenstates of $\sigma_x=X$ and $\sigma_y=Y$.}
  \label{fig:heterodyne-states}
\end{figure}

The second form of the Kraus operators in \cref{eq:kraus-hetro} is equivalent to
finding the Kraus operators of the primary probe qubit for the heterodyne
measurement model on the left side of \cref{fig:heterodyne-meas}.  One sees from this second form
that the $\sigma_x$ and $\sigma_y$ measurements on the two probe qubits are
equivalent to projecting the primary probe qubit onto one of the following four
states:
\begin{equation}
  \ket{\phi_{\pm,\tpm}}\defined
  \ket{\phi_\pm(\pm\tpm\pi/4)}
  =\frac{1}{\sqrt{2}}\left(\ket{g}\pm e^{\mp\tpm i\pi/4}\ket{e}\right)
  =\frac{1}{\sqrt{2}}\left(\ket{g}+\frac{1}{\sqrt2}(\pm1\tmp i)\ket{e}\right)\,.
  \label{eq:het-meas-states}
\end{equation}
These four states are depicted in \cref{fig:heterodyne-states}; they carry
two bits of information, which are the results of the $\sigma_x$ and $\sigma_y$
measurements in the beamsplitter measurement model.  The four states not being
orthogonal, they must be subnormalized by the factor of $\sqrt2$ that appears
in \cref{eq:kraus-hetro} to obtain legitimate Kraus operators.

\begin{figure}[ht!]
  \begin{center}
    \includegraphics[width=\columnwidth]{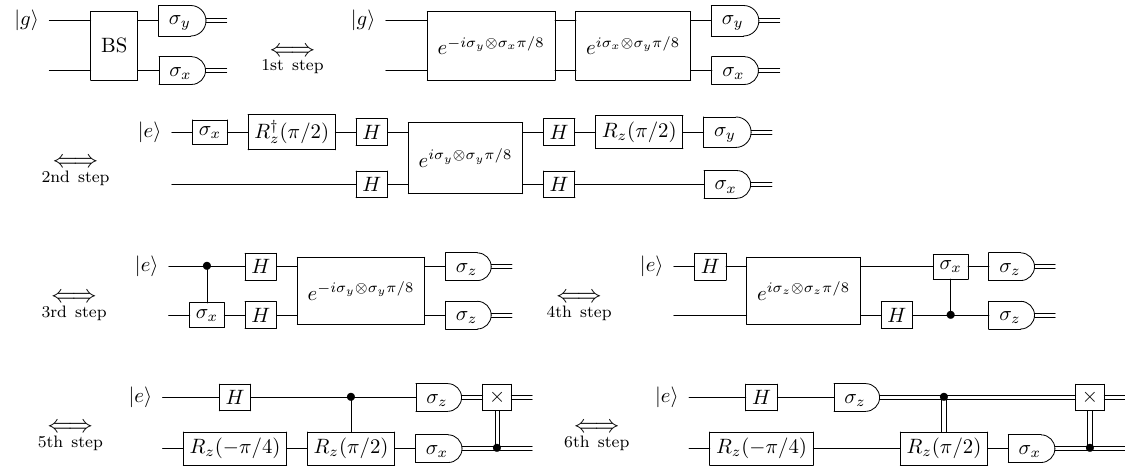}
  \end{center}
  \caption{Circuit-identity conversion of the original heterodyne measurement
    circuit of \cref{fig:heterodyne-meas}, which involves a beamsplitter on two
    probe qubits followed by $\sigma_x$ and $\sigma_y$ measurements on the two
    qubits, to a fair-coin flip, mediated by the ancillary (top) probe qubit,
    that chooses between measurements of $X(\varphi/4)$ and $X(-\varphi/4)$ on
    the primary (bottom) probe qubit.  The first step writes the beamsplitter
    unitary $\mathrm{BS}$ as a product of the two commuting unitaries in
    \cref{eq:BSfactored}.  The second step makes major changes.  It first
    discards the second piece of the beamsplitter unitary because that piece
    commutes with the measurements and thus has no effect on outcome
    probabilities.  It then changes the initial state of the ancillary probe
    qubit to $\ket e$ by including a bit flip $\sigma_x$; surrounds the
    beamsplitter unitary on the top wire with $\pi/2$ rotations about $z$ at the
    expense of changing the top-wire $\sigma_x$ in the beamsplitter unitary to
    $-\sigma_y$, \ie, $\sigma_x=-R_z(\pi/2)\sigma_y R_z\dg(\pi/2)$; and finally,
    surrounds the beamsplitter unitary with Hadamard gates,
    $H=(\sigma_x+\sigma_z)/\sqrt2$, on both wires, without changing the
    beamsplitter unitary because $H\sigma_y H=-\sigma_y$.  The third step
    discards the first $z$ rotation on the top wire because it only introduces
    an irrelevant phase change; pushes the $\sigma_x$ on the top wire to the end
    of the circuit, with the sign of the second $\sigma_y$ in the beamsplitter
    unitary changing along the way and with the gate ultimately being discarded
    because it becomes a $\sigma_y$ gate preceding the $\sigma_y$ measurement;
    converts the measurements to $\sigma_z$ measurements by using
    $\sigma_z=H\sigma_x H=HR_z\dg(\pi/2)\sigma_yR_z(\pi/2)H$; and finally
    introduces an initial CNOT gate, which does nothing since it acts on the
    initial state $\ket e$ on the top wire.  The fourth step pushes the CNOT
    through to the end of the circuit: the two Hadamards reverse the direction
    of the CNOT, putting the control on the bottom wire and the target on the
    top wire; pushing this CNOT through the beamsplitter unitary transforms the
    $\sigma_y\otimes\sigma_y$ to $-\sigma_x\otimes\sigma_z$.  After this move,
    the bottom Hadamard is pushed through the beamsplitter unitary, further
    converting the $\sigma_x\otimes\sigma_z$ to $\sigma_z\otimes\sigma_z$.  The
    fifth step converts the beamsplitter unitary to a rotation and a controlled
    rotation using
    $e^{i\sigma_z\otimes\sigma_z\pi/8}=\exp[-i\sigma_z\otimes\frac12(\Id-\sigma_z)\pi/4]e^{i\sigma_z\otimes\Id\pi/8}$;
    pushes the CNOT through the $\sigma_z$ measurements to become a classical
    controlled operation that does nothing if the outcome of the bottom
    measurement is the eigenvalue $+1$ ($\ket e$) and multiplies the result of
    the top measurement by $-1$ if the outcome of the bottom measurement is the
    eigenvalue $-1$ ($\ket g$); and finally pushes the Hadamard on the bottom
    wire through the measurement, converting it to a measurement of $\sigma_x$.
    The sixth step converts the controlled rotation into a classically
    controlled rotation of the bottom qubit, controlled on the outcome of the
    $\sigma_z$ measurement on the top qubit; this final circuit embodies the
    fair-coin flip version of the heterodyne measurement.  The apparently
    irrelevant CNOT introduced in the third step is actually crucial.  When
    pushed to the end of the circuit, it makes the outcome of the coin flip the
    parity of the original measurements of $\sigma_x$ and $\sigma_y$; the parity
    thus chooses between the measurement of $X(\pi/4)$ and $X(-\pi/4)$ on the
    primary probe qubit.  The classical version of this CNOT at the end of the
    circuit is there to get strict equivalence to the original circuit; it
    returns the classical bit carried by the top wire to the outcome of the
    $\sigma_y$ measurement in the original circuit.}
  \label{fig:alt-heterodyne-meas}
\end{figure}

We conclude that as far as the primary probe qubit is concerned, the heterodyne
measurement can be regarded as flipping a fair coin to determine whether one
measures $X(\pi/4)=(\sigma_x+\sigma_y)/\sqrt2$ or
$X(-\pi/4)=(\sigma_x-\sigma_y)/\sqrt2$.  The $\pm1$ eigenstates of $X(\pi/4)$
are $\ket{\phi_\pm(\pi/4)}$; $\ket{\phi_+(\pi/4)}=\ket{\phi_{+\widetilde+}}$ has
eigenvalue~$+1$, and $\ket{\phi_-(\pi/4)}=\ket{\phi_{-\widetilde-}}$ has
eigenvalue~$-1$.  The $\pm1$ eigenstates of $X(-\pi/4)$ are
$\ket{\phi_\pm(-\pi/4)}$; $\ket{\phi_+(-\pi/4)}=\ket{\phi_{+\widetilde-}}$ has
eigenvalue~$+1$, and $\ket{\phi_-(-\pi/4)}=\ket{\phi_{-\widetilde+}}$ has
eigenvalue~$-1$.  Notice that the fair coin that decides between these two
measurements is the parity of the measurements of $\sigma_x$ and $\sigma_y$ in
the heterodyne circuit of \cref{fig:heterodyne-meas}.
\Cref{fig:alt-heterodyne-meas} goes through the circuit identities that convert
the heterodyne circuit involving measurements on two probe qubits to one that is
a coin flip that chooses between the two measurement bases on the primary probe
qubit.

We now turn to deriving the explicit form of the conditional heterodyne
difference equation,
\begin{align}\label{eq:hetdiffeq}
  \Delta\rho_{\pm,\tpm}
  &\defined\frac{K_{\pm,\tpm}\,\rho\,K_{\pm,\tpm}\dg}{\tr{\rho E_{\pm,\tpm}}}-\rho\,.
\end{align}
This time we need two binary random variables to account for the dependence on
measurement outcome:
\begin{align}
  \Delta R_x:(\pm,\tpm)\mapsto\pm\sqrtdt\,,\\
  \Delta R_y:(\pm,\tpm)\mapsto\tpm\sqrtdt\,.
\end{align}
We want to write the equation in terms of innovations again, so we need the
probability distribution of measurement outcomes in order to calculate
$\Expt{\Delta R_x}$ and $\Expt{\Delta
R_y}$:
\begin{align}
  \pr(\pm,\tpm)=\tr{\rho E_{\pm,\tpm}}
  =\frac{1}{4}\Bigg(1\pm\sqrtdt\tr{\rho\frac{c+c\dg}{\sqrt2}}\tpm\sqrtdt\tr{\rho\frac{ic+(ic)\dg}{\sqrt2}}\Bigg)\,.
  \label{eq:het-out-prob}
\end{align}
The marginal probabilities, given by
\begin{align}
  \pr(\pm)
  &=\sum_{\tpm}\pr(\pm,\tpm)
  =\frac{1}{2}\Bigg(1\pm\sqrtdt\tr{\rho\frac{c+c\dg}{\sqrt{2}}}\Bigg)\,, \\
  \pr(\tpm)
  &=\sum_{\pm}\pr(\pm,\tpm)
  =\frac{1}{2}\Bigg(1\tpm\sqrtdt\tr{\rho\frac{ic+(ic)\dg}{\sqrt{2}}}\Bigg)\,,
  \label{eq:het-marg-prob}
\end{align}
allow us to calculate expectation values,
\begin{align}
  \Expt{\Delta R_x} &=\sum_\pm\pm\sqrtdt\Pr(\pm)=
  \dt\tr{\rho\frac{c+c\dg}{\sqrt{2}}}\,, \\
  \Expt{\Delta R_y} &=\sum_{\tpm}\tpm\sqrtdt\Pr(\tpm)=
  \dt\tr{\rho\frac{ic+(ic)\dg}{\sqrt{2}}}\,.
  \label{eq:het-meas-expt}
\end{align}
We can also find the correlation matrix,
\begin{align}
\Expt{(\Delta R_x)^2}&=\sum_\pm\dt\Pr(\pm)=\dt\,,\\
\Expt{(\Delta R_y)^2}&=\sum_{\tpm}\dt\Pr(\tpm)=\dt\,,\\
\Expt{\Delta R_x\Delta R_y}&=\sum_{\pm,\tpm}\pm\tpm\dt\Pr(\pm,\tpm)=0\,.
\end{align}
The first nonvanishing cross-moment of $\Delta R_x$ and $\Delta R_y$
is $\Expt{(\Delta R_x)^2(\Delta R_y)^2}=\dt^2$.  This means that we
should think of $\Delta R_x\Delta R_y=\pm\tpm\dt$ as a stochastic
term of order $\dt$, and this is too small to survive the limit
$\dt\rightarrow0$ (only stochastic terms of order $\sqrt{\dt}$ survive
this limit).

Returning now to the difference equation~(\ref{eq:hetdiffeq}), we find,
to linear order in $\dt$,
\begin{align}\label{eq:hetdiffeq2}
\begin{split}
  \Delta\rho_{\pm,\tpm}
  &=\dt\,\D{c}\rho
  +\frac{1}{\sqrt2}\big(\Delta\innovation_x\Hc{c}\rho+\Delta\innovation_y\Hc{ic}\rho\big) \\
  &\quad-\frac{1}{2}\Delta R_x\Delta R_y\Big(\tr{\rho\big(ic+(ic)\dg\big)}\Hc{c}\rho+\tr{\rho(c+c\dg)}\Hc{ic}\rho\Big)\,,
\end{split}
\end{align}
where we introduce the innovations for the two random processes,
\begin{align}
\Delta\innovation_x&\defined\Delta R_x-\Expt{\Delta R_x}\,, \\
\Delta\innovation_y&\defined\Delta R_y-\Expt{\Delta R_y}\,.
\end{align}
The innovations are zero-mean random processes, with variance
$\Expt{(\Delta\innovation_x)^2}=\Expt{(\Delta\innovation_x)^2}=\dt$.  Since, as
we discussed above, the term proportional to $\Delta R_x\Delta R_y$ is a zero-mean
stochastic term of order $\dt$ (and thus vanishes in the continuous-time limit),
we drop it, leaving us with the difference equation
\begin{align}\label{eq:hetdiffeq3}
  \Delta\rho_{\pm,\tpm}
  =\dt\,\D{c}\rho
  +\frac{1}{\sqrt2}\Big(\Delta\innovation_x\Hc{c}\rho+\Delta\innovation_y\Hc{ic}\rho\Big)\,.
\end{align}

When we take the continuous-time limit, the innovations
$\Delta\innovation_{x,y}$ become $\sqrt\gamma\,\df W_{x,y}$, where $\df W_{x,y}$
are independent Weiner processes, \ie, $\Expt{\df W_{x,y}}=0$ and $\Expt{\df
W_j\df W_k}=dt\,\delta_{jk}$.  The resulting SME is
\begin{align}
  \df\rho_{\mathrm{Het}}=\df t\,\D{\sqrt{\gamma}\,c}\rho
  + \frac{1}{\sqrt{2}}\Big(\df W_x\,\Hc{\sqrt{\gamma}\,c}\rho+\df W_y\,\Hc{i\sqrt{\gamma}\,c}\rho\Big)\,.
  \label{eq:vacsmehet}
\end{align}
The unconditional master equation, obtained by averaging over the
Weiner processes, is, of course, the vacuum master equation~(\ref{eq:vacme}).

Notice that the heterodyne SME~(\ref{eq:vacsmehet})
has the same form as homodyne SME~(\ref{eq:vacsmehomo}),
except that the former has two Weiner processes acting independently in the
place where the latter has just one.  This is a consequence of the
heterodyne measurement's having provided information about two quadrature
components of the system, $c+c\dg$ and $i(c-c\dg)$.
The relationship between heterodyne and homodyne SMEs has been discussed previously in the literature~\cite{GisiPerc92,WiseMilb93a}.

\subsection{Summary of qubit-probe measurement schemes}\label{sec:meas-summary}

\begin{table}[htp]
\begin{center}
\begin{tabular}{|c|c|c|c|}
\hline
Initial state & Measurement basis       & Kraus operators & SME\\
\hline
$\ket{g}$     & $\ket{e},\ket{g}$                                    &
  $K_e=\sqrtdt c,\,K_g=\Id-\half\dt\,c\dg c$                           &
  Jump \\
\hline
$\ket{g}$     & $\ket{\phi_\pm}=\frac{1}{\sqrt{2}}(\ket{g}\pm\ket{e}$      &
  $K_\pm =\frac{1}{\sqrt{2}}(\Id\pm\sqrtdt c-\half\dt\,c\dg c)$        &
  Homodyne $X$ \\
\hline
$\ket{g}$     & $\ket{\phi_{\pm}(\varphi)}=
  \frac{1}{\sqrt{2}}(\ket{g}\pm e^{-i\varphi}\ket{e})$ &
  $K_{\pm}(\varphi)=
  \frac{1}{\sqrt{2}}(\Id\pm\sqrtdt e^{i\varphi}c-\half\dt\,c\dg c)$ &
  Homodyne $X(\varphi)$ \\
\hline
$\ket{g}$     & $\ket{\phi_{\pm,\tpm}}=\frac{1}{\sqrt{2}}\left(\ket{g}+
  \frac{1}{\sqrt{2}}(\pm1\tmp i)\ket{e}\right)$ &
  $K_{\pm,\tpm}=\frac{1}{2}\left(\Id+\sqrtdt\frac{1}{\sqrt{2}}(\pm1\tpm i)c-
  \half\dt\,c\dg c\right)$ &
  Heterodyne \\
\hline
\end{tabular}
\end{center}
\caption {Input state, measurement basis, Kraus operators,
and type of resulting stochastic master equation~(SME); $X(\varphi)$
is the arbitrary probe quadrature defined in \cref{eq:Xvarphi}.}
\label{tab:kraus-sme}
\end{table}

To end this section, we briefly summarize the results.  Throughout this section,
we kept the probe initial state fixed as the ground state~$\ket g$, and we kept
the interaction unitary fixed as that in \cref{eq:disc-int-qubit-unitary}.  What
changed from one subsection to the next was the kind of measurement on the probe
qubits.  Section~\ref{sec:photon} analyzed measurements of the probe qubits in
$Z$ basis, which is analogous to photon-counting measurements for probe fields;
this resulted in a SME that is identical to the photon-counting SME.
Section~\ref{sec:homodyne} considered measurement of the probes in the $X$
basis, which is analogous to homodyne measurements on probe fields; this
resulted in a stochastic master equation that is  identical to the homodyne SME.
Section~\ref{sec:heterodyne} derived the stochastic master equation for a
generalized measurement on the probe qubits that is analogous to heterodyne
measurement on a probe field; the SME is identical to the heterodyne SME.  The
results for vacuum photon-counting, homodyne, and heterodyne measurements are
summarized in Table~\ref{tab:kraus-sme}, as well as the comparable information
for homodyne measurement of an arbitrary quadrature.

\section{Quantum trajectories for Gaussian probe-qubit states}\label{sec:gaussian-states}

In this section, we generalize the results of the previous section by addressing
the following question: Can we extend our qubit bath model, so successful in
capturing the behavior of vacuum stochastic dynamics, to describe more general
Gaussian stochastic dynamics?  By Gaussian we mean that the probe field is in a
state with a Gaussian Wigner function.  Gaussian baths are capable of describing
combinations of mean fields (probe field in a coherent state), thermal
fluctuations (probe field in a thermal state), and quadrature
correlation/anticorrelation (probe field in a squeezed state).  These baths have
been thoroughly studied in the literature.  Wiseman and Milburn, who did much of
the primary work in~\cite{WiseMilb93,WiseMilb93a,Wise94a}, summarize the results
in Wiseman's thesis~\cite{Wise94} and in their joint book~\cite{WiseMilb10}.
Important related work exists on simulation
methods~\cite{GardParkZoll92,DumParkZoll92} and the mathematical formalism
behind these descriptions~\cite{Goug03,NechPell09,AttaPell10,HellHoneKoum02}.

To handle the case of a vacuum probe field in terms of qubit probes, it is
sufficient, we found, to have a fixed initial probe state $\ket g$ and the fixed
interaction unitary~(\ref{eq:disc-int-qubit-unitary}).  To handle the general
Gaussian case in terms of qubit probes, we must allow a variety of initial probe
states, including mixed states, as one does with fields, but we also find it
necessary to allow modifications to the interaction
unitary~(\ref{eq:disc-int-qubit-unitary}).  The reason is that a qubit has
nowhere near as much freedom in states as even the Gaussian states of a field
mode; thus, for example, to handle a squeezed bath in terms of qubits, we have
to modify the interaction unitary to handle the quadrature-dependent noise that
for a field bath comes from putting the field in a squeezed state.

To guide our generalization procedure, we recall in
\cref{sec:gaussian-setup-fields} some facts about the standard field-mode
analysis; we also review how standard input-output formalism of quantum optics
emerges from this analysis.  We then proceed in \cref{sec:gaussian-setup-qubits}
to the translation to probe qubits.  Throughout these discussions, we label
field operators with the letters $b$ and $B$, and we label the analogous
operators for probe qubits with $a$ and $A$.

\subsection{Gaussian problem for probe fields and input-output formalism}
\label{sec:gaussian-setup-fields}

For a probe field divided up into the discrete temporal modes of \cref{sec:qbittraj},
we can introduce the quantum noise increment,
\begin{align}
  \Delta B_n&\defined\sqrtdt b_n=\sqrt\gamma\int_{t_n}^{t_{n+1}}\df s\,b(s)\,.
  \label{eq:field-white-noise-increment}
\end{align}
Gaussian bath statistics of the field are captured by the first and second
moments of these increments:
\begin{subequations} \label{eq:probe-increment-bath-stats}
  \begin{align}
    \expt{\Delta B_n}&=\expt{b_n}\sqrtdt=\alpha_n\dt\,,\quad\beta_n=\sqrt\gamma\,\alpha_n\,, \\
    \expt{\Delta B_n\dg\Delta B_n}&=\expt{b_n^\dagger b_n}\dt=N\dt\,, \\
    \expt{\Delta B_n^2}&=\expt{b_n^2}\dt=M\dt\,, \\
    \expt{[\Delta B_n,\Delta B_n\dg]}&=\expt{[b_n,b_n\dg]}\dt=\dt\,,
  \end{align}
\end{subequations}
where $\beta_n$ is the mean probe field, $N$ is related to the mean number of
thermal photons, and $M$ is related to the amount of squeezing.  These
interpretations are made more precise in
Secs.~\ref{sec:coh-states}--\ref{sec:squeezing}.  The
parameters $N$ and $M$ satisfy the inequality
\begin{align}\label{eq:gauss-param-ineq}
  \abs{M}^2&\leq N(N+1)\,,
\end{align}
which ensures that the field state is a valid Gaussian quantum state.  Noise
increments for different time segments are uncorrelated, in accordance with
the Markovian nature of Gaussian noise.

Much of the quantum-optics literature works directly with the quantum Weiner
process or infinitesimal quantum noise increment, $\df B(t)$, which is defined
as an appropriate limit of $\Delta B_n$,
\begin{align}\label{eq:field-white-noise}
  \df B(t)
  &\defined\int_t^{t+\df t}b(s)\,\df s
  =\lim_{\Delta t\to\df t}\sqrt{\Delta t}\,b_n
  =\lim_{\Delta t\to\df t}\Delta B_n/\sqrt\gamma\,,
\end{align}
where the limiting form assumes $t=t_n$.  Equation~(\ref{eq:field-white-noise})
is analogous to the relationship between classical white noise $\xi(s)$ and the
Weiner process~$dW_t\defined\int_t^{t+\df t}\xi(s)\,\df s$.  The Gaussian bath
statistics of an instantaneous field mode are described by the first and second
moments of $\df B(t)$,
\begin{subequations}\label{eq:bath-stats}
\begin{align}
  \expt{\df B(t)} &= \beta(t)\,\df t\,,\quad\beta(t)=\sqrt\gamma\,\alpha(t)\,, \\
  \expt{\df B\dg(t) \df B(t)} &= N\,\df t\,, \\
  \expt{\df B(t)^2} &= M\,\df t\,, \\
  \expt{[\df B(t),\df B\dg(t)]} &= \df t\,.
\end{align}
\end{subequations}

As a first step in our generalization to qubits below, we consider the unconditional
master equation for general Gaussian baths in the continuous limit (taken from
Eq.~4.254 of~\cite{WiseMilb10}):
\begin{align}
\begin{split}
  \df\rho=\df t\,\Big(
  &\big[\beta^*(t)\sqrt{\gamma}\,c-\beta(t)\sqrt{\gamma}\,c\dg,\rho\big]
  +(N+1)\D{\sqrt{\gamma}\,c}\rho
  +N\D{\sqrt{\gamma}\,c\dg}\rho \\
  &+\half M^*[\sqrt{\gamma}\,c,[\sqrt{\gamma}\,c,\rho]]
  +\half M[\sqrt{\gamma}\,c\dg,[\sqrt{\gamma}\,c\dg,\rho]]\Big)\,.
  \label{eq:gen-gauss-master-eq}
\end{split}
\end{align}
Notice that, as is well known, the terms linear in $c$, \ie, those
proportional to $\beta$, are a commutator that corresponds to Hamiltonian
evolution; indeed, this Hamiltonian is the system evolution one gets if one
replaces the bath by its mean field, neglecting quantum effects entirely. Just
as we discussed for an external Hamiltonian in Sec.~\ref{sec:qbittraj}, these
mean fields must vary much slower than $1/\Omega$. In comparing
\cref{eq:gen-gauss-master-eq} and other results between our paper
and~\cite{WiseMilb10}, it is important to be aware of the distinction between
our definitions of the $c$ operators.  We choose $c$ to be a dimensionless
system operator, whereas in Wiseman and Milburn, it contains an implicit factor
of $\sqrt{\gamma}$, which is pointed out in the paragraph below their
Eq.~(3.155).

Stochastic master equations for a general Gaussian bath, conditioned on the
sorts of measurements considered in Sec.~\ref{sec:qubit-vac}, are significantly more
complicated than the unconditional master equation~(\ref{eq:gen-gauss-master-eq})
and are the subject of Secs.~\ref{sec:coh-states}--\ref{sec:squeezing}.

Before getting to the qubit-probe model, however, we pause to review how
this is related to the input-output formalism of quantum optics.  From the
interaction unitary~(\ref{eq:disc-int-unitary}), we can calculate how the
probe-field operators for each time segment change in the Heisenberg
picture.  To make the distinction clear, we now label all the Heisenberg
probe operators before the interaction as $\Delta B_{n}^{\rm in} $.  The
output operators are obtained by unitarily evolving the input operators
\begin{align}
\Delta B_n^{\rm out}
=\sqrtdt b_n^{\rm out}
={U_I^{(n)\dag}} \Delta B_n^{\rm in}\,U_I^{(n)}
=\Big(\Id-\frac12\dt[c,c\dg]\Big)\otimes\sqrtdt b_n^{\rm in}
+ c\,\dt+\BigO(\dt^2)
= \Delta B_n^{\rm in}  + c\,\dt  +  \BigO(\dt^{3/2})\,,
\label{eq:IO_b}
\end{align}
which shows that the output field is the scattered input field plus radiation
from the system.  We can calculate the number of quanta in the output probe field,
\begin{align}
  \Delta N^\text{out}
  =\frac{\Delta B_n^{\text{out}\dag}\Delta B_n^\text{out}}{\dt}= b_n^{{\rm out}\dag}b_n^{\rm out}
=\Big(\Id-\dt[c,c\dg]\Big)\otimes {b_n^{\rm in}}\dg b_n^{\rm in}
+\Big(c\otimes \Delta B_n^{{\rm in}\dag}+c\dg\otimes\Delta B_n^{\rm in}\Big)
+c\dg c\,\dt  +  \BigO(\dt^{3/2})\,.
\label{eq:IO-flux}
\end{align}
In the literature these are known as input-output relations.  Analyses using
input-output relations were first used in quantum optics to analyze the noise
added as a bosonic mode is amplified~\cite{Haus62,Cave82} and, most importantly,
in the pioneering description of linear damping by Yurke and
Denker~\cite{YurkDenk84}.  The input-output relations display clearly how the
probe field is changed by scattering off the system.  Although we work in the
interaction picture in this paper, one can see the input-output relations at
work indirectly in our results.  Specifically, the conditional expectation of
the measurement result at the current time step [see, e.g.,
Eqs.~(\ref{eq:exptDeltaN}) and (\ref{eq:homodyne_expect}) and similar equations
below] is the trace of the relevant output operator with the initial field state
and a conditional system state.

Experienced practitioners of input-output theory might express concern about the
term proportional to $[c,c\dg]$ in \cref{eq:IO-flux}, but not to worry.  When
one takes the expectation of this equation in vacuum or a coherent state, the
commutator term becomes too high an order in $\dt$  and thus can be ignored.
For thermal and squeezed baths, \cref{eq:IO-flux} is irrelevant since we can't
sensibly perform photon counting on such fields due to the field's infinite
photon flux (which can be identified in our model as the finite photon-detection
probability in each infinitesimal time interval).

\subsection{Gaussian problem for probe qubits}
\label{sec:gaussian-setup-qubits}

To make the correspondence to our qubit-probe model, we define a
qubit quantum-noise increment analogous to the probe-field quantum noise
increment~(\ref{eq:field-white-noise-increment}):
\begin{align}
  \Delta A_n&\defined\sqrtdt a_n\,.
  \label{eq:qubit-white-noise-increment}
\end{align}
In Sec.~\ref{sec:qubit-vac}, we consistently chose $a_n$ to be the qubit
lowering operator, but in this section, we find it useful to allow more
general possibilities.  We remind the reader that in the picture of time
increments $\Delta t$, we work with the dimensionless time interval
$\dt=\gamma\Delta t$, not with $\Delta t$ itself.  The main way this might
cause confusion is that if we introduce a continuous-time noise increment
$dA=\sqrt{\df t}\,a_n$ for a qubit probe---or use the continuous-time field
increment $\df B(t)=\sqrt{dt}\,b_n$---we have to remember the factor of
$\sqrt\gamma$ in $\Delta A_n=\sqrt\gamma\sqrt{\Delta t}\,a_n$.

For probe qubits prepared in the (possibly mixed) state $\sigma$ (distinguished
from the similarly notated Pauli operators by subscripts or lack thereof), we
write the qubit bath statistics as
\begin{subequations} \label{eq:qubit-bath-stats}
  \begin{align}
    \expt{\Delta A_n}_\sigma&=\alpha_n\dt\,,\quad\beta_n=\sqrt\gamma\,\alpha_n\,, \\
    \expt{\Delta A_n\dg\Delta A_n}_\sigma&=N\dt\,, \\
    \expt{\Delta A_n^2}_\sigma&=M\dt\,, \\
    \expt{[\Delta A_n,\Delta A_n\dg]}_\sigma&=\dt\,.
  \end{align}
\end{subequations}
For the choice $a_n\defined\sigma_-^{(n)}$ that we used in
Sec.~\ref{sec:qubit-vac}, with vacuum probe state $\sigma=\oprod{g}{g}$, these
relations are satisfied with $\alpha=N=M=0$.  Notice that with a slight abuse of
notation, which we have already used and which can be excused because we only
want to get the scaling with $\gamma$ right, we have $\beta\,
dt=\expt{dA}=\expt{\Delta A}/\sqrt\gamma=\alpha\dt/\sqrt\gamma=\sqrt\gamma\,\alpha
dt$, which implies that $\beta_n=\sqrt\gamma\,\alpha_n$, as displayed above.

Replacing the explicit $\sigma_-$ in the interaction
unitary~(\ref{eq:disc-int-qubit-unitary-unsimp}) with the more general qubit
operator $a$ (and thus $\sigma_+$ with $a\dg$), we get a new interaction
unitary,
\begin{align}\label{eq:UIa}
\begin{split}
U_I&=\Id\otimes\Id
  +\sqrtdt\left( c\otimes a_n\dg - c\dg\otimes a_n\right)
  +\frac{1}{2}\dt\left(c\otimes a_n\dg - c\dg\otimes a_n\right)^2 \\
  &=\Id\otimes\Id
  +\left(c\otimes\Delta A_n\dg - c\dg\otimes\Delta A_n\right)
  +\frac{1}{2}\left(c\otimes\Delta A_n\dg - c\dg\otimes\Delta A_n\right)^2\,,
\end{split}
\end{align}
which we use throughout the remainder of this section, specifying the operator
$a_n$ appropriately for each case we consider.  Using this new interaction
unitary, we find that the expectation values~(\ref{eq:qubit-bath-stats}) are the
only properties of the bath that influence the unconditional master equation,
\begin{align}
\label{eq:gen-qubit-master-eq}
\begin{split}
  \Delta\rho_n&\defined\Tr\probe\big[U_I(\rho_n\otimes\sigma_n)U_I\dg\big]-\rho_n \\
  &=\big[\expt{\Delta A_n}_\sigma^* c-\expt{\Delta A_n}_\sigma c\dg,\rho_n\big]
  +\Big(\expt{\Delta A_n\dg\Delta A_n}_\sigma+
  \expt{[\Delta A_n,\Delta A_n\dg]}_\sigma\Big)
  \D{c}\rho_n+\expt{\Delta A_n\dg\Delta A_n}_\sigma\D{c\dg}\rho_n \\
  &\qquad\quad+\half\expt{\Delta A_n^2}_\sigma^*[c,[c,\rho_n]]+\half\expt{\Delta A_n^2}_\sigma[c\dg,[c\dg,\rho_n]]\,.
\end{split}
\end{align}
Here $\Tr\probe$ denotes a trace over the $n$th probe qubit; in the interaction
unitary and the master equation, we only keep terms to linear order in $\dt$ or,
equivalently, quadratic order in $\Delta A_n$.  This tells us that satisfying
Eqs.~(\ref{eq:qubit-bath-stats}) is a necessary and sufficient condition for
reproducing the Gaussian master equation with our qubit model. Since a SME
implies a master equation, \cref{eq:qubit-bath-stats} is also a necessary
condition for reproducing the corresponding conditional evolution, \ie, the
Gaussian SMEs, with our qubit model.  This serves as a guiding principle for
exploring nonvacuum probes in the qubit model.

Notice that we could develop an input-output formalism for probe qubits,
analogous to that for fields in Eqs.~(\ref{eq:IO_b}) and~(\ref{eq:IO-flux}).
Since $a_n$ and $a_n^\dag$ do not satisfy the canonical bosonic commutation
relations, however, the qubit input-output relation will not have the same form
as the field relations~(\ref{eq:IO_b}) and~(\ref{eq:IO-flux}). Another
complication in the qubit input-output formalism is the dependence of $a_n$ on
the Gaussian field state we want to model, which results in a state-dependent
input-output relation. This complication shows up in the field input-output
relations as well, and so isn't unique to our qubit model. Everything would work
out right once we included the probe initial state and the appropriate
measurement, but these complications mean that the qubit input-output formalism
does not have the simple interpretation we can attach to the vacuum field
version, so we do not develop it here.

\subsection{Coherent states and mean-field stochastic master equation}
\label{sec:coh-states}

One way to extend the qubit model presented so far is to to generalize
to nonvacuum Gaussian pure states, the simplest of which is a coherent state.
For a field probe, we create a coherent state with a wave-packet mean field
$\beta(t)$ by applying to the vacuum the continuous-time displacement operator \cite{BlowLoudPhoe90},
\begin{align}\label{eq:D-field}
D[\beta(t)]\defined\exp\!\left(\int\df t\,[\beta(t)b\dg(t)-\beta^*(t)b(t)]\right)\,.
\end{align}
To use this continuous-time displacement operator, it is often convenient to
write it as a product of displacement operators for the field modes $b_n$ of the
time increments, during each of which the mean field is assumed to be
essentially constant, yielding
\begin{align}
D[\beta(t)]=\prod_n D(\alpha_n)\,,
\end{align}
where
\begin{align}\label{eq:D-field-mode}
D(\alpha_n)\defined e^{\sqrt{\Delta t}(\beta_nb\dg_n-\beta^*_nb_n)}
=e^{\sqrtdt(\alpha_n b\dg_n-\alpha^*_n b_n)}
=e^{\alpha_n\Delta B\dg_n-\alpha^*_n\Delta B_n}\,,
\end{align}
is the displacement operator for the $n$th field mode $b_n$ and
$\beta_n=\beta(t_n)=\sqrt\gamma\,\alpha_n$.  Applying this displacement operator
to vacuum creates a product coherent state, in which the field mode $b_n$ for
the $n$th time increment is a coherent state with mean number of photons
$|\beta_n|^2\Delta t=\gamma\,|\alpha_n|^2\dt$.  Thus, in the continuous-time
limit, the mean rate at which photons encounter the system is
$|\beta(t)|^2=\gamma\,|\alpha(t)|^2$.

Up till this point in this section, we have retained the subscript $n$ that
labels each time increment, but from here on, as in Sec.~\ref{sec:qubit-vac}, we
omit this label because it is just a nuisance when dealing with the time
increments one at a time.  We only note that the omitted $n$ dependence is
necessary to describe a time-changing mean field
$\alpha(t)=\beta(t)/\sqrt\gamma$.

To translate from field modes to qubits, we let $a=\sigma_-$, as in
Sec.~\ref{sec:qubit-vac}, and we introduce a qubit analogue of a displacement
operator for a probe qubit,
\begin{align}\label{eq:qubit-displacement-operator}
    D(\alpha)
    \defined e^{\sqrtdt(\alpha\sigma_+-\alpha^*\sigma_-)}
    =e^{\alpha\Delta A\dg-\alpha^*\Delta A}\,.
\end{align}
This operator doesn't act much like the field displacement operator for large
displacements, but because we are working with small time increments, we can
assume that $\alpha\sqrtdt$ is small and expand the displacement operator as
\begin{align}\label{eq:qubit-displacement-operator-expand}
\begin{split}
    D(\alpha)
    &=\Id+\alpha\Delta A\dg-\alpha^*\Delta A
    +\half\big(\alpha\Delta A\dg-\alpha^*\Delta A\big)^2+\BigO(\dt^{3/2}) \\
    &=\Id+\alpha\Delta A\dg-\alpha^*\Delta A
    -\half|\alpha|^2\big(\Delta A \Delta A\dg+\Delta A\dg\Delta A\big)
    +\BigO(\dt^{3/2})\,,
\end{split}
\end{align}
where the final form uses $\Delta A^2=0=(\Delta A\dg)^2$ since $a=\sigma_-$.
Throughout we work to linear order in $\dt$, without bothering to indicate
explicitly that the next-order terms are $\BigO(\dt^{3/2})$. The aficionado
might notice \cref{eq:qubit-displacement-operator-expand} is related to the
quantum stochastic differential equation for the displacement (or ``Weyl'')
operator; see Eq.~(4.11) of~\cite{BoutHandJame07}.

Applying the displacement operator to the ground state $\ket g$ gives the
normalized probe coherent states,
\begin{align}
\begin{split}
  \label{eq:qubit-coherent-state}
  \ket{\alpha}&\defined D(\alpha)\ket{g} \\
  &=(1-\half|\alpha|^2\dt)\ket{g}+\alpha\sqrtdt\ket{e} \\
  &=(1-\half|\alpha|^2\dt)\big(\ket g+\alpha\sqrtdt\ket e\big)\,.
\end{split}
\end{align}
The state~(\ref{eq:qubit-coherent-state}) is analogous to a field-mode
coherent state because it reproduces the mean-field bath statistics (and
therefore the unconditional master equation):
\begin{subequations}\label{eq:coherent-bath-stats}
\begin{align}
  \expt{\Delta A}_{\alpha}&=\alpha\,\dt\,, \\
  \expt{\Delta A\dg \Delta A}_{\alpha}&=0\,, \\
  \expt{\Delta A^2}_{\alpha}&=0\,, \\
  \expt{[\Delta A,\Delta A\dg]}_{\alpha}&=\dt\,.
\end{align}
\end{subequations}

In calculating the difference equation for any kind of measurement on the probe
qubits, we necessarily use normalized post-measurement system states.  Since we
normalize the post-measurement state we can work with an unnormalized probe
initial state, because the magnitude of the probe initial state cancels out when
the post-measurement state is normalized.  In particular, it is convenient to
work here with an unnormalized version of the coherent states,
\begin{align}
  \ket{\alpha}=\ket{g}+\alpha\sqrtdt\ket{e}\,,
\end{align}
keeping in mind that the resulting Kraus operators are off by a factor
of $1-\half|\alpha|^2\dt$ and POVM elements and probabilities of measurement
outcomes are off by a factor of $1-|\alpha|^2\dt$.

We focus now on the case of performing photon counting on the probes, \ie,
a measurement in the basis $\{\ket g,\ket e\}$.  This results in Kraus operators,
\begin{subequations}\label{eq:coherent-kraus-operators}
\begin{align}
  K_g&=\sand{g}{U_I}{\alpha}
  =\Id-\dt(\alpha c\dg+\half c\dg c)\,, \\
  K_e&=\sand{e}{U_I}{\alpha}=\sqrtdt(\alpha\Id+c)\,,
\end{align}
\end{subequations}
which are analogous to Eqs.~4.53 and~4.55 in~\cite{WiseMilb10} (in
comparing, recall that a Hamiltonian term can be added in trivially).
As we observed for the vacuum case in
Eqs.~(\ref{eq:vac-homodyne-kraus-operators}) and~(\ref{eq:kraus-hetro}), both
the homodyne and heterodyne Kraus operators are linear combinations of the
photon-counting Kraus operators.

Following our treatment of the vacuum case for
photon counting in Sec.~\ref{sec:photon}, we now find
a difference equation
\begin{align}\label{eq:diffeq-pc-alpha}
\begin{split}
  \Delta\rho_{\Delta N}
  &=\Delta N\,\mathcal{G}[\alpha\Id+c]\rho-\dt\,\Hc{\alpha c\dg+\half c\dg c}\rho\\
  &=\dt\big([\alpha^*c-\alpha c\dg,\rho]+\mathcal{D}[c]\rho\big)+\Delta\innovation_D\G{\alpha\Id+c}\rho\,,
\end{split}
\end{align}
where $\Delta N$ is the bit-valued random variable introduced in Sec.~\ref{sec:photon}, \ie, $\Delta N=0$
for outcome~$g$ and $\Delta N=1$ for outcome~$e$, and $\Delta\innovation_D=\Delta N-\Expt{\Delta N}$
is the photon-counting innovation~(\ref{eq:innovationD}).  Taking the continuous-time limit
gives the Gaussian SME with a mean field for the case of direct detection,
\begin{align}\label{eq:coh-SME-physics}
\begin{split}
  \df\rho_D
  &=dN\,\mathcal{G}[\beta\Id+\sqrt\gamma\,c]\rho
  -\df t\,\Hc{\beta\sqrt{\gamma}\,c\dg+\half\gamma c\dg c}\rho\\
  &=\df t\,\Big(
  \big[\beta^*\sqrt{\gamma}\,c-\beta\sqrt\gamma\,c\dg,\rho\big]+\mathcal{D}[\sqrt{\gamma}\,c]\rho
  \Big)
  +\df\innovation_D\mathcal{G}[\beta\Id+\sqrt\gamma\,c]\rho\,,
\end{split}
\end{align}
where $\beta=\sqrt\gamma\,\alpha$; this result is also found in~\cite{WiseMilb10}.
The driving terms due to the mean field are those of a Hamiltonian
$i\sqrt\gamma\,(\beta^*c-\beta c\dg)$, as we would get if we replaced the
probe operators with their mean values.  The unconditional master equation for a
mean-field probe follows from retaining only the deterministic part of
\cref{eq:coh-SME-physics} and agrees with \cref{eq:gen-gauss-master-eq} when
we set $N=M=0$.

An equivalent method for dealing with a bath with a mean field, more attuned to
the approach we use later in this section, is discussed at the end of
Sec.~\ref{sec:squeezing}
and is sketched in \cref{fig:sq-therm-mf-field}.

\subsection{Thermal states}

Having dealt with a pure state that carries a mean field, we turn now to
Gaussian states that have more noise than vacuum, \ie, thermal baths.  A
thermal state at temperature $T$ is defined by
\begin{equation}
  \sigma_\text{th}\defined\frac{e^{-H/k_BT}}{\tr{e^{-H/k_BT}}}\;,
  \label{eq:gen_thermal}
\end{equation}
For a field mode at frequency $\Omega$, the Hamiltonian is $H=\hbar\omega(a\dg a+\half)$,
and the corresponding thermal state is given by
\begin{align}\label{eq:osc_thermal}
  \sigma_\text{th}=\frac{1}{N+1}\sum_{m=0}^\infty\left(\frac{N}{N+1}\right)^m\oprod{m}{m}\,,
\end{align}
where
\begin{align}\label{eq:Nthermal}
  N\defined\frac{1}{e^{\hbar\omega/k_BT}-1}
\end{align}
is the mean number of photons.

The thermal state for a qubit probe is diagonal
in the basis $\{\ket g,\ket e\}$ with the ratio of excited-state population
to ground-state population being $N/(N+1)$:
\begin{align}
  \sigma_\text{th} = \frac{N+1}{2N+1}\oprod{g}{g}+\frac{N}{2N+1}\oprod{e}{e}\,.
  \label{eq:qb-thermal}
\end{align}
This state has an obvious problem, however, since if we choose
$a=\sigma_-$ ($\Delta A=\sqrtdt\sigma_-$), we find that $\expt{\Delta A\dg\Delta
A}_\text{th}/\dt=\sigma_+\sigma_-=\sand{e}{\sigma_\text{th}}{e}=N/(2N+1)$.
Indeed, no qubit state has more than one excitation in it, and the thermal
state~(\ref{eq:qb-thermal}) has at most half an excitation.  It is easy to deal
with this problem, however, by introducing an \emph{effective\/} qubit field
operator,
\begin{align}
  a_\text{th}=\sqrt{2N+1}\,\sigma_-\,,
  \label{eq:therm-qubit-field-op}
\end{align}
which goes into the qubit increment $\Delta A=\sqrtdt a_\text{th}$.
This increases the strength of the coupling of the qubit probes to the
system in a way that yields the desired bath statistics,
\begin{subequations}\label{eq:thermal-bath-stats}
\begin{align}
  \expt{\Delta A}_\text{th}&=0\,, \\
  \expt{\Delta A\dg\Delta A}_\text{th}&=N\,\dt\,, \\
  \expt{\Delta A^2}_\text{th}&=0\,, \\
  \expt{[\Delta A,\Delta A\dg]}_\text{th}&=\dt\,.
\end{align}
\end{subequations}
A glance at the interaction unitary~(\ref{eq:UIa}) shows that the rescaled coupling
strength is $\gamma_N=(2N+1)\gamma$, \ie,
\begin{align}\label{eq:UIathermal}
U_{\text{th},I}=
  \Id\otimes\Id
  +\sqrt{2N+1}\sqrtdt\left(c\otimes\sigma_+ - c\dg\otimes\sigma_-\right)
  +\frac{1}{2}(2N+1)\dt\left(c\otimes\sigma_+ - c\dg\otimes\sigma_-\right)^2\,.
\end{align}

The power delivered by this idealized broadband thermal bath is infinite, so
photon counting yields nonsensical results.  Instead, we consider homodyne
detection on the bath, \ie, measurement in the basis~(\ref{eq:Xbasis}),
which avoids the infinite-power problem.  Because the
probe state is a mixture of two pure states, $\oprod{g}{g}$ and $\oprod{e}{e}$,
we need Kraus operators corresponding to each combination of probe pure state
and measurement outcome in order to calculate the unnormalized updated state
[see~\cref{eq:anc-kraus-ops-mixed}]:
\begin{align}
K_{\pm g}&=\sqrt{\frac{N+1}{2N+1}}\sand{\phi_\pm}{U_{\text{th},I}}{g}
    =\frac{1}{\sqrt2}\sqrt{\frac{N+1}{2N+1}}\!\left(
    \Id\pm\sqrtdt\sqrt{2N+1}\,c-\frac12\dt(2N+1)c\dg c\right)\,, \\
K_{\pm e}&=\sqrt{\frac{N}{2N+1}}\sand{\phi_\pm}{U_{\text{th},I}}{e}
    =\pm\frac{1}{\sqrt2}\sqrt{\frac{N}{2N+1}}\!\left(
    \Id\mp\sqrtdt\sqrt{2N+1}\,c\dg-\frac12\dt(2N+1)cc\dg\right)\,.
\end{align}
The $\pm$ at the head of the expression for $K_{\pm e}$ can be
ignored, since Kraus operators always appear in a quadratic
combination involving the Kraus operator and its adjoint.
We are interested in the state after a measurement that yields
the result $\pm$, and this means summing over the two possibilities
for the initial state of the probe,
\begin{align}\label{eq:rhopmthermal}
\rho_\pm=\frac{K_{\pm g}\rho K\dg_{\pm g}+K_{\pm e}\rho K\dg_{\pm e}}{\tr{\rho E_\pm}}\,,
\end{align}
where
\begin{align}
E_\pm\defined K\dg_{\pm g}K_{\pm g}+K\dg_{\pm e}K_{\pm e}
=\frac12\bigg(\Id\pm\sqrtdt\frac{c+c\dg}{\sqrt{2N+1}}\bigg)\,.
\end{align}
is the POVM element for the outcome $\pm$.

The resulting difference equation for the system state is
\begin{align}
\begin{split}\label{eq:st-delrho-thermal}
  \Delta\rho_\pm
  &=\left(\pm\sqrtdt-\dt\frac{\tr{\rho(c+c\dg)}}{\sqrt{2N+1}}\right)
  \left((N+1)\frac{\mathcal{H}[c]\rho}{\sqrt{2N+1}}-N\frac{\mathcal{H}[c\dg]\rho}{\sqrt{2N+1}}\right) \\
  &\qquad+\dt(N+1)\D{c}\rho+\dt N\D{c\dg}\rho\vphantom{\bigg(} \\
  &=\dt\Big((N+1)\D{c}\rho+N\mathcal{D}[c\dg]\rho\Big)
  +\frac{\Delta\innovation_H}{\sqrt{2N+1}}\mathcal{H}\big[(N+1)c-Nc\dg\big]\rho\,,
\end{split}
\end{align}
where we use the same random process $\Delta R=\pm\sqrtdt$ and
innovation $\Delta\innovation_H=\Delta R-\Expt{\Delta R}$ as
for the vacuum \hbox{SME} for homodyning.  In the continuous-time limit,
the difference equation becomes
\begin{align}
  \label{eq:thermal-sme}
  \df\rho=\df t\,\Big((N+1)\D{\sqrt{\gamma}\,c}\rho+N\mathcal{D}[\sqrt{\gamma}\,c\dg]\rho\Big)
  +\frac{\df W}{\sqrt{2N+1}}\Hc{(N+1)\sqrt{\gamma}\,c-N\sqrt{\gamma}\,c\dg}\rho\,,
\end{align}
where $dW$ is the Weiner process that is the limit of the innovation.  This
result agrees with Eqs.~4.253 and 4.254 of~\cite{WiseMilb10} when we set $M=0$
in those equations.  The unconditional thermal master equation retains only the
deterministic part of \cref{eq:thermal-sme} and agrees with
\cref{eq:gen-gauss-master-eq} when we set $\beta=0$ and $M=0$.

The strategy of increasing the coupling strength clearly allows us to handle the
thermal-state SME, but it is worth spelling out in a
little more detail how that works, \ie, how we are able to mimic a field mode
that has all energy levels occupied in a thermal state with a qubit that has
only two levels.  Because the thermal state for a field mode is diagonal in the
number basis, the terms from $U_I(\rho\otimes\sigma_{\text{th}})U_I\dg$ that
survive tracing out the probe field are those balanced in $b_m$ and $b_m\dg$:
\begin{align}
-\half\Delta t\,c\dg c\,\gamma\tr{\sigma_{\text{th}}b\dg_n b_n}
-\half\Delta t\,cc\dg\,\gamma\tr{\sigma_{\text{th}}b_n b\dg_n}\,.
\end{align}
The normally ordered expression with $b_n\dg b_n$ corresponds to the system
absorbing an excitation from the bath, while the antinormally ordered expression
with $b_nb_n\dg$ corresponds to the system emitting an excitation into the bath.

Focusing just on the coupling strength for these two processes, the relevant
expressions are
\begin{align}
\begin{split}
  \gamma\tr{\sigma_{\text{th}}b_n\dg b_n}
  &=\gamma\sum_{m=0}^\infty\Pr\left(m\middle|N\right)m \\
  &=\frac{N}{2N+1}\sum_{m=0}^\infty\Pr\left(m\middle|N\right)\frac{\gamma(2N+1)m}{N} \\
  &=\frac{N}{2N+1}\gamma_N \,,
\end{split} \\
\begin{split}
  \gamma\tr{\sigma_{\text{th}}b_nb_n\dg}
  &=\gamma\sum_{m=0}^\infty\Pr\left(m\middle|N\right)(m+1) \\
  &=\frac{N+1}{2N+1}\sum_{m=0}^\infty\Pr\left(m\middle|N\right)\frac{\gamma(2N+1)(m+1)}{N+1} \\
  &=\frac{N+1}{2N+1}\gamma_N\,,
\end{split}
\end{align}
where
\begin{align}
  \Pr\left(m\middle|N\right)=\frac{1}{N+1}\left(\frac{N}{N+1}\right)^m
\end{align}
is the thermal probability for $m$ photons given mean number~$N$ and
\begin{align}
  \gamma_N&\defined(2N+1)\gamma
\end{align}
is a rescaled interaction strength.  The terms have been written so as to
suggest the following: absorption occurs with overall probability $N/(2N+1)$ and
effective interaction strength $\gamma(2N+1)m/N$, which depends on the number of
photons $m$ in the field mode, and emission occurs with probability
$(N+1)/(2N+1)$ and effective interaction strength $\gamma(2N+1)(m+1)/(N+1)$. The
absorption and emission probabilities are, respectively, proportional to the
absorption and total (spontaneous plus stimulated) emission rates given by the
Einstein $A$ and $B$ coefficients for a collection of two-level atoms in thermal
equilibrium with an optical cavity at temperature $T$
\cite[Sec.~1.2.2]{ChiaGarr08}. Since $\expt{m}=N$, both of the effective
interaction strengths average to the rescaled interaction strength $\gamma_N$.
This is what allows us to replace the effective interaction strengths by their
average and pretend that only two bath levels undergo absorption and emission.

It is worth noting here what happens if we measure the rotated quadrature
component $X(\varphi)$ of \cref{eq:Xvarphi} instead of $X=\sigma_x$, \ie, if we
measure in the basis of \cref{eq:phipmvarphi}.  The Kraus operators become
\begin{align}\label{eq:anc-kraus-ops-mixed-varphi}
K_{\pm g}&=\sqrt{\frac{N+1}{2N+1}}\sand{\phi_\pm(\varphi)}{U_{\text{th},I}}{g}
    =\frac{1}{\sqrt2}\sqrt{\frac{N+1}{2N+1}}\!\left(
    \Id\pm\sqrtdt\sqrt{2N+1}\,e^{i\varphi}c-\frac12\dt(2N+1)c\dg c\right)\,, \\
K_{\pm e}&=\sqrt{\frac{N}{2N+1}}\sand{\phi_\pm(\varphi)}{U_{\text{th},I}}{e}
    =\pm\frac{e^{i\varphi}}{\sqrt2}\sqrt{\frac{N}{2N+1}}\!\left(
    \Id\mp\sqrtdt\sqrt{2N+1}\,e^{-i\varphi}c\dg-\frac12\dt(2N+1)cc\dg\right)\,.
\end{align}
The $\pm e^{i\varphi}$ at the head of the expression for $K_{\pm e}$ can be
ignored, since a Kraus operator always appears in combination with its adjoint.
Thus all the results for homodyne measurement of $X(\varphi)$ follow from those
for homodyne measurement of $\sigma_x$ by replacing $c$ by $e^{i\varphi}c$.  In
particular, the difference equation and the limiting SME are given by
\begin{align}
\label{eq:st-delrho-thermal-varphi}
  \Delta\rho_\pm
  &=\dt\Big((N+1)\D{c}\rho+N\mathcal{D}[c\dg]\rho\Big)
  +\frac{\Delta\innovation_H}{\sqrt{2N+1}}\mathcal{H}\big[(N+1)e^{i\varphi}c-Ne^{-i\varphi}c\dg\big]\rho\,,\\
  \label{eq:thermal-sme-varphi}
  \df\rho
  &=\df t\,\Big((N+1)\D{\sqrt{\gamma}\,c}\rho+N\mathcal{D}[\sqrt{\gamma}\,c\dg]\rho\Big)
  +\frac{\df W}{\sqrt{2N+1}}\Hc{(N+1)\sqrt{\gamma}\,e^{i\varphi}c-N\sqrt{\gamma}\,e^{-i\varphi}c\dg}\rho\,.
\end{align}

\subsection{Pure and thermal squeezed states}
\label{sec:squeezing}

When we turn our attention to squeezed baths, the use of qubit probes
immediately presents a new challenge.  This comes from the obvious fact that if
we choose $a\propto\sigma_-$, as in all previous work in this paper, the second
moment that quantifies squeezing, $\expt{\Delta A_n^2}_{\text{sq}}=M\dt$, cannot
be nonzero for any choice of qubit probe state since $\sigma_-^2=0$.  To
surmount this obstacle, it is clear that we should make a different choice for
$a$; fortunately, once one has formulated the problem properly, the right choice
becomes obvious, although it has not been considered previously.

To see how to proceed, consider first the case of field modes in pure squeezed
vacuum,
\begin{align}\label{eq:squeezed-vacuum}
\ket{\phi_{\text{sq}}}\defined S(r,\mu)\ket{\text{vac}}\,,
\end{align}
which is generated from vacuum by the \emph{squeeze operator},
\begin{align}
  \label{eq:squeeze-operator}
  S(r,\mu)\defined\exp\Big[\half r\big(e^{-2i\mu}b^2-e^{2i\mu}{b\dg}^2\big)\Big]\,.
\end{align}
The squeeze operator conjugates the field annihilation operator $b$ according to
\begin{align}\label{eq:squeezing-transformation}
  S\dg(r,\mu)\,b\,S(r,\mu)=b\cosh r-e^{2i\mu}b\dg\sinh r
  \defines b_{\text{sq}}\,,
\end{align}
yielding new field operators~$b_{\text{sq}}$.  Using this transformation, it is
easy to see that $\expt{b}_{\text{sq}}=\expt{b_{\text{sq}}}_{\text{vac}}=0$ and
\begin{align}\label{eq:Npuresqueezed}
  N&=\expt{b\dg b}_{\text{sq}}=\expt{b\dg_{\text{sq}}b_{\text{sq}}}_{\text{vac}}=\sinh^2\!r\,, \\
  M&=\expt{b^2}_{\text{sq}}=\expt{b^2_{\text{sq}}}_{\text{vac}}=-e^{2i\mu}\sinh r\cosh r\,.
  \label{eq:Mpuresqueezed}
\end{align}
We stress that for all our results on a pure squeezed bath, Eqs.~(\ref{eq:Npuresqueezed})
and~(\ref{eq:Mpuresqueezed}) are the expressions
we use to relate the squeezing parameters $r$ and $\mu$ to the bath parameters $N$ and $M$
of \cref{eq:bath-stats}.  Notice that for this case of pure squeezed bath, the
inequality~(\ref{eq:gauss-param-ineq}) is saturated.  In this subsection, we find it
useful to let $M_R=(M+M^*)/2$ and $M_I=-i(M-M^*)/2$ denote the real and imaginary parts
of $M$.

\begin{figure}[ht!]
  \begin{center}
    \includegraphics[width=\columnwidth]{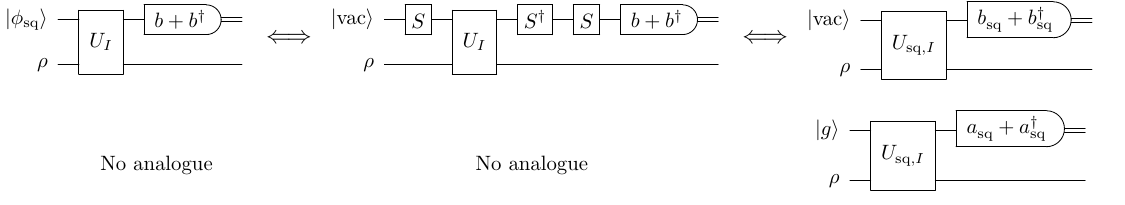}
  \end{center}
  \caption{The top row shows the transformation of the field-mode squeezed-bath
  circuit into a form where the squeezed noise, instead of being described by an
  initial squeezed state of a field mode, is described by squeezed field
  operators in the interaction unitary and in the measured observable.  In the
  left-most circuit, the field mode starts in the squeezed
  vacuum~(\ref{eq:squeezed-vacuum}); it interacts with the system via the joint
  unitary $U_I$ of \cref{eq:disc-int-unitary}; finally, it is subjected to a
  (homodyne) measurement of the observable $b+b\dg$.  The middle circuit
  introduces squeeze operators so that the field mode starts in vacuum, and the
  joint unitary and the measurement are ready to be transformed.  The third
  circuit shows the result of the transformation: the field mode starts in
  vacuum; it interacts with the system via the joint unitary $U_{\text{sq},I}$,
  in which the field-mode creation and annihilation operators in
  \cref{eq:disc-int-unitary} are replaced by the transformed operators
  $b_{\text{sq}}$ and $b\dg_{\text{sq}}$ of \cref{eq:squeezing-transformation};
  finally, the observable $b_{\text{sq}}+b\dg_{\text{sq}}$ is measured on the
  field mode.  The bottom row shows the corresponding squeezed-bath circuit for
  a qubit probe; this is a direct translation of the rightmost field-mode
  circuit to a qubit probe in the manner we are accustomed to.  The middle and
  leftmost circuits are not available to qubits, because the two-dimensional
  Hilbert space of the probe qubit cannot accommodate squeezed vacuum or a
  squeeze operator.  The qubit model involves a probe that starts in the ground
  state $\ket g$; interaction of the system and the probe qubit is described by
  the interaction unitary $U_{\text{sq},I}$ of \cref{eq:UIasqueezed}, which is
  obtained by substituting $a_{\text{sq}}$ of \cref{eq:qubit-sq-pure-field-op}
  for $a$ in the interaction unitary~(\ref{eq:UIa}); and finally, a measurement
  of the observable $a_{\text{sq}}+a\dg_{\text{sq}}$ on the qubit.}
  \label{fig:heis-sq-pure-field}
\end{figure}

The transformation~(\ref{eq:squeezing-transformation}) is the key to translating
from field modes to qubits.  What the transformation allows us to do is to model
squeezed noise in terms of vacuum noise that has a quadrature-dependent coupling
to the system.  In this section, we again focus on homodyne measurements of the
probe; just as for thermal states, this is because of the infinite photon intensity
of the infinitely broadband squeezed states we are considering.  As part of the
overall transformation, the homodyne measurement is also transformed to measurement
of another observable.   The transformation and the translation from field modes
to qubits are depicted and described in detail in terms of circuits in
\cref{fig:heis-sq-pure-field}.

The conclusion is that we can model squeezed noise in terms of qubits by starting
the probe in the ground state $\ket g$ and having it interact with the system
via an interaction unitary obtained from \cref{eq:UIa} by substituting
\begin{align}\label{eq:qubit-sq-pure-field-op}
\begin{split}
  a_{\text{sq}}
  &\defined\sigma_-\cosh r-e^{2i\mu}\sigma_+\sinh r \\
  &=\frac{\cosh r-e^{2i\mu}\sinh r}{2}\sigma_x-i\frac{\cosh r+e^{2i\mu}\sinh r}{2}\sigma_y
\end{split}
\end{align}
in place of $a$.  The resulting interaction unitary is
\begin{align}\label{eq:UIasqueezed}
\begin{split}
U_{\text{sq},I}&=\Id\otimes\Id
  +\sqrtdt\left(c\otimes a\dg_{\text{sq}} - c\dg\otimes a_{\text{sq}}\right)
  +\frac{1}{2}\dt\left(c\otimes a\dg_{\text{sq}} - c\dg\otimes a_{\text{sq}}\right)^2 \\
  &=\Id\otimes\Id
  +\sqrtdt\left(c_{\text{sq}}\otimes\sigma_+ - c\dg_{\text{sq}}\otimes\sigma_-\right)
  +\frac{1}{2}\dt\left(c_{\text{sq}}\otimes\sigma_+ - c\dg_{\text{sq}}\otimes\sigma_-\right)^2\,,
\end{split}
\end{align}
where
\begin{align}\label{eq:csq}
c_{\text{sq}}\defined c\cosh r+e^{2i\mu}c\dg\sinh r
\end{align}
is a species of squeezed system operator.

The qubit operators reproduce the general pure-state, but zero-mean-field Gaussian
bath statistics for $\Delta A_{\text{sq}}=\sqrtdt a_{\text{sq}}$:
\begin{subequations}\label{eq:squeezed-bath-stats}
\begin{align}
  \expt{\Delta A_{\text{sq}}^{}}_g &=0\,, \\
  \expt{\Delta A\dg_{\text{sq}}\Delta A_{\text{sq}}^{}}_g &=\sinh^2\!r\,\dt=N\,\dt\,, \\
  \expt{\Delta A_{\text{sq}}^2}_g &=-e^{2i\mu}\sinh r\cosh r\,\dt=M\,\dt\,, \\
  \expt{[\Delta A_{\text{sq}}^{},\Delta A\dg_{\text{sq}}]}_g &=\dt\,.
\end{align}
\end{subequations}
Thus we know that the qubit model generates the desired unconditional system
evolution.

A helpful way to think about this transformation is as a modification of the
coupling of the system to the bath.  Just as for thermal states, where we were
able to make up for a limited number of excitations in the bath by increasing
the interaction strength, here we compensate for the limitation that the qubit
ground state has equal uncertainties in $\sigma_x$ and $\sigma_y$ by modifying
the originally symmetric coupling to $a=\sigma_-=(\sigma_x-i\sigma_y)/2$ to the
asymmetric ``squeezed'' coupling embodied in the operator $a_{\text{sq}}$ of
\cref{eq:qubit-sq-pure-field-op}.  One sees the effect of the coupling strengths
most plainly when $\mu=0$, in which case $a_{\text{sq}}=(\sigma_x
e^{-r}-i\sigma_y e^r)/2$; \ie, the coupling of $\sigma_x$ to the system is
reduced by the squeeze factor $e^{-r}$, and the coupling of $\sigma_y$ is
increased by the same factor.  The change in coupling strengths isn't the only
twist, however.  The Pauli operators $\sigma_x=X$ and $\sigma_y=Y$ are
transformed under the squeezing transformation into
\begin{align}\label{eq:Xsq}
X_{\text{sq}}&\defined a_{\text{sq}}+a\dg_{\text{sq}}
=\sigma_x(\cosh r-\cos2\mu\sinh r)+\sigma_y\sin2\mu\sinh r\,,\\
Y_{\text{sq}}&\defined ia_{\text{sq}}-ia\dg_{\text{sq}}
=\sigma_x\sin2\mu\sinh r\,\sigma_x+\sigma_y(\cosh r+\cos2\mu\sinh r)\,.
\label{eq:Ysq}
\end{align}
These operators have the same commutator as $\sigma_x$ and $\sigma_y$, \ie,
$[X_{\text{sq}},Y_{\text{sq}}]=[\sigma_x,\sigma_y]=2i\sigma_z$, but unlike
$\sigma_x$ and $\sigma_y$, they are correlated in vacuum when $\sin2\mu\ne0$:
\begin{align}
  \expt{\half(X_{\text{sq}}Y_{\text{sq}}+Y_{\text{sq}}X_{\text{sq}})}_{\text{vac}}
  =2\sin2\mu\,\sinh r\cosh r=-2M_I\,.
\end{align}
This vacuum correlation is how our qubit model captures the correlation
between quadrature components in the squeezed state of a field mode.

\begin{figure}[ht!]
  \begin{center}
    \includegraphics{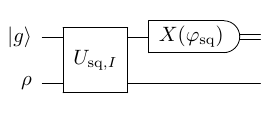}
  \end{center}
  \caption{Final qubit circuit for pure squeezed noise and a homodyne measurement
  of $\sigma_x$.  In our model of this situation, the probe, initially in the
  ground state~$\ket g$, interacts with the system via the unitary $U_{\text{sq},I}$
  of \cref{eq:UIasqueezed} and then is subjected to a measurement of the spin component
  $X(\varphi_{\text{sq}})$.  This model is identical to that of \cref{fig:heis-sq-pure-field}
  except that the measurement of $a_{\text{sq}}+a\dg_{\text{sq}}=\sqrt L X(\varphi_{\text{sq}})$
  is replaced by the equivalent measurement of $X(\varphi_{\text{sq}})$.}
  \label{fig:qubit-sq-pure-field}
\end{figure}

As mentioned above, we focus on homodyne detection, which in our transformed
and translated scheme, corresponds to measuring the observable $X_{\text{sq}}$
of \cref{eq:Xsq}, but $X_{\text{sq}}$ is not a normalized spin component.  We can,
however, write $X_{\text{sq}}$ as
\begin{align}
X_{\text{sq}}=\sqrt{L}X(\varphi_{\text{sq}})\,,
\end{align}
where
\begin{align}
  L\defined1+2\sinh^2\!r-2\cos2\mu\,\sinh r\cosh r
  \label{eq:L-def}
\end{align}
and $X(\varphi_{\text{sq}})$ is the normalized spin component of \cref{eq:Xvarphi},
with the phase angle defined by
\begin{align}\label{eq:varphisq}
e^{i\varphi_{\text{sq}}}\defined\frac{\cosh r-e^{-2i\mu}\sinh r}{\sqrt L}\,.
\end{align}
Since the factor $\sqrt L$ in $X_{\text{sq}}$ changes only the eigenvalues, not
the eigenvectors of the measured observable, we can say that we are measuring
the spin component $X(\varphi_{\text{sq}})$, instead of
$a_{\text{sq}}+a\dg_{\text{sq}}$; either way, we are measuring in the basis
$\ket{\phi_\pm(\varphi_{\text{sq}})}$.  Making this change brings our qubit
model for homodyne measurement on pure squeezed noise into its final form,
depicted in \cref{fig:qubit-sq-pure-field}.

The qubit model is now identical to the vacuum qubit model for measurement of
an arbitrary spin component, so we can obtain the results for the present case
by appropriating the results for vacuum homodyning, replacing $c$ with $c_{\text{sq}}$
of \cref{eq:csq} and choosing the homodyne angle to be $\varphi_{\text{sq}}$
of \cref{eq:varphisq}.  Formulas helpful in making this replacement are given
in App.~\ref{sec:ptsqb}.  The resulting Kraus operators are
\begin{align}
\begin{split}
K_{\pm}&=\sand{\phi_\pm(\varphi_{\text{sq}})}{U_{\text{sq},I}}{g} \\
  &=\frac{1}{\sqrt{2}}
  \Big(\Id\pm\sqrtdt e^{i\varphi_{\text{sq}}}c_{\text{sq}}-\half\dt\,c\dg_{\text{sq}}c_{\text{sq}}\Big)\\
  &=\frac{1}{\sqrt{2}}\left(\Id\pm\frac{\sqrtdt}{\sqrt{L}}\Big((N+M^*+1)c
  -(N+M)c\dg\Big)-\frac{1}{2}\dt\,\Big((N+1)c\dg c+Ncc\dg-M^*c^2-Mc^{\dagger\,2}\Big)\right)\,,
  \label{eq:sq-pure-kraus-ops}
\end{split}
\end{align}
with corresponding POVM elements
\begin{align}
E_\pm
=K_\pm\dg K_\pm
=\frac12\!\left(\Id\pm\frac{\sqrtdt}{\sqrt L}(c+c\dg)\right)\,.
\end{align}

Likewise, the conditional difference equation is
\begin{align}\label{eq:diffeq-pure-squeezed}
\begin{split}
  \Delta\rho_\pm
  &=\dt\,\D{c_{\text{sq}}}\rho
  +\Delta\innovation_H\Hc{e^{i\varphi_{\text{sq}}}c_{\text{sq}}}\rho \\
  &=\dt\left((N+1)\D{c}\rho+N\mathcal{D}[c\dg]\rho+
  \frac12 M^*\comm{c}{\comm{c}{\rho}}+\frac12 M[c\dg,[c\dg,\rho]\right)
  +\frac{\Delta\innovation_H}{\sqrt{L}}\Hc{(N+M^*+1)c-(N+M)c\dg}\rho \,,
\end{split}
\end{align}
and the SME is
\begin{align}\label{eq:pure-squeezed-SME-csqform}
  \df\rho=\df t\,\D{\sqrt{\gamma}\,c_{\text{sq}}}\rho
  +\df W\,\Hc{\sqrt{\gamma}\,c_{\text{sq}}\,e^{i\varphi_{\text{sq}}}}\rho\,,
\end{align}
which  becomes
\begin{align}\label{eq:pure-squeezed-SME}
\begin{split}
  \df\rho&=\df t\left((N+1)\D{\sqrt{\gamma}\,c}\rho+N\mathcal{D}[\sqrt{\gamma}\,c\dg]\rho+
  \frac{1}{2}M^*\comm{\sqrt{\gamma}\,c}{\comm{\sqrt{\gamma}\,c}{\rho}}+
  \frac12M[\sqrt{\gamma}\,c\dg,[\sqrt{\gamma}\,c\dg,\rho]\right) \\
  &\quad+\frac{\df W}{\sqrt{L}}\Hc{\left(N+M^*+1 \right)\sqrt{\gamma}\,c
  -(N+M)\sqrt{\gamma}\,c\dg}\rho\,.
\end{split}
\end{align}
This SME is presented as Eqs~4.253 and~4.254 in~\cite{WiseMilb10}; see also
Sec.~4.4.1 of Wiseman's thesis~\cite{Wise94}.

\begin{figure}[ht!]
\begin{center}
  \subfloat[]{
    \includegraphics[width=\columnwidth]{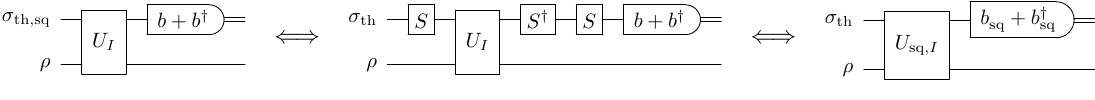}
    \label{fig:heis-sq-therm-field}
  }\\
  \subfloat[]{
     \includegraphics{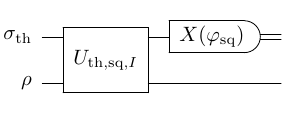}
     \label{fig:qubit-sq-therm-field}
  }
  \end{center}
  \caption{(a)~Field-mode circuits for a squeezed thermal bath.  The circuit on the left, in which
  the field mode begins in squeezed thermal state $\sigma_{\text{th,sq}}$, is transformed so that the
  effect of the squeezing appears not in the initial field-mode state, but in the interaction
  unitary and in the observable that is measured on the field mode.  In the circuit on the right,
  the interaction unitary $U_{\text{sq},I}$ is obtained by replacing the field-mode creation
  and annihilation operators in \cref{eq:disc-int-unitary} with the transformed operators
  $b_{\text{sq}}$ and $b\dg_{\text{sq}}$ of \cref{eq:squeezing-transformation}; the measured
  observable is obtained from the same replacement.
  (b)~Qubit model for squeezed thermal bath.  The unitary
  $U_{\text{th,sq},I}$ is understood to be derived from the interaction
  unitary~(\ref{eq:UIa}) by substituting $a_{\text{th,sq}}$ of
  \cref{eq:qubit-sq-therm-field-op} in place of~$a$  or, equivalently,
  from the thermal interaction unitary~(\ref{eq:UIathermal}) by
  substituting in place of $\sigma_-$ the squeezed operator $a_{\text{sq}}$ of
  \cref{eq:qubit-sq-pure-field-op}.  The homodyne measurement is a
  measurement of $X_{\text{sq}}=a_{\text{sq}}+a\dg_{\text{sq}}$, but rescaled to be
  a measurement of the normalized spin component $X(\varphi_{\text{sq}})$,
  with the homodyne angle $\varphi_{\text{sq}}$ determined by \cref{eq:varphisq}.}
  \label{fig:sq-therm-field}
\end{figure}

Generalizing to a squeezed thermal bath for a field mode, for which the
inequality~(\ref{eq:gauss-param-ineq}) is not saturated, proceeds by
representing a squeezed thermal bath as a thermal state~(\ref{eq:osc_thermal})
to which the squeeze operator~(\ref{eq:squeeze-operator}) has been applied:
\begin{align}
  \sigma_{\text{th},\text{sq}}&=S(r,\mu)\sigma_{\text{th}}S\dg(r,\mu)\,.
\end{align}
For this squeezed thermal state, the bath parameters $N$ and $M$ are
functions both of the squeezing parameters $r$ and $\mu$ and of the
``thermal excitation number'' $N_{\text{th}}$:
\begin{subequations}\label{eq:sq-th}
\begin{align}\label{eq:thermal-excitations}
  N_{\text{th}}&\defined\expt{b_n\dg b_n}_{\text{th}}\,, \\
  N&=\expt{b\dg b}_{\text{th,sq}}=\expt{b\dg_{\text{sq}}b_{\text{sq}}}_{\text{th}}
  =(2N_{\text{th}}+1)\sinh^2\!r+N_{\text{th}}\,, \label{eq:Nthermalsqueezed} \\
  M&=\expt{b^2}_{\text{th,sq}}=\expt{b^2_{\text{sq}}}_{\text{th}}
  =-(2N_{\text{th}}+1)e^{2i\mu}\sinh r\cosh r\,. \label{eq:Mthermalsqueezed}
\end{align}
\end{subequations}
Making this squeezed thermal state the initial state of the field
mode gives the circuit for a squeezed thermal bath.  Transforming the
squeezing to appear not in the initial state, but in the interaction
unitary and the homodyne measurement, is the same as for a squeezed-vacuum
input and is depicted in \cref{fig:heis-sq-therm-field}.  We emphasize
that for the case of squeezed thermal bath,
Eqs.~(\ref{eq:sq-th}) are the expressions we use to derive the bath
parameters $N$ and $M$ from the thermal parameter $N_{\text{th}}$ and squeezing
parameters $r$ and $\mu$.

We need only to combine our work on thermal baths and pure squeezed baths to translate the
rightmost circuit in \cref{fig:heis-sq-therm-field} to a qubit probe.  The result of
the translation is depicted in \cref{fig:qubit-sq-therm-field}.  The initial state of the
probe qubit is the thermal state~(\ref{eq:qb-thermal}), now written as
\begin{align}\label{eq:qb-thermal-two}
 \sigma_{\text{th}}=\frac{N_{\text{th}}+1}{2N_{\text{th}}+1}\oprod{g}{g}
  +\frac{N_{\text{th}}}{2N_{\text{th}}+1}\oprod{e}{e}\,.
\end{align}
The interaction unitary $U_I$ is translated by letting $a$ in \cref{eq:UIa}
be the squeezed annihilation operator~(\ref{eq:qubit-sq-pure-field-op}), further
modified by being rescaled by the thermal coupling factor $\sqrt{2N_{\text{th}}+1}$,
\ie, $\Delta A_{\text{th,sq}}=\sqrtdt a_{\text{th,sq}}$, where
\begin{align}\label{eq:qubit-sq-therm-field-op}
  a_{\text{th},\text{sq}}\defined
  \sqrt{2N_{\text{th}}+1}\,a_{\text{sq}}
  =\sqrt{2N_{\text{th}}+1}\,\big(\sigma_-\cosh r-e^{2i\mu}\sigma_+\sinh r\big)\,.
\end{align}
The resulting interaction unitary is
\begin{align}\label{eq:UIathermalsqueezed}
\begin{split}
U_{\text{th,sq},I}
  &=\Id\otimes\Id
  +\sqrt{2N_{\text{th}}+1}\sqrtdt\left(c\otimes a\dg_{\text{sq}}-c\dg\otimes a_{\text{sq}}\right)
  +\frac{1}{2}(2N_{\text{th}}+1)\dt\left(c\otimes a\dg_{\text{sq}}-c\dg\otimes a_{\text{sq}}\right)^2 \\
  &=\Id\otimes\Id
  +\sqrt{2N_{\text{th}}+1}\sqrtdt\left(c_{\text{sq}}\otimes\sigma_+ - c\dg_{\text{sq}}\otimes\sigma_-\right)
  +\frac{1}{2}(2N_{\text{th}}+1)\dt\left(c_{\text{sq}}\otimes\sigma_+ - c\dg_{\text{sq}}\otimes\sigma_-\right)^2\,,
\end{split}
\end{align}
where $c_{\text{sq}}$ is the squeezed system operator of \cref{eq:csq}.
The homodyne measurement of $\sigma_x$ is transformed to a measurement of
$a_{\text{sq}}+a\dg_{\text{sq}}=\sqrt L X(\varphi_{\text{sq}})$; as we
discussed previously for a pure squeezed bath, we can regard this as a
measurement of the normalized spin component~$X(\varphi_{\text{sq}})$, with the phase
angle defined by \cref{eq:varphisq}.

The qubit operators $\Delta A_{\text{th,sq}}=\sqrtdt a_{\text{th,sq}}$ reproduce the general,
but zero-mean-field Gaussian bath statistics:
\begin{subequations}\label{eq:thermal-squeezed-bath-stats}
\begin{align}
  \expt{\Delta A_{\text{th,sq}}^{}}_{\text{th}} &=0\,, \\
  \expt{\Delta A\dg_{\text{th,sq}}\Delta A_{\text{th,sq}}^{}}_{\text{th}}
    &=\big[(2N_{\text{th}}+1)\sinh^2\!r+N_{\text{th}}\big]\dt=N\,\dt\,, \\
  \expt{\Delta A_{\text{sq}}^2}_{\text{th}} &=-(2N_{\text{th}}+1)e^{2i\mu}\sinh r\cosh r\,\dt=M\,\dt\,, \\
  \expt{[\Delta A_{\text{sq}}^{},\Delta A\dg_{\text{sq}}]}_{\text{th}} &=\dt\,.
\end{align}
\end{subequations}

For the conditional evolution we note that the circuit in \cref{fig:qubit-sq-therm-field}
is the same as that for a thermal probe with no squeezing, subjected to a homodyne
measurement specified by the angle $\varphi_{\text{sq}}$ of \cref{eq:varphisq}, and
with the system operator $c$ replaced by the squeezed system operator $c_{\text{sq}}$ of
\cref{eq:csq}.  In particular, just as the thermal bath with no squeezing has the two
pairs of Kraus operators in \cref{eq:anc-kraus-ops-mixed-varphi}, the squeezed thermal
bath gives the two pairs of Kraus operators,
\begin{align}
\begin{split}\label{eq:sq-therm-g-kraus-operators}
K_{\pm g}&=\sqrt{\frac{N_{\text{th}}+1}{2N_{\text{th}}+1}}
    \sand{\phi_\pm(\varphi_{\text{sq}})}{U_{\text{th,sq},I}}{g}\\
    &=\frac{1}{\sqrt2}\sqrt{\frac{N_{\text{th}}+1}{2N_{\text{th}}+1}}
    \!\left(
    \Id\pm\sqrtdt\sqrt{2N_{\text{th}}+1}\,e^{i\varphi_{\text{sq}}}c_{\text{sq}}
    -\frac12\dt(2N_{\text{th}}+1)c_{\text{sq}}\dg c_{\text{sq}}
    \right)\\
    &=\frac{1}{\sqrt2}\sqrt{\frac{N_{\text{th}}+1}{2N_{\text{th}}+1}}
    \bigg(
    \Id\pm\frac{\sqrtdt}{\sqrt{L'}}\Big((N+N_{\text{th}}+M^*+1)c-(N-N_{\text{th}}+M)c\dg\Big)\\
    &\phantom{\frac{1}{\sqrt2}\sqrt{\frac{N_{\text{th}}+1}{2N_{\text{th}}+1}}}\qquad
    -\frac12\dt\Big((N+N_{\text{th}}+1)c\dg c+(N-N_{\text{th}})c c\dg-M^*c^2-M{c\dg}^2\Big)
    \bigg)\,,
\end{split} \\
\begin{split}\label{eq:sq-therm-e-kraus-operators}
K_{\pm e}&=\sqrt{\frac{N_{\text{th}}}{2N_{\text{th}}+1}}
    \sand{\phi_\pm(\varphi_\text{sq})}{U_{\text{th,sq},I}}{e} \\
    &=\pm\frac{e^{i\varphi_\text{sq}}}{\sqrt2}\sqrt{\frac{N_{\text{th}}}{2N_{\text{th}}+1}}
    \!\left(
    \Id\mp\sqrtdt\sqrt{2N_{\text{th}}+1}\,e^{-i\varphi_{\text{sq}}}c_{\text{sq}}\dg
    -\frac12\dt(2N_{\text{th}}+1)c_{\text{sq}}c_{\text{sq}}\dg
    \right)\\
    &=\pm\frac{e^{i\varphi_\text{sq}}}{\sqrt2}\sqrt{\frac{N_{\text{th}}}{2N_{\text{th}}+1}}
    \bigg(
    \Id\pm\frac{\sqrtdt}{\sqrt{L'}}\Big((N-N_{\text{th}}+M^*)c-(N+N_{\text{th}}+M+1)c\dg\Big)\\
    &\phantom{\pm\frac{e^{i\varphi_\text{sq}}}{\sqrt2}\sqrt{\frac{N_{\text{th}}}{2N_{\text{th}}+1}}}\qquad
    -\frac12\dt\Big((N+N_{\text{th}}+1)cc\dg+(N-N_{\text{th}})c\dg c-M^*c^2-M{c\dg}^2\Big)
    \bigg)\,.
\end{split}
\end{align}
The corresponding POVM elements for the measurement outcomes $\pm$ are
\begin{align}
E_\pm\defined K\dg_{\pm g}K_{\pm g}+K\dg_{\pm e}K_{\pm e}
=\frac12\bigg(\Id\pm\frac{\sqrtdt}{\sqrt{L'}}(c+c\dg)\bigg)\,.
\end{align}
In these results we introduce
\begin{align}\label{eq:Lprime}
L'\defined(2N_{\text{th}}+1)L
=(2N_{\text{th}}+1)(1+2\sinh^2\!r-2\cos2\mu\,\sinh r\cosh r)
=2N+2M_R+1
\end{align}
[see Eqs.~(\ref{eq:Ldef}) and~(\ref{eq:Lprimedef})].

Updating the system state to find the conditional difference equation is done
using \cref{eq:rhopmthermal}, with the result
\begin{align}
\label{eq:sqth-difference-eq}
\begin{split}
  \Delta\rho_\pm
  &=\dt\Big(
  \big(N_{\text{th}}+1\big)\D{c_{\text{sq}}}\rho
  +N_{\text{th}}\mathcal{D}[c_{\text{sq}}\dg]\rho\Big)
  +\frac{\Delta\innovation_H}{\sqrt{2N_{\text{th}}+1}}
  \mathcal{H}\big[(N_{\text{th}}+1)e^{i\varphi_{\text{sq}}}c_{\text{sq}}
  -N_{\text{th}}e^{-i\varphi_{\text{sq}}}c\dg_{\text{sq}}\big]\rho \\
   &=\dt\!\left((N+1)\mathcal{D}[c]\rho+N\mathcal{D}[c\dg]\rho
   +\frac12 M^*[c,[c,\rho]]+\frac12 M[c\dg,[c\dg,\rho]]\right)
   +\frac{\Delta\innovation_H}{\sqrt{L'}}\Hc{(N+M^*+1)c-(N+M)c\dg}\rho\,.
\end{split}
\end{align}
The final result is identical to that given in \cref{{eq:diffeq-pure-squeezed}}
for the case of a pure squeezed bath (since $L'=2N+M_R+1$, which is what $L$ is
in the case of pure squeezed bath), except that now $N$ and $M$ need only
satisfy the inequality~(\ref{eq:gauss-param-ineq}), rather than saturating it.
This means that the corresponding SME has the form of
\cref{eq:pure-squeezed-SME}, but with $L$ replaced by $L'$.  Notice that in the
final form of the difference equation and the SME, all explicit reference to $N_{\text{th}}$
disappears, whereas the Kraus operators do depend explicitly on $N_{\text{th}}$; this is
because the Kraus operators involve projections onto the two possible
initial states, $\ket g$ and $\ket e$, of the probe qubit, whereas the
difference equation and  the SME only involve appropriate averages over these
two possibilities.

\begin{figure}[ht!]
\begin{center}
  \subfloat[]{
    \includegraphics[width=\columnwidth]{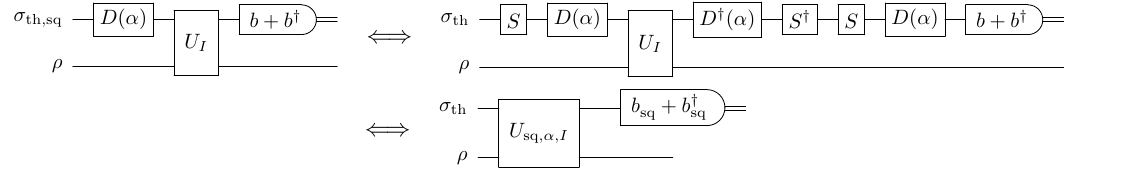}
    \label{fig:heis-sq-therm-mf-field}
  }\\
  \subfloat[]{
     \includegraphics{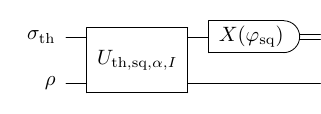}
     \label{fig:qubit-sq-therm-mf-field}
  }
  \end{center}
  \caption{(a)~Field-mode circuits for a squeezed thermal bath with a mean field.
  The field-mode displacement operator $D(\alpha)$ is defined in \cref{eq:D-field-mode}.
  The first circuit, in which the field mode begins in squeezed thermal state
  $\sigma_{\text{th,sq}}$, which is then displaced, is transformed so that the effect
  of the squeezing and the displacement appears in the interaction unitary and in the
  observable that is measured on the field mode.  In the last circuit, the interaction unitary
  $U_{\text{sq},\alpha,I}$ is obtained from the joint unitary~(\ref{eq:disc-int-unitary})
  by replacing $b$ with $b_{\text{sq}}+\alpha\sqrtdt$, where $b_{\text{sq}}$ is
  the squeezed annihilation operator of \cref{eq:squeezing-transformation}; the measured
  observable is obtained from the same replacement, except that the displacement can
  be ignored on the grounds that it does not change the measured basis, only the eigenvalues
  in that basis.
  (b)~Qubit model for squeezed thermal bath with a mean field.  The interaction
  unitary $U_{\text{th,sq},\alpha,I}$ is understood to be derived from the interaction
  unitary~(\ref{eq:UIa}) by substituting $a_{\text{th,sq},\alpha}$ of
  \cref{eq:qubit-sq-thermal-field-mf-op} in place of~$a$.  The homodyne
  measurement is a measurement of $X_{\text{sq}}=a_{\text{sq}}+a\dg_{\text{sq}}$,
  but rescaled to be a measurement of the normalized spin component
  $X(\varphi_{\text{sq}})$, with the homodyne angle $\varphi_{\text{sq}}$
  determined by \cref{eq:varphisq}.}
  \label{fig:sq-therm-mf-field}
\end{figure}

The case of a squeezed thermal bath is nearly the most general case of a
Gaussian bath, with pure-squeezed and unsqueezed-thermal baths emerging
as special cases.  The only thing unaccounted for in the squeezed-thermal
case is a mean field.  It is easy to add a mean field to the current paradigm;
the procedure for doing so is sketched in \cref{fig:sq-therm-mf-field}.
The key point is to modify the interaction unitary~(\ref{eq:UIa}) by replacing
$a$ with the operator
\begin{align}\label{eq:qubit-sq-thermal-field-mf-op}
  a_{\text{th,sq},\alpha}&\defined a_{\text{th,sq}}+\alpha\sqrtdt\,\Id
  =\sqrt{2N_{\text{th}}+1}\,\big(\sigma_-\cosh r-e^{2i\mu}\sigma_+\sinh r\big)+\alpha\sqrtdt\,\Id\,.
\end{align}
The effect of this is to add to the interaction
unitary~(\ref{eq:UIathermalsqueezed}) the additional term $\dt(\alpha^* c-\alpha
c\dg)$.  When this interaction unitary evolves a density operator $\rho$, the
additional term leads to a commutator, $\dt[\alpha^* c-\alpha c\dg,\rho]$, which
is, of course, just the commutator that describes the mean-field evolution in
the difference equation; it becomes the mean-field commutator $\df
t\,[\beta^*\sqrt{\gamma}\,c-\beta\sqrt{\gamma}\,c\dg,\rho]$ in the \hbox{SME}.
The mean-field terms appear only in the deterministic part of the SME and do not
affect the conditional evolution.

Perhaps as interesting as the success of the qubit technique we develop here is
the failure of a variety of other techniques, which generally are unable to
capture the conditional SME correctly. A sampling of these other techniques,
which typically involve either more than one probe qubit in each time segment or
probes with more than two levels, is discussed in App.~\ref{sec:mixed-squeezed}.

\section{Strong interactions at random times}\label{sec:interaction-rand}

In this section, we consider a different variety of continuous measurement.
Instead of probing the system with a continuous stream of weakly interacting
probes, we take inspiration from \cite{Milb14} and send a sequence of strongly
interacting probes distributed randomly in time. We show that this technique for
``diluting'' sequential interactions, \ie, by making the interactions occur less
frequently as opposed to more weakly, bears some resemblance to the cases
discussed before, but ultimately yields an inequivalent trajectory picture.
A slight variation on the dilution scheme is seen, however, to provide an
interpretation of inefficient detection equivalent to the commonly employed
model of attenuating the field incident on a photodetector with a beamsplitter.

\begin{figure}[ht!]
  \begin{center}
    \includegraphics[width=4.5truein]{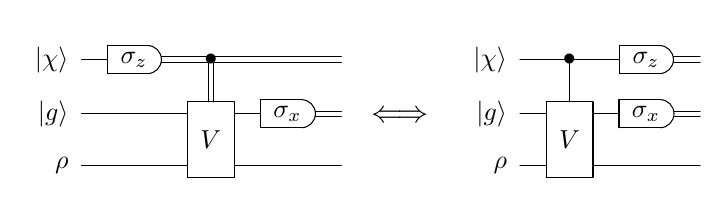}
  \end{center}
  \caption{Poisson measurement circuit.  The interaction between the system
    qubit and the probe qubit is the controlled unitary $V$ of \cref{eq:V}.
    Given that primary (lower) probe qubit begins in the ground state $\ket g$
    and that it is measured in the eigenbasis $\ket{\phi_\pm}$ of $\sigma_x$,
    the result is a measurement of $\sigma_z$ on the system.  The ancilla qubit
    on the top begins in the state $\ket\chi=\sqrt{1-\lambda\Delta
    t}\ket{e}+\sqrt{\lambda\Delta t}\ket{g}$.  In the left circuit, the ancilla
    qubit is subjected to a measurement of $\sigma_z$, the result of which
    controls classically the application of the joint unitary $V$.  The joint
    unitary is thus applied with probability $\lambda\Delta t$; in the
    continuous-time limit, this yields a sequence of measurements of $\sigma_z$
    on the system, which are Poisson-distributed in time with rate $\lambda$.
    In the right circuit, the classical control is moved through the measurement
    of $\sigma_z$ to become a quantum control, with the measurement of
    $\sigma_z$ telling one whether the joint unitary $V$ was applied to the
    system and the probe.}
  \label{fig:rand-meas-ckt}
\end{figure}

Probe dilution generalizes Sec.~\ref{sec:qubit-vac} by considering a
qualitatively different interaction, while fixing an initial probe state, and
thus complements the generalization carried out in
Sec.~\ref{sec:gaussian-states}, where the focus was on varying the initial probe
state.  The probabilistic interaction of a single qubit probe during a time
interval $\Delta t$ can be modeled as a joint unitary interaction (not
necessarily weak) between the system and the probe that is controlled off an
ancilla qubit; this interaction is followed by strong measurements of the
primary probe and the ancilla, as illustrated in \cref{fig:rand-meas-ckt}.
Measurement of the ancilla determines if the interaction between the system and
the probe occurred or not; thus, depending on the result of the ancilla
measurement, the probe measurement might or might not reveal information about
the system.  By choosing the state of the ancilla so that the probability of the
interaction occurring is $\lambda\Delta t$, the continuous limit can be thought
of as a long sequence of single-shot measurements occurring at
Poisson-distributed times with constant rate $\lambda$.

The probe/system interaction we consider is generated by an interaction
Hamiltonian of the form~(\ref{eq:interaction-picture-hamiltonian-discrete}),
except that we convert to qubit probes by making the standard replacement
$b_n\rightarrow\sigma_-$.  To achieve strong interactions, we now allow $\Delta
t$ to be as large as or even larger than $1/\gamma$.  A glance at the derivation
of the interaction
Hamiltonian~(\ref{eq:interaction-picture-hamiltonian-discrete}) in
Sec.~\ref{sec:qbittraj} shows that for such strong interactions, we cannot
neglect the sideband modes in each time segment.  Thus we cannot consistently
use such a strong interaction in the case of a probe field, but instead should
think of the interaction as coming directly from the interaction with a
sequence of qubit probes, perhaps two-level atoms.  That said, the interaction
Hamiltonian we assume for each probe time segment is
\begin{align}
H&\defined i\sqrt{\frac{\gamma}{\Delta t}}\left( c\otimes\sigma_+-c\dg\otimes\sigma_-\right)\,,
\end{align}
with corresponding joint unitary for the time segment,
\begin{align}\label{eq:V1}
V(\theta)=e^{-iH\Delta t}
=\exp\big[\half\theta\big(c\otimes\sigma_+-c\dg\otimes\sigma_-\big)\big]\,,
\end{align}
where we define $\theta\defined2\sqrt{\gamma\Delta t}$.  To ensure a strong interaction,
we hold $\theta$ constant in the limit $\Delta t\rightarrow0$.

When dealing with these strongly interacting probes, we are not justified in
truncating the Taylor expansion of the exponential.  This makes it difficult
to draw general conclusions for all kinds of systems and all coupling
operators~$c$, so we retreat to investigating a particular example to illustrate
what happens.  We assume the system is a qubit that is coupled to the
probe through the operator $c=\sigma_z$.  Thus the strong probe/system
interaction and subsequent measurement on the probe yield a measurement
of $\sigma_z$ on the system.  The interaction unitary~(\ref{eq:V1}) becomes
\begin{align}\label{eq:Vtheta}
\begin{split}
V(\theta)&=e^{\sigma_z\otimes(\sigma_+-\sigma_-)\theta/2}\\
&=e^{i\sigma_z\otimes\sigma_y\theta/2}\\
&=\Id\otimes\Id\cos(\theta/2)+i\sigma_z\otimes\sigma_y\sin(\theta/2)\\
&=\oprod{e}{e}\otimes R_y(-\theta)+\oprod{g}{g}\otimes R_y(\theta)\,,
\end{split}
\end{align}
where $R_y(\theta)\defined e^{-i\sigma_y\theta/2}$ is a rotation of the
probe qubit by angle $\theta$ about the $y$ axis of the Bloch sphere.  This
interaction is conveniently thought of as a controlled operation, with the system
qubit as control and the probe qubit as target.  If the system is in the excited
state $\ket e$, the probe qubit is rotated by $-\theta$ about $y$; if the system
is in the ground state state $\ket g$, the probe qubit is rotated by $\theta$
about $y$.

We assume that the probe starts in the ground state and that after the
interaction with the system, it is subjected to a measurement of $\sigma_x$, \ie,
a measurement in the basis $\ket{\phi_\pm}$ of \cref{eq:Xbasis}.  Under these
circumstances, it is easy to see that the measurement on the probe yields
information about the $z$ component of the system's spin.  Indeed, it becomes
an ideal measurement of $\sigma_z$ if we choose $\theta=\pi/2$, so that the
interaction unitary is
\begin{align}\label{eq:V}
V\defined V(\pi/2)
=\frac{1}{\sqrt2}\big(\Id\otimes\Id+i\sigma_z\otimes\sigma_y\big)
=\oprod{e}{e}\otimes R_y(-\pi/2)+\oprod{g}{g}\otimes R_y(\pi/2)\,.
\end{align}
This leaves us in the \emph{good-measurement limit\/} defined by~\cite{Milb14}.
One sees the perfect correlation between system and probe after the interaction
in
\begin{align}
  V\big(\alpha\ket{g}+\beta\ket{e}\big)\otimes\ket{g}
  &=\frac{1}{\sqrt{2}}
  \big(\alpha\ket{g}\otimes\ket{\phi_-}+\beta\ket{e}\otimes\ket{\phi_+}\big)\,.
  \label{eq:strong-int-corr}
\end{align}

To make the strong measurement events correspond to a Poisson process, we
control the unitary $V$ off an additional ancillary probe qubit that is initialized
in the state
\begin{align}
\ket{\chi}=\sqrt{1-\lambda\Delta t}\ket{e}+\sqrt{\lambda\Delta t}\ket{g}\,.
\end{align}
The controlled unitary is defined in the same way as the previously encountered
CNOT as
\begin{align}\label{eq:CV}
\mathrm{C}V\defined\Id\otimes\Id\otimes\oprod{e}{e}+V\otimes\oprod{g}{g}\,.
\end{align}
The ancilla qubit is measured in the eigenbasis of $\sigma_z$.  The unitary $V$
is applied to the system and the primary probe qubit when the result is $\ket g$,
which occurs with probability $\lambda\Delta t$.  The protocol for a single time segment
is summarized in the circuit diagrams of \cref{fig:rand-meas-ckt}.

The Kraus operators for obtaining result $\pm$ on the probe qubit and result $g$ or $e$ on the ancilla
qubit are
\begin{subequations} \label{eq:poiss-kraus}
\begin{align}
  K_{\pm,e}&=\big(\bra{\phi_\pm}\otimes\bra{e}\big)\mathrm{C}V\big(\ket g\otimes\ket\chi\big)
  =\sqrt{1-\lambda\Delta t}\sand{\phi_\pm}{\Id\otimes\Id}{g}=\sqrt{\frac{1-\lambda\Delta t}{2}}\Id\,, \\
  K_{\pm,g}&=\big(\bra{\phi_\pm}\otimes\bra{g}\big)\mathrm{C}V\big(\ket g\otimes\ket\chi\big)
  =\sqrt{\lambda\Delta t}\sand{\phi_\pm}{V}{g}=\sqrt{\lambda\Delta t}\frac{\Id\pm\sigma_z}{2}\,.
\end{align}
\end{subequations}
The ancilla outcome, $g$ or $e$, is not something for which we actually record
a measurement outcome, its purpose being merely to mock up a Poisson process, so we
coarse-grain over those measurement outcomes to find the unnormalized
conditional-state updates:
\begin{align}
\label{eq:pois-updates}
\begin{split}
K_{\pm,e}\rho K\dg_{\pm,e}+K_{\pm,g}\rho K\dg_{\pm,g}
&=\frac12(1-\lambda\Delta t)\rho+\lambda\Delta t\frac{\Id\pm\sigma_z}{2}\rho\frac{\Id\pm\sigma_z}{2}\\
&=\frac12\Big(\rho\pm\half\lambda\Delta t\,\big(\rho\sigma_z+\sigma_z\rho\big)
+\half\lambda\Delta t\,\big(\sigma_z\rho\sigma_z-\rho\big)\Big)\,.
\end{split}
\end{align}

It is informative to compare this model of occasional strong measurements
with our previously discussed model of continuous weak measurements.  To do so,
notice that if we remove the ancilla qubit from the circuit in \cref{fig:rand-meas-ckt}
and replace the strong interaction unitary $V$ with the analogous weak interaction
unitary $U_I$ of \cref{eq:disc-int-qubit-unitary}, letting $c=\sigma_z$, we
are left with the homodyne measurement model analyzed in Sec.~\ref{sec:homodyne}.
The Kraus operators for this model are given by \cref{eq:vac-homodyne-kraus-operators}
with $c=\sigma_z$, \ie,
\begin{align}
  K_\pm
  =\frac{1}{\sqrt{2}}\left(\Id\pm\sqrt{\gamma\Delta t}\,\sigma_z-\half\gamma\Delta t\,\Id\right)\,;
\end{align}
The corresponding unnormalized conditional-state updates are
\begin{align}
  \label{eq:sup-probe-updates}
  K_\pm\rho K_\pm\dg
  =\frac{1}{2}\left(\rho\pm\sqrt{\gamma\Delta t}\,\big(\rho\sigma_z+\sigma_z\rho\big)
  +\gamma\Delta t\,\big(\sigma_z\rho\sigma_z-\rho\big)\right)\,.
\end{align}

The unnormalized conditional state-update rules
\cref{eq:pois-updates,eq:sup-probe-updates} reveal an important distinction
between infrequent strong interactions and constant weak interactions. The
$\pm(\rho\sigma_z+\sigma_z\rho)$ terms correspond to conditional stochastic
evolution, while the $\sigma_z\rho\sigma_z-\rho$ terms correspond to the
unconditional average evolution. In the case of infrequent strong interactions,
the conditional and unconditional terms have the same scaling with respect to
the time interval $\Delta t$, in contrast to the different scalings of these
two terms in the case of continuous weak interactions.  This means, in particular,
that no matter what the Poisson rate $\lambda$ is, the stochastic steps corresponding
to the conditional evolution are of order $\Delta t$, not the $\sqrt{\Delta t}$
of a continuous weak interaction, and thus have vanishing effect on
the trajectory in the continuum limit.  The strong measurements, which project
the system onto an eigenstate of $\sigma_z$, occur too infrequently to have any
effect on the conditional evolution.  The $\pm$ measurement outcomes are pure
noise, providing no information about the system, and the conditional and
unconditional evolutions are identical.

Something more interesting happens when we combine weak interaction strength with
measurements that don't always happen.  If we assume that
$\theta=2\sqrt{\gamma\Delta t}=2\sqrtdt$  is small, then we are back in the
domain of weak interactions.  If we also define $\eta\defined\lambda\Delta t$
to be a finite number between $0$ and $1$, then we are in a situation where
ancilla outcome $g$, which occurs with probability $\eta$, leads to the weak
interaction, and ancilla outcome $e$, which occurs with probability $1-\eta$,
leads to no interaction.  This corresponds to an inefficient weak measurement.
It is easy to see that the Kraus operators are
\begin{subequations}\label{eq:inefficient-meas-kraus}
\begin{align}
  K_{\pm,e}&=\sqrt{\frac{1-\eta}{2}}\Id\,, \\
  K_{\pm,g}&=\sqrt{\frac{\eta}{2}}
  \left(\Id\pm\sqrtdt\,\sigma_z-\half\dt\Id\right)\,.
\end{align}
\end{subequations}
For the unnormalized conditional updates, we have
\begin{align}
\label{eq:inefficient-meas-updates}
\begin{split}
K_{\pm,e}\rho K\dg_{\pm,e}+K_{\pm,g}\rho K\dg_{\pm,g}
=\frac{1}{2}\left(\rho\pm\eta\sqrtdt\big(\rho\sigma_z+\sigma_z\rho\big)
  +\eta\dt\big(\sigma_z\rho\sigma_z-\rho\big)\right)\,.
\end{split}
\end{align}
Not surprisingly, these are the same as the updates~(\ref{eq:sup-probe-updates}),
except for the additional factors of $\eta$ in front of both the conditional and
unconditional parts of the evolution.  This means that we can read the
SME directly off the vacuum homodyne SME~(\ref{eq:vacsmehomo}), by incorporating
a factor of $\eta$ on the right-hand side:
\begin{align}
  \df\rho=\df t\,\eta\,\D{\sqrt{\gamma}\sigma_z}\rho
  +\df W\,\eta\,\Hc{\sqrt{\gamma}\sigma_z}\rho\,.
  \label{eq:eta-hom-sme}
\end{align}

By defining a renormalized coupling strength $\bar\gamma\defined\eta\gamma$, we
can put the SME~(\ref{eq:eta-hom-sme}) in the form
\begin{align}
 \df\rho=\df t\,\D{\sqrt{\bar\gamma}\sigma_z}\rho
  +\df W\sqrt\eta\,\Hc{\sqrt{\bar\gamma}\sigma_z}\rho\,.
  \label{eq:inefficient-hom-sme}
\end{align}
This equation is in the form of a homodyne measurement using detectors with
efficiency $\eta$ \cite[see Eq.~4.238]{WiseMilb10}, where the efficiency is the
probability of the detector's recording a count in the presence of an excitation.
Relative to the model in \cite{WiseMilb10}, the additional renormalization of the
coupling strength in \cref{eq:inefficient-hom-sme} is because our model has a probabilistic
interaction followed by a perfect measurement instead of a deterministic interaction
followed by a measurement with sub-unity efficiency.

\section{Examples: simulation and code}\label{sec:examples-and-code}

A helpful strategy for gaining intuition about the unconditional master equations
and SMEs discussed above is to visualize the kinds of evolution they describe.
The authors have published a software package in Python~\cite{GrosPySME17}
designed to make such visualizations easy to produce.  This package has been
used to create many of the visualizations below and includes documentation
that walks through formulating the quantum problem in a way that facilitates
application of known stochastic-integration techniques \cite{KloePlat92}.

\begin{figure}[ht!]
  \begin{center}
    \includegraphics[scale=.75]{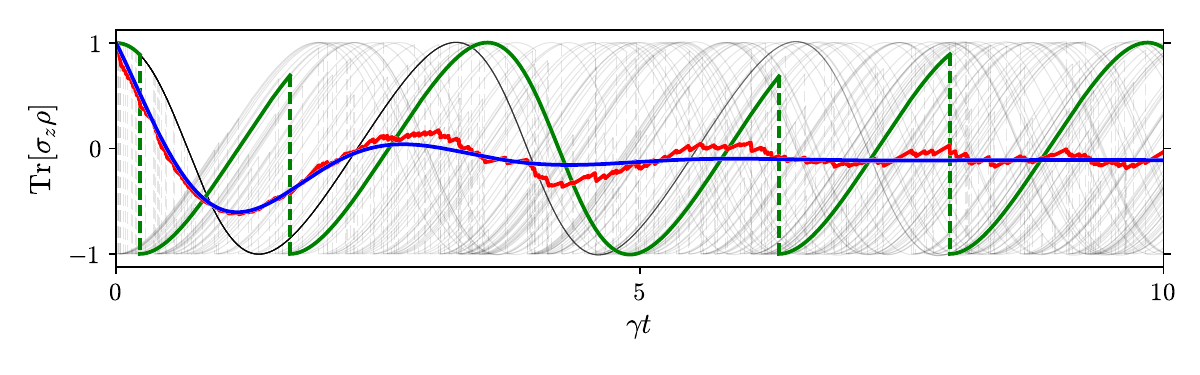}
  \end{center}
  \caption{Comparison of photon-counting conditional evolution to unconditional
  open-system dynamics. The smooth blue curve in the foreground is the
  unconditional evolution. The jagged red curve that closely follows the
  unconditional evolution is the ensemble average of 64 photon-counting
  trajectories, which are also displayed as wispy grayscale traces in the
  background. A single trajectory from that ensemble is highlighted in green,
  exhibiting discontinuous evolution at times when photons were detected
  (represented by dashed vertical lines) connected by smooth, nonlinear
  modifications to ordinary Rabi oscillations arising from the backaction of the
  ``no photon'' measurement result.}
  \label{fig:photon-counting-plot}
\end{figure}

We begin by considering a photon-counting measurement described by
\cref{eq:vacsmedirect}. Our example system is a two-level atom, coupled to some
one-dimensional continuum of modes of the electromagnetic field (perhaps a
waveguide) that are initially in the vacuum state, with coupling described by
the operator $c=\sigma_-$. We additionally include a classical field
driving Rabi oscillations between the two energy levels of the atom, as
described by a system Hamiltonian $H_{\rm ext}=\gamma\sigma_x$. The coupling to the
waveguide induces decoherence, so we expect the evolution of the system,
ignorant of the state of the waveguide, to exhibit damped Rabi oscillations.
This unconditional evolution is given by \cref{eq:vacme}, and when we solve for
the evolution, as shown by the smooth blue curve in the foreground of
\cref{fig:photon-counting-plot}, that is exactly what we see.

If we put a photon detector at the end of the waveguide, we maintain full
information about the two-level atom. Therefore, we don't expect to see
decoherence, but rather jumps in the system evolution when we detect photons in
the waveguide. When we solve for a particular instance of the stochastic
evolution, as highlighted by the discontinuous green curve in
\cref{fig:photon-counting-plot}, we see the jumps corresponding to photon
detection, as well as a deformation of pure Rabi oscillations that arises from
the backaction of the ``no-photon'' result from our photon detector.

\begin{figure}[ht!]
  \begin{center}
    \includegraphics[width=\columnwidth]{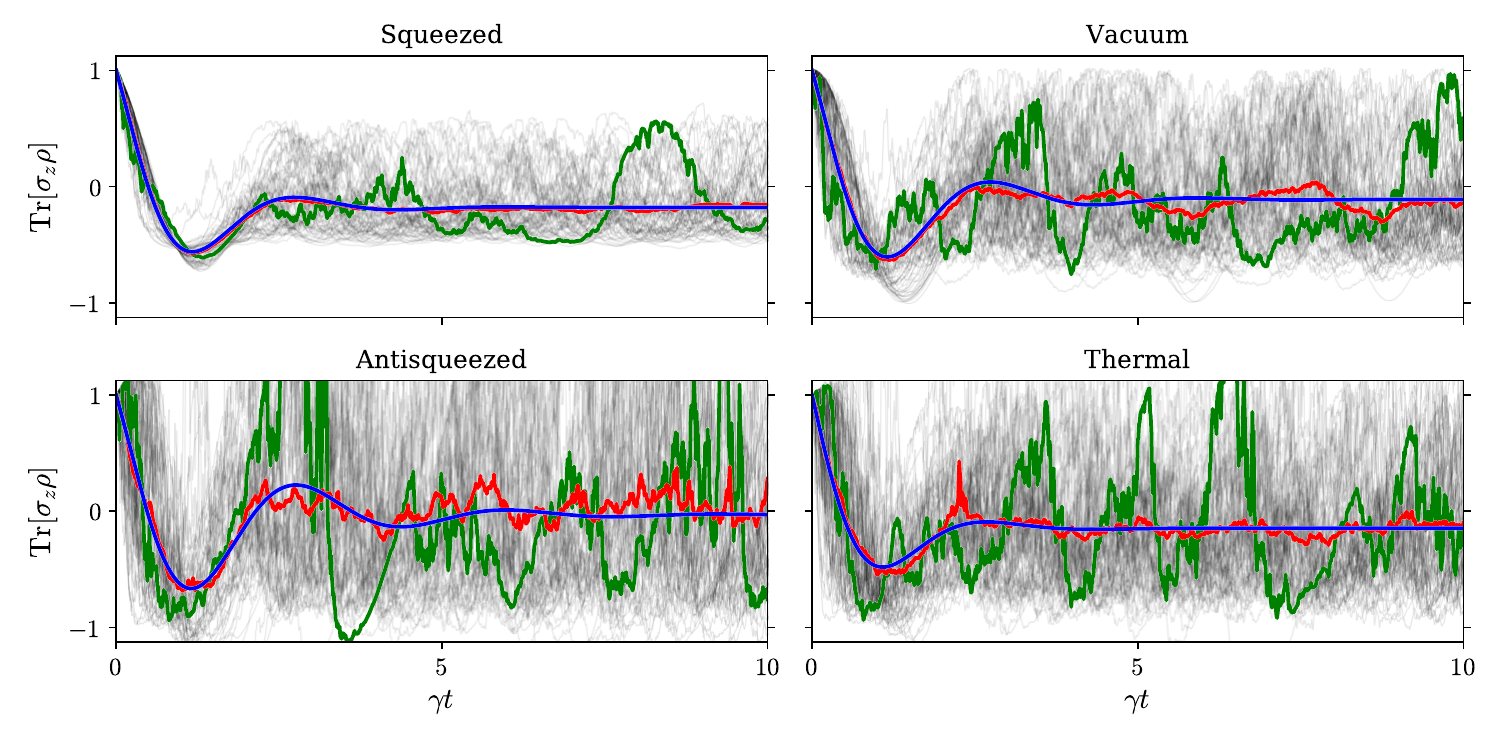}
  \end{center}
  \caption{Comparison of homodyne trajectories and unconditional evolutions for
  squeezed, vacuum, antisqueezed, and thermal baths. The interpretation of the
  various curves is analogous to \cref{fig:photon-counting-plot}: the smooth
  blue curves in the foreground are the unconditional dynamics, the jagged red
  curves that approximate those dynamics are ensemble averages over the 64
  trajectories plotted in the background in faint grayscale, and one member of
  that ensemble is highlighted in green. As one might expect, fluctuations in
  the homodyne trajectories decrease (increase) for the squeezed (antisqueezed,
  thermal) bath relative to the vacuum.  Astute observers might notice that the
  conditional expectation values sometimes exceed the range of the observable's
  spectrum.  While it might be tempting to attribute this to some fundamental
  property of the interaction, these excesses are in fact artefacts of the
  finite integration step size and indicate an unphysical density matrix with negative
  eigenvalues. The unconditional evolutions exhibit less (more)
  damping for antisqueezed (squeezed, thermal) baths relative to the vacuum.
  This is due to a combination of how much information from the system is being
  lost to the environment (more for squeezed, less for antisqueezed) and how
  much noise from the environment is polluting the system (thermal noise in the
  case of the thermal bath).}
  \label{fig:gauss-evol-comp}
\end{figure}

If we instead monitor the waveguide with homodyne measurements, the system
undergoes qualitatively different evolution. The system state never jumps, but
vacuum fluctuations appear as jagged white-noise effects in the trajectory. One
instance of a stochastic homodyne trajectory is highlighted in green in the
upper-right-hand corner of \cref{fig:gauss-evol-comp}. The other plots in
\cref{fig:gauss-evol-comp} provide some intuition regarding what effects the
squeezing and thermalization of the bath have on the system.

The expression that various SMEs are ``unravelings'' of
the unconditional master equation is meant to communicate that averages of
larger and larger ensembles of trajectories converge to the unconditional
evolution. These ensemble averages are presented in
\cref{fig:photon-counting-plot,fig:gauss-evol-comp} as the jagged red curves
closely hugging the smooth, blue curves of the unconditional evolution. For
finite ensembles of trajectories, one still sees vestigial qualities of the
underlying stochastic evolution, although the average trajectories are visibly
converging to the unconditional evolution.


\section{Discussion and Conclusion}\label{sec:discuss-conclusion}

We have now completed our introduction to trajectory theory, using only a
minimal understanding of quantum optics to construct an interaction unitary and
iterated quantum circuit. From that starting point, we built up the Gaussian
theory using tools from finite-dimensional quantum information and computation.
We found quantum circuits to illuminate the variety of conditions required to
make the Markov approximations. We also found that all the relevant qualities of
Gaussian bath modes can be compressed into the single transition of a qubit
probe, capturing the effects of mean fields with infinitesimal rotations,
thermal excitations with enhanced coupling, and squeezing with modified
correlations between system operators $c$ and $c\dg$ in the interaction. We also
explored an alternative means of producing weak interactions with an
environment, noting that stretches of isolation punctuated by strong probe
interactions can yield unconditional open-system dynamics similar to that
of continuous weak interactions, but the conditional stochastic dynamics
remain irreconcilably distinct.

Our program of qubit-izing continuous quantum measurements could be extended in
several ways. For instance, one can imagine defining a wavepacket single-photon
creation operator for $N$ qubits as
\begin{align}
  B\dg(\xi) = \sum_{n=1}^N \xi_i \sigma_+^{(n)}\quad \text{where}\quad \sum_n |\xi_n|^2=1\,.
\end{align}
Applying this to the multiqubit ``vacuum'' yields the single-photon state
\begin{align}
\ket{1_\xi}=B\dg(\xi)\ket{ggg\ldots g}=\xi_1\ket{egg\ldots g}+\xi_2\ket{geg\ldots g}+\xi_3\ket{gge...g}+\cdots+\xi_N\ket{ggg\ldots e}.
\end{align}
Quantum trajectories for the single-photon~\cite{GougJameNurd11,GougJameNurd12}
and $N$-photon case~\cite{BaraComb17} could be derived and studied, just as we did
for the Gaussian cases.
D\c{a}browska \emph{et al.} have done recent work along these lines~\cite{Dabr17,DabrSarbChru17}.

In addition to investigating different kinds of states, interactions, and
measurements, we could also imagine trying to reproduce
Carmichael's~\cite{Carm93} and Gardiner's~\cite{Gard93} cascaded quantum systems
formalism (and its generalization \cite{GougJame09,CombKercSaro17}).  In the
Markov approximation and the limit of zero delay between successive systems, we
could imagine using an input-output formalism for probe qubits, of the sort
alluded to at the end of \cref{sec:gaussian-setup-qubits}, to connect successive
systems, or we could work directly in the interaction picture as we have done
throughout this paper.

Such extensions hold the promise of identifying parsimonious descriptions of the
associated theory, but ultimately, we believe our approach can produce fruit
beyond economy of description.  The tools we emphasized encourage a particular
way of thinking about continuous measurements, trajectories, and feedback that
we believe is conducive to the development of improved forms of quantum control.

\vspace{1cm}
{\em Acknowledgments\/}: The authors thank Ben Baragiola and Rafael Alexander.
JG, JC, and CMC were supported in part by National Science Foundation
Grant Nos.~PHY-1521016 and~PHY-1314763 and by Office of Naval Research Grant
No.~N00014-11-1-0082.  JC was supported in part by the Australian Research
Council through a Discovery Early Career Researcher Award (DE160100356)
and by the Perimeter Institute for Theoretical Physics.  Research at the
Perimeter Institute is supported by the Government of Canada through the
Department of Innovation, Science and Economic Development Canada and by
the Province of Ontario through the Ministry of Research, Innovation and
Science.  GJM and JC were supported via the Centre of Excellence in Engineered
Quantum Systems (EQuS), Project Number~CE110001013.

\appendix
\section{Pure and thermal squeezed baths}\label{sec:ptsqb}

In this Appendix we list relations that are useful for deriving the results
for pure and thermal squeezed baths.  The expressions involve the squeezed
system operator~$c_{\text{sq}}$ of \cref{eq:csq} and the homodyne angle
$\varphi_{\text{sq}}$ of \cref{eq:varphisq}.  When written in terms of the
squeezing parameters, $r$ and $\mu$, the expressions are indifferent to
whether the bath is pure or thermal; when written in terms of the bath
parameters, $N$, $M$, and $N_{\text{th}}$, the expressions use
Eqs.~(\ref{eq:Nthermalsqueezed}) and~(\ref{eq:Mthermalsqueezed}), applicable
for a squeezed thermal bath, to convert from the squeezing parameters to the
bath parameters.  To specialize the same expressions to a pure squeezed bath,
one sets $N_{\text{th}}=0$.

Now the formulas:
\begin{align}
\cosh^2\!r&=\frac{N+N_{\text{th}}+1}{2N_{\text{th}}+1}\,,\\
\sinh^2\!r&=\frac{N-N_{\text{th}}}{2N_{\text{th}}+1}\,,\\
e^{2i\mu}\sinh r\cosh r&=-\frac{M}{2N_{\text{th}}+1}\,,
\end{align}
\begin{align}\label{eq:Ldef}
L&=1+2\sinh^2\!r-2\cos2\mu\,\sinh r\cosh r=\frac{2N+2M_R+1}{2N_{\text{th}}+1}=\frac{L'}{2N_{\text{th}}+1}\,,\\
L'&=(2N_{\text{th}}+1)L=2N+2M_R+1\;,\label{eq:Lprimedef}
\end{align}
\begin{align}
\begin{split}
e^{i\varphi_{\text{sq}}}c_{\text{sq}}
&=\frac{c\,(\cosh^2\!r-e^{-2i\mu}\sinh r\cosh r)+c\dg(e^{2i\mu}\sinh r\cosh r-\sinh^2\!r)}{\sqrt L}\\
&=\frac{c\,(N+N_{\text{th}}+M^*+1)-c\dg(N-N_{\text{th}}+M)}{\sqrt{2N_{\text{th}}+1}\sqrt{L'}}\,,
\end{split}
\end{align}
\begin{align}
\begin{split}
c_{\text{sq}}\rho c\dg_{\text{sq}}
&=c\rho c\dg\cosh^2\!r+c\dg \rho c\,\sinh^2\!r
+c\rho c\,e^{-2i\mu}\sinh r\cosh r+c\dg\rho c\dg e^{2i\mu}\sinh r\cosh r \\
&=\frac{c\rho c\dg(N+N_{\text{th}}+1)+c\dg \rho c\,(N-N_{\text{th}})-c\rho c\,M^*-c\dg\rho c\dg M}{2N_{\text{th}}+1}\,,
\end{split}\\
\begin{split}
c\dg_{\text{sq}}c_{\text{sq}}
&=c\dg c\,\cosh^2\!r+c c\dg\sinh^2\!r
+c^2e^{-2i\mu}\sinh r\cosh r+{c\dg}^2 e^{2i\mu}\sinh r\cosh r \\
&=\frac{c\dg c\,(N+N_{\text{th}}+1)+c c\dg(N-N_{\text{th}})-c^2\,M^*-{c\dg}^2 M}{2N_{\text{th}}+1}\,,
\end{split}\\
\D{c_{\text{sq}}}\rho
&=\frac{1}{2N_{\text{th}}+1}
\!\left(
(N+N_{\text{th}}+1)\mathcal{D}[c]\rho+(N-N_{\text{th}})\mathcal{D}[c\dg]\rho
+\frac12 M^*[c,[c,\rho]]+\frac12 M[c\dg,[c\dg,\rho]]
\right)\,,
\end{align}
\begin{align}
\begin{split}
c\dg_{\text{sq}}\rho c_{\text{sq}}
&=c\dg\rho c\,\cosh^2\!r+c\rho c\dg\sinh^2\!r
+c\rho c\,e^{-2i\mu}\sinh r\cosh r+c\dg\rho c\dg e^{2i\mu}\sinh r\cosh r \\
&=\frac{c\dg\rho c\,(N+N_{\text{th}}+1)+c\rho c\dg(N-N_{\text{th}})-c\rho c\,M^*-c\dg\rho c\dg M}{2N_{\text{th}}+1}\,,
\end{split}\\
\begin{split}
c_{\text{sq}}c\dg_{\text{sq}}
&=c c\dg\cosh^2\!r+c\dg c\,\sinh^2\!r
+c^2\,e^{-2i\mu}\sinh r\cosh r+{c\dg}^2 e^{2i\mu}\sinh r\cosh r \\
&=\frac{c c\dg(N+N_{\text{th}}+1)+c\dg c\,(N-N_{\text{th}})-c^2 M^*-{c\dg}^2 M}{2N_{\text{th}}+1}\,,
\end{split}\\
\D{c\dg_{\text{sq}}}\rho
&=\frac{1}{2N_{\text{th}}+1}
\!\left(
(N+N_{\text{th}}+1)\mathcal{D}[c\dg]\rho+(N-N_{\text{th}})\mathcal{D}[c]\rho
+\frac12 M^*[c,[c,\rho]]+\frac12 M[c\dg,[c\dg,\rho]]
\right)\,.
\end{align}

\section{Mixed squeezed states}\label{sec:mixed-squeezed}

To evaluate potential qubit models for mixed squeezed states, it is convenient
to derive necessary and sufficient conditions on the combination of bath
state~$\sigma$, probe operator~$a$, and measured observable for reproducing the
stochastic evolution for homodyne detection, much as
Eqs.~(\ref{eq:qubit-bath-stats}) provide necessary and sufficient conditions on
the bath state and probe operator for reproducing the unconditional evolution.
Since the mixed nature of the bath generally introduces mixing into the system
even when the bath is monitored, it is necessary to have at least four Kraus
operators; this allows for two measurement outcomes and a conditional evolution
for each of those outcomes, which is coarse-grained over two different Kraus
evolutions corresponding to the other two outcomes.  These Kraus operators can
arise from a four-level probe in a pure initial state
$\ket{\Phi}$ and observable eigenvectors $\ket{\pm\tpm}$ (where the $\tpm$
degree of freedom is coarse-grained over to reflect incomplete information) or from
a pair of two-level probes in a mixed initial state
$\lambda_{\tplus}\oprod{\psi_{\tplus}}{\psi_{\tplus}}+
\lambda_{\tminus}\oprod{\psi_{\tminus}}{\psi_{\tminus}}$ and observable
eigenvectors $\ket{\pm}$. When we use these arrangements with the interaction
unitary
\begin{align}
  U_I(a)&\defined\exp\left[\sqrtdt\left(c\otimes a\dg
  -c\dg\otimes a\right)\right]\,,
  \label{eq:squeeze-interaction-unitary}
\end{align}
where $a$ is a finite-dimensional operator analogous to the field annihilation
operator $b$, we parametrize the resulting Kraus operators up to $\BigO(\dt)$ as
\begin{align}
\begin{split}
  K_{\pm\tpm}
  &=\expt{U_I}_{\pm\tpm} \\
  &=\alpha_{\pm\tpm}+\sqrtdt\!\left(\beta_{0\pm\tpm}\,c
  +\beta_{1\pm\tpm}\,c\dg \right)
  +\dt\!\left(\gamma_{0\pm\tpm}\,c^2
  +\gamma_{1\pm\tpm}\,cc\dg
  +\gamma_{2\pm\tpm}\,c\dg c
  +\gamma_{3\pm\tpm}\,{c\dg}^2 \right)\,.
  \label{eq:gen-kraus-operators}
\end{split}
\end{align}
Here we introduce the notation that for any operator $A$,
$\expt{A}_{\pm\tpm}\defined\bra{\pm\tpm}A\ket{\Phi}$ for the case of a pure bath with
four measurement outcomes and
$\expt{A}_{\pm\tpm}\defined\sqrt{\lambda_{\tpm}}\bra{\pm}A\ket{\psi_{\tpm}}$ for a
mixed bath with two measurement outcomes.  With this notation, the various
parameters become
\begin{subequations}\label{eq:gen-kraus-op-coeffs}
\begin{align}
  \alpha_{\pm\tpm}&=\vphantom{\frac12}\expt{\Id}_{\pm\tpm}\,, \\
  \beta_{0\pm\tpm}&=\vphantom{\frac12}\big\langle a^\dagger\big\rangle_{\pm\tpm}\,, &
  \beta_{1\pm\tpm}&=-\big\langle a\big\rangle_{\pm\tpm}\,, \\
  \gamma_{1\pm\tpm}&=-\frac12\big\langle a^\dagger a\big\rangle_{\pm\tpm}\,, &
  \gamma_{2\pm\tpm}&=-\frac12\big\langle aa^\dagger\big\rangle_{\pm\tpm}\,, \\
  \gamma_{0\pm\tpm}&=\frac12\big\langle {a^\dagger}^2\big\rangle_{\pm\tpm}\,, &
  \gamma_{3\pm\tpm}&=\frac12\big\langle a^2\big\rangle_{\pm\tpm}\,.
\end{align}
\end{subequations}
Just like the case of a thermal bath, the updated state is calculated by
coarse-graining over one of the two binary variables in the measurement
outcome, giving
\begin{equation}
  \rho_\pm=\frac{K_{\pm\tplus}^{}\rho K_{\pm\tplus}\dg+K_{\pm\tminus}^{}\rho K_{\pm\tminus}\dg}
  {\Tr[(E_\pm\rho]}\,,
  \label{eq:gen-kraus-state-update}
\end{equation}
where
$E_\pm=K_{\pm\tplus}\dg K_{\pm\tplus}+K_{\pm\tminus}\dg\rho K_{\pm\tminus}$.

If we calculate the conditional difference equation from the parametrized Kraus
operators~(\ref{eq:gen-kraus-operators}), we can match terms to the
squeezed-bath conditional difference equation~(\ref{eq:sqth-difference-eq}) and
come up with a set of constraints on the Kraus parameters.  For the $\alpha$
parameters, we get
\begin{subequations}
\begin{align}
\alpha_{\pm\tplus}&=\frac{1}{\sqrt2}\cos\phi_\pm\,, \\
\alpha_{\pm\tminus}&=\frac{1}{\sqrt2}\sin\phi_\pm\,,
\end{align}
\end{subequations}
where we use the phase freedom inherent in the Kraus operators to make
$\alpha_{\pm\tpm}\ge0$, \ie, $0\leq\phi_\pm\leq\pi/2$.  The constraints
for the $\beta$ parameters are
\begin{subequations}
\begin{align}
\label{eq:beta_0}
\left\vert\beta_{0\pm\tplus}\right\vert^2+\left\vert\beta_{0\pm\tminus}\right\vert^2&=\frac{N+1}{2}\,, \\
\label{eq:beta_1}
\left\vert\beta_{1\pm\tplus}\right\vert^2+\left\vert\beta_{1\pm\tminus}\right\vert^2&=\frac{N}{2}\,, \\
\beta_{1\pm\tplus}\beta_{0\pm\tplus}^*+\beta_{1\pm\tminus}\beta_{0\pm\tminus}^*&=-\frac{M}{2}\,, \\
\beta_{0\pm\tplus}\cos\phi_\pm+\beta_{0\pm\tminus}\sin\phi_\pm&=\pm\frac{N+M^*+1}{\sqrt{2L'}}\,, \\
\beta_{1\pm\tplus}\cos\phi_\pm+\beta_{1\pm\tminus}\sin\phi_\pm&=\mp\frac{N+M}{\sqrt{2L'}}\,,
\end{align}
\end{subequations}
where $L'=2N+M+M^*+1$ as in \cref{eq:Lprime}.  The nature of the constraints on
the $\gamma$ variables always allows appropriate values to be found, given any
solution of the above equations:
\begin{subequations}
\begin{align}
\gamma_{1\pm\tpm}&=-\frac{N}{4\sin(\phi_\pm+\pi/4)}\,, \\
\gamma_{2\pm\tpm}&=-\frac{N+1}{4\sin(\phi_\pm+\pi/4)}\,, \\
\gamma_{3\pm\tpm}&=-\frac{M}{4\sin(\phi_\pm+\pi/4)}\,, \\
\gamma_{0\pm\tpm}&=-\frac{M^*}{4\sin(\phi_\pm+\pi/4)}\,.
\end{align}
\end{subequations}

\subsection{Araki-Woods}

One technique that presents itself for the squeezed thermal case is the
Araki-Woods construction, employed in \cite{Goug03} to treat general Gaussian
states with a vacuum-based technique. This construction transforms the bath
statistics from the probe state to the field operator in a slightly different
manner than the successful technique and puts the probe into a two-mode vacuum.
The form of the updated field operators is
\begin{align}
  b_{\text{AW}}&\defined x(\Id\otimes b)+y(b\dg\otimes\Id)+z(b\otimes\Id)\,,
\end{align}
corresponding to a qubit model with probe initial state $\ket{gg}$ and updated qubit
field operators
\begin{equation}
  a_{\text{AW}}^{}\defined x(\Id\otimes\sigma_-)+y(\sigma_+\otimes\Id)+z(\sigma_-\otimes\Id)\,,
  \label{eq:araki-woods-operator}
\end{equation}
where the constants $x$, $y$, and $z$ are defined as
\begin{subequations}\label{eq:araki-woods-consts}
\begin{align}
  x &\defined\sqrt{N+1-\abs{M}^2/N}\,, \\
  y &\defined\sqrt{N}\,, \\
  z &\defined M/\sqrt{N}\,.
\end{align}
\end{subequations}
The above definitions are slightly different from those presented in \cite{Goug03}, as
we have suppressed the mean field term, it being trivial to include such a term, as is
described at the end of \cref{sec:squeezing}, and changed the ordering of the
subsystems to reflect our notational conventions.

When the bath is in a pure state, $\abs{M}^2=N(N+1)$ and thus $x=0$. Since the
only term in $a_{\text{AW}}^{}$ that involves the second  is proportional to
$x$, for a pure bath we only need to consider the first field mode.  In this
case, using $a_{\text{AW}}^{}$ in the interaction unitary and measuring
$a_{\text{AW}}^{}+a_{\text{AW}}\dg$ gives the Kraus operators
\cref{eq:sq-pure-kraus-ops} and therefore produces the correct stochastic
evolution.

Unfortunately, even though the Araki-Woods discrete field operators give the
appropriate bath statistics and thus the correct unconditional evolution even
for $\abs{M}^2<N(N+1)$, they do not satisfy the constraints given above to
produce the correct conditional evolution.  In particular, the Araki-Woods Kraus
operators give us
\begin{subequations}
\begin{align}
  \cos\phi_\pm&=\sin\phi_\pm=\frac{1}{\sqrt{2}}\,,\\
  \beta_{0\pm\tplus}+\beta_{0\pm\tminus}
  &=\pm\frac{N+M^*+1}{\sqrt{L^\prime}}\sqrt{\frac{N+2M_R+\abs{M}^2/N}{2N+2M_R+1}}\,.
\end{align}
\end{subequations}
This only satisfies the SME constraints when $\vert M\vert^2=N(N+1)$, \ie,
when the bath is in a pure state.

One indication that something might break in the mixed case is that the field homodyne
observable, $b_{\text{AW}}^{}+b_{\text{AW}}\dg$, ought to have degeneracy in
eigenvalues, since in the field picture the thermalization of the field can be
interpreted as entanglement with an auxiliary mode that is not measured (\ie,
we measure $(b+b\dg)\otimes\Id$), leading to degenerate subspaces for each
eigenvalue of $b+b\dg$.  In the qubit picture, the homodyne observable
$a_{\text{AW}}^{}+a_{\text{AW}}\dg$ has the spectrum
\begin{align}
  \lambda_{\pm\tpm}&=\tpm x\pm\abs{y+z}\,,
\end{align}
which is degenerate only in the case $x=0$, since
$\abs{z}\leq\sqrt{y^2+1}\leq y$ from \cref{eq:gauss-param-ineq}. This condition
is met only when the bath is pure.

\subsection{Two-qubit setup analogous to two-mode squeezing}

To manufacture thermal statistics, we might think to consider the thermal state
of a bath mode as the marginal state of a two-mode squeezed state.  Then, much
as we did in the pure-state case, we could transfer all squeezing from the bath
state to the field operators and on to analogous qubit operators.  Using the
two-mode squeeze operator,
\begin{align}
  S^{(1,2)}(r_{\text{th}})&\defined\exp\left[r_{\text{th}}\left(b\otimes b
  -b\dg\!\otimes b\dg\right)\right]\,,
  \label{eq:two-mode-squeeze}
\end{align}
gives us the squeezed field operator
\begin{align}
  b_{\text{sq}}&=
  {S^{(12)}}\dg\!(r_{\text{th}})\,{S^{(1)}}\dg\!(r,\phi)\,(b\otimes\Id)\,
  S^{(1)}(r,\phi)\,S^{(12)}(r_{\text{th}}) \\
  &=\cosh r\cosh r_{\text{th}}(b\otimes\Id)
  -e^{2i\phi}\cosh r\sinh r_{\text{th}}(\Id\otimes b\dg)
  -e^{2i\phi}\sinh r\cosh r_{\text{th}}(b\dg\otimes\Id)
  +\sinh r\sinh r_{\text{th}}(\Id\otimes b)\,,
\end{align}
which translates to the squeezed qubit operators
\begin{align}
  a_{\text{sq}}&=\cosh r\cosh r_{\text{th}}(\sigma_-\otimes\Id)
  -e^{2i\phi}\cosh r\sinh r_{\text{th}}(\Id\otimes\sigma_+)
  -e^{2i\phi}\sinh r\cosh r_{\text{th}}(\sigma_+\otimes\Id)
  +\sinh r\sinh r_{\text{th}}(\Id\otimes\sigma_-)\,.
\end{align}
The quadrature operator for homodyne measurement,
$a_{\text{sq}}+a_{\text{sq}}\dg$, has a problem similar to that of the Araki-Woods
quadrature operator in that its eigenvalues are nondegenerate:
\begin{align}
  \lambda_{1\pm}&=\pm\frac{1}{\sqrt{2}e^r}\sqrt{-e^{4r}e^{2r_{\text{th}}}
  \cos 2\phi+e^{4r}e^{2r_{\text{th}}}+e^{2r_{\text{th}}}\cos 2\phi
  +e^{2r_{\text{th}}}}\,, \\
  \lambda_{2\pm}&=\pm\frac{1}{\sqrt{2}e^re^{r_{\text{th}}}}
  \sqrt{-e^{4r}\cos 2\phi+e^{4r}+1+\cos 2\phi}\,.
\end{align}

This construction fails to satisfy the Kraus-operator constraints on thermal stochastic
evolution even in the absence of squeezing ($N>0$, $M=0$), confirming our suspicion
based on the nondegeneracy of the observable eigenvalues.

\subsection{Two-qubit squeezed-thermal state}

Another simple two-qubit setup uses the probe annihilation operator
\begin{align}
  a_{\text{sq}}&=\sqrt{\frac{2N+1}{2}}(\Id\otimes\sigma_-+\sigma_-\otimes\Id)
  \end{align}
in conjunction with the interaction unitary $U(a_{\text{sq}})$ from
\cref{eq:squeeze-interaction-unitary} and the initial probe state
\begin{align}
  \sigma_{\text{sq}}&=\frac{1}{2N+1}\begin{pmatrix}
  N & 0 & 0 & M \\
  0 & 0 & 0 & 0 \\
  0 & 0 & 0 & 0 \\
  M^* & 0 & 0 & N+1\end{pmatrix}\,,
  \label{eq:two-qubit-squeezed-state}
\end{align}
which we consider as it is a positive state precisely when the parameters $M$
and $N$ satisfy the usual constraint $\vert M\vert^2\leq(N+1)N$ (obtaining
purity only when $\vert M\vert^2=(N+1)N$).  By observation or by calculation, we
see that the bath density operator has rank two, \ie, has support only on a qubit
subspace.

The consequences of this model are analogous to those of the Araki-Woods
construction: unconditional statistics are reproduced, but the stochastic
evolution is incorrect when measuring $a_{\text{sq}}+a_{\text{sq}}\dg$.
The difficulty again appears to be a lack of degeneracy in the eigenvalues of the
$a_{\text{sq}}+a_{\text{sq}}\dg$.  Specifically, this setup reproduces the correct bath
statistics, but the observable $a_{\text{sq}}+a_{\text{sq}}\dg$ has three
distinct eigenvalues (unique positive and negative eigenvalues with a twofold
degeneracy corresponding to an eigenvalue of 0) instead of degenerate positive
and negative subspaces as we expect from the field case.

Na\"ively pairing half of the zero subspace with both the positive and negative
outcomes yields a SME in the pure case very close to the correct result, except
that the Wiener process $\df W$ is divided by $\sqrt{2(2N+1)}$ instead of
$\sqrt{L'}$; this corresponds to doing homodyne detection on a pure squeezed
bath with detectors having subunity efficiency.  Likewise, setting $M=0$ for a
thermal bath with no squeezing yields a thermal SME with an extra factor of
$1/\sqrt{2}$ in the stochastic term, again analogous to subunity detection
efficiency.

This inefficiency makes sense given that we na\"ively lumped distinguishable
measurement outcomes together. Unfortunately, this model doesn't provide clear
alternative recipes with which to construct a SME, so we don't consider this
model any further.

\subsection{Qutrit}

One can also mock up a bath with three-level probes and field operators that
give the correct unconditional evolution. We define the qutrit probe
annihilation operator to be
\begin{equation}
\label{eq:qt_low_op}
a:=\sqrt{2N+1}\big(\ket{0}\bra{1}+\ket{1}\bra{2}\big)\,.
\end{equation}
The thermal and squeezed qutrit probe state we choose, following the reasoning
by which we arrived at \cref{eq:two-qubit-squeezed-state}, is
\begin{equation}
\sigma_\text{sq}:=\frac{1}{2N+1}
\begin{pmatrix}
N & 0 & M \\
0   & 0 & 0 \\
M^* & 0 & N+1
\end{pmatrix}\,.
\end{equation}
In our matrix representations we have ordered the rows and columns starting from
the top and left with $\ket{2}$ and decreasing to $\ket{0}$ as we move to the
bottom and right.

The combination of the above lowering operator and state gives the correct
unconditional master equation. Three-level systems present even more difficulty
in understanding what to do with the conditional evolution, however, as the
restriction to three Kraus operators means only one of the measurement results
can be coarse-grained over multiple (two) Kraus operators, leaving the other
measurement result only associated with a single Kraus operator and therefore
producing no statistical mixing of the system state.

\bibliography{contin_mes_refs}

\end{document}